\documentclass[a4paper,11pt]{article}

\usepackage{jheppub}
\usepackage{amssymb,amsfonts,amsmath}
\usepackage{dsfont}

\newcommand{\cN}{{\cal N}}

\title{Massless higher-spin supermultiplets in 5D harmonic superspace}

\author[a]{Evgeny I. Buchbinder,}
\author[a]{Sergei M. Kuzenko,}
\author[a]{and Igor B. Samsonov}

\affiliation[a]{Department of Physics M013, The University of Western Australia,\\
35 Stirling Highway, Perth W.A. 6009, Australia}

\emailAdd{evgeny.buchbinder@uwa.edu.au}
\emailAdd{sergei.kuzenko@uwa.edu.au}
\emailAdd{igor.samsonov@uwa.edu.au}

\abstract{
Off-shell actions for massless higher-spin $\mathcal{N}=1$ supermultiplets in five dimensions (5D) are constructed in harmonic superspace in terms of unconstrained prepotentials. Cubic couplings of these supermultiplets to the $q$-hypermultiplet are derived. We also provide a detailed derivation of the unconstrained prepotential for the 5D Weyl supermultiplet which has been used within the conformal superspace approach to $\mathcal{N}=1$ supergravity-matter systems. 
}

\begin{document}
\maketitle
\section{Introduction}

In the literature, there are several gauge-invariant formulations for massive higher-spin fields, see \cite{Pashnev:1989, Klishevich:1997pd, Klishevich:1998yt, Zinoviev:2001, Buchbinder:2006nu, Metsaev, Buchbinder:2008ss, Lindwasser} for an incomplete list of publications. Using such formulations, one may obtain on-shell supersymmetric massive higher-spin models by combining bosonic and fermionic models, as was achieved by Zinoviev and collaborators \cite{Zinoviev:2007js, Buchbinder:2015mta, Buchbinder:2017izy, Buchbinder:2019dof}. This approach is not efficient, however, for constructing off-shell gauge-invariant actions for massive supermultiplets. As is well known, off-shell supersymmetry has been crucial for deriving the most general supergravity-matter systems with four supercharges in $d=4$ and eight supercharges in $d=4,5,6$ dimensions, see \cite{WB, GIOS, FVP, Lauria:2020rhc} for  comprehensive reviews. Therefore one may expect that off-shell gauge-invariant formulations for massive supermultiplets may facilitate the problem of constructing consistent interactions for such fields. Is there a systematic procedure to derive such formulations? 

It is known that gauge-invariant actions for massive higher spins in $d$ dimensions may be derived by Kaluza-Klein reduction from massless gauge theories in $(d+1)$ dimensions. The corresponding conceptual setup was sketched in \cite{Aragone:1987dtt}. In the bosonic case, this setup was realized by Pashnev\footnote{The work by Pashnev \cite{Pashnev:1989} was submitted to the journal {\it Theoretical and Mathematical Physics} in August 1987, and it was published in March 1989. The work by Aragon, Deser and Yang \cite{Aragone:1987dtt} was submitted to {\it Nuclear Physics B} in April 1987. It appears that Pashnev was not aware of \cite{Aragone:1987dtt}.}  \cite{Pashnev:1989} who constructed the first gauge-invariant model for the massive integer spin-$s$ field in Minkowski space ${\mathbb M}^d$ by dimensional reduction, ${\mathbb M}^{d+1} \to {\mathbb M}^d \times S^1$, of Fronsdal's action \cite{Fronsdal:1978} (see also \cite{Curtright:1979uz}) for the massless spin-$s$ field in $(d+1)$ dimensions. Dimensional reduction of Fronsdal's bosonic \cite{Fronsdal:1978, Curtright:1979uz} and fermionic \cite{FF} actions was used to generate gauge-invariant higher-spin models in many publications \cite{Rindani:1988gb, Rindani:1989ym, Buchbinder:2008ss, Asano:2019smc, Lindwasser}. In the integer-spin case, the models derived in \cite{Klishevich:1997pd, Buchbinder:2008ss, Asano:2019smc, Lindwasser} may be seen to be equivalent to the Pashnev theory, as demonstrated in \cite{Fegebank:2024yft}. These non-supersymmetric results indicate that Kaluza-Klein reduction of massless superfield models will provide a systematic approach to generate gauge-invariant off-shell formulations for massive higher-spin supermultiplets. 

Off-shell massless higher-spin supermultiplets were constructed in four and three dimensions. In the four-dimensional case, the $\cN=1$ supermultiplets were first derived in Minkowski superspace \cite{KSP,KS} (see \cite{BK} for a pedagogical review) and then in anti-de Sitter (AdS) superspace \cite{KuzenkoSibiryakov94, Buchbinder:2018nkp}. The $\cN=1$ supersymmetric models proposed in \cite{KuzenkoSibiryakov94} were subsequently used to derive off-shell $\cN=2$ higher-spin supermultiplets in AdS${}_4$ \cite{Gates:1996my}.\footnote{Building on the models introduced in \cite{KuzenkoSibiryakov94}, the universal linearized action for massless gauge multiplets of arbitrary superspin in AdS superspace was proposed in Ref.~\cite{Gates:1996xs} as a gauge theory of unconstrained superfields taking their values in the commutative algebra of analytic functions over a one-sheeted hyperboloid in ${\mathbb R}^{3,1}$. The action is invariant under $\cN = 2$ supersymmetry transformations which form a closed algebra off the mass shell.} The higher-spin models introduced in Ref.~\cite{Gates:1996my} are invariant under $\cN = 2$ supersymmetry transformations which form a closed algebra in AdS${}_4$, but this algebra becomes open in a flat limit.\footnote{The higher-spin $\cN=2$ supermultiplets of \cite{Gates:1996my} are obtained by combining two massless $\cN=1$ supersymmetric models corresponding to superspins $s$ and $s+1/2$, with $s$ a positive integer, and thus they cannot be considered as genuine off-shell $\cN=2$ supermultiplets. The closure of the algebra of $\cN=2$ supersymmetry transformations in these models is a specific feature of the AdS${}_4$ supersymmetry, and it is also manifested in the case of $\cN=2$ supersymmetric nonlinear sigma models in AdS${}_4$ realized in terms of $\cN=1$ chiral superfields \cite{Butter:2011zt, Butter:2011kf}.} Genuine off-shell massless $\cN=2$ supermultiplets were constructed in \cite{Buchbinder:2021ite} within the harmonic superspace approach. 

In three dimensions, massless higher-spin supersymmetric models do not have any propagating degrees of freedom. However, these models can be combined with linearized superconformal higher-spin actions\footnote{Linearized $\cN$-extended superconformal gauge-invariant actions have the following general form 
\cite{Buchbinder:2019yhl}:
$ S^{( n |\cN )} [ H_{\alpha(n)} ] = ( {\rm i}^n /2)    \int {\rm d}^{3}x  {\rm d}^{2 \cal N} \theta  \, H^{\alpha(n)} 
{W}_{\alpha(n)}(H )$, with $n>0$. Here $H_{\alpha(n)} = H_{(\alpha_1 \dots \alpha_n) }$ is the unconstrained superconformal gauge prepotential and $W_{\alpha(n)}$ is the linearized super-Cotton tensor,  a unique gauge-invariant primary descendant of $H_{\alpha(n)}$. In the non-supersymmetric ($\cN=0$) case,  $W_{\alpha(n)}$ was constructed in \cite{PopeTownsend} for even $n$ and \cite{Kuzenko:2016qdw} for odd $n$. In the $\cN=1$ and $\cN=2$ cases, $W_{\alpha(n)}$ was constructed in \cite{Kuzenko:2016qdw, Kuzenko:2016qwo}  and \cite{KO}, respectively.
Explicit expressions for $W_{\alpha(n)}$ for $\cN>2$ were given in \cite{Buchbinder:2019yhl}.
} 
to generate off-shell topologically massive higher-spin supersymmetric theories, following the philosophy of topologically massive gauge theories \cite{WS,JS,DJT1,DJT2,DeserKay}. Recently, there has been much progress in the construction of such off-shell theories with various supersymmetries. It is based on the approaches developed in \cite{Kuzenko:2016qwo} for ${\cN}=1$ and in \cite{KO} for ${\cN}=2$ Poincar\'e supersymmetries. They were later extended to $\rm AdS_3$ for the following cases: $\cN=1$ \cite{Kuzenko:2018lru, HK19, Kuzenko:2021hyd}, $\cN=(1,1)$ \cite{HKO} and $\cN=(2,0)$ \cite{HK18}. Linearized actions for $\cal N$-extended higher-spin superconformal gravity were derived in \cite{Buchbinder:2019yhl}.\footnote{In four dimensions, linearized superconformal higher-spin actions were derived in \cite{KMT,KP, Kuzenko:2020jie} for $\cN=1$ and \cite{Kuzenko:2021pqm} for $\cN>1$, including the discovery of $\cN=2$ superconformal gravitino multiplet \cite{HKR}. The $\cN=2$ results of \cite{Kuzenko:2021pqm, HKR} were recast in terms of projective \cite{Kuzenko:2024vms} and harmonic  \cite{BIZ24,Ivanov:2024bsb} prepotentials.}

The massless 4D $\cN=1$ supersymmetric higher-spin models \cite{KSP,KS, KuzenkoSibiryakov94, Buchbinder:2018nkp} may be used to construct new off-shell gauge-invariant formulations for massive higher-spin supermultiplets in three dimensions. In order to obtain off-shell gauge-invariant actions for massive higher-spin supermultiplets in four dimensions, it is necessary to start with off-shell massless higher-spin models in five dimensions. To the best of our knowledge such models have not been constructed yet.\footnote{There exists an off-shell formulation for massive higher-spin $\cN=1$ supermultiplets without gauge invariance \cite{Koutrolikos:2020tel}. It is natural to think of this construction as a supersymmetric extension of the Singh-Hagen models \cite{Singh:1974, SinghHagen2}.} The main goal of this paper is to derive them building upon the 4D harmonic superspace construction of \cite{Buchbinder:2021ite}. 

The rest of the paper is organized as follows. In Sec.~\ref{Sec:sugra-overview} we give a short overview of existing superfield approaches to 5D supergravity and discuss some key results which represent a background for further sections. In Sec.~\ref{Sec:Supergravity}, we introduce prepotentials for 5D Poincar\'e supergravity in the harmonic superspace framework. We use these prepotentials for construction of the superfield action for linearized supergravity in flat Minkowski superspace and consider a minimal hypermultiplet coupling to this supergravity background. This allows us to derive conserved current superfields for the $q$-hypermultiplet model. This section serves as a starting point for the introduction of superfield prepotentials with the corresponding gauge transformations for higher-spin supermultiplets which are given in Sec.~\ref{Sec:higher-spins}. There we construct gauge-invariant quadratic actions for higher-spin supermultiplets in harmonic superspace and derive the corresponding superfield equations of motion. Next, we build a cubic interaction vertex for these supermultiplets with the $q$-hypermultiplet. This vertex allows us to derive conserved higher currents in the $q$-hypermultiplet model. Section \ref{Sec:Summary} is devoted to a discussion of other interesting open problems which may be addressed by building upon the results of the present paper. In Appendix \ref{AppA} we give a summary of our superspace notation and conventions, while Appendix \ref{App:component-structure} contains technical details of the derivation of the component structure of higher-spin prepotentials in the Wess-Zumino gauge. Finally, in Appendix \ref{AppendixC}, we present prepotentials and gauge transformations for 5D conformal supergravity.


\section{Comments on 5D superfield supergravity}
\label{Sec:sugra-overview}

General $\cN=1$ supergravity-matter systems in five dimensions are naturally described using the superconformal tensor calculus \cite{Ohashi1,Ohashi2,Ohashi3,Ohashi4,Bergshoeff1,Bergshoeff2,Bergshoeff3} (see also \cite{Lauria:2020rhc} for a pedagogical review\footnote{In the literature, $\cN=1$ supersymmetry in five dimensions is often denoted $\cN=2$, see e.g. \cite{Lauria:2020rhc, Ohashi1,Ohashi2,Ohashi3,Ohashi4,Bergshoeff1,Bergshoeff2,Bergshoeff3}, by analogy  with the 4D $\cN=2$ case to indicate that the number of supercharges is equal to eight.}) in which hypermultiplets are either on-shell or involve a gauged central charge. These hypermultiplet realizations cannot be used to provide an off-shell formulation for the most general locally supersymmetric sigma model. Such a sigma model formulation is known to require one to use off-shell hypermultiplets possessing an infinite number of auxiliary fields, which make them extremely difficult to work with at the component level. This technical problem was solved within the superspace approach to 5D $\cN=1$ supergravity-matter systems  \cite{KT-M_5D2,KT-M_5D3,KT-M08} by putting forward the concept of covariant projective multiplets. These supermultiplets are a curved superspace extension of the 4D $\cN=2$ and 5D $\cN=1$ superconformal projective multiplets \cite{Kuzenko2006,K07}. The latter reduce to the off-shell projective multiplets pioneered by Lindstr\"om and Ro\v{c}ek \cite{KLR,LR1,LR2} in the 4D $\cN=2$ super-Poincar\'e case and  generalized to the cases of 5D $\cN=1$ Poincar\'e and AdS supersymmetries in \cite{KL} and \cite{Kuzenko:2007aj}, respectively. 

The powerful virtues of the superconformal tensor calculus and of the superfield supergravity are naturally combined in the framework of 5D $\cN=1$ conformal superspace \cite{Butter:2014xxa} which was obtained by gauging the 5D superconformal algebra in superspace. This work reproduced practically all off-shell constructions derived at the time and provided a plethora of new results. In particular, it offered covariant superspace realizations for all off-shell formulations of 5D $\cN=1$ supergravity with the Weyl supermultiplet coupled to two compensating supermultiplets.\footnote{Numerous applications of the 5D conformal superspace approach have appeared in Refs.~\cite{Hutomo:2022hdi, Gold:2023dfe, Gold:2023ymc, Gold:2023ykx, Gold:2025ttt}.} The supergravity equations of motion were obtained in Ref.~\cite{Butter:2014xxa} by making use of the fact that the 5D $\cN=1$ Weyl multiplet is described in superspace by a single unconstrained real prepotential, $\mathfrak U$, similarly to the 4D $\cN=2$ case \cite{HST,KT,Butter:2010sc}. Given a supersymmetric theory of (matter) superfields $\varphi^i$ coupled to the Weyl multiplet, the supercurrent of this theory is a (dimension-3 primary) real scalar superfield defined by 
\begin{align}
{\cal J} = \frac{\Delta }{\Delta {\mathfrak U}}   S[ \varphi ] ~,
\end{align}
where $S[ \varphi ] $ is the action and
$\Delta / {\Delta {\mathfrak U}}$ denotes a covariantized variational derivative with respect to $\mathfrak U$, see Ref.~\cite{Butter:2014xxa} and Appendix \ref{App:C3} for more details.

In the case of 4D $\cN=2$ supergravity, the unconstrained prepotential $\mathfrak U$ for the Weyl multiplet originates most elegantly \cite{KT,Butter:2010sc} within the harmonic superspace approach \cite{GIKOS,Galperin:1987em,Galperin:1987ek}. The same approach has recently been used to construct linearized off-shell actions for 4D $\cN=2$ massless higher-spin supermultiplets \cite{Buchbinder:2021ite,Buchbinder:2022kzl,Buchbinder:2022vra} since the supergravity prepotentials in the harmonic superspace possess a natural generalization to higher-spin supermultiplets. Therefore, in the next section, we develop prepotentials for linearized $\cN=1$ Poincar\'e supergravity in 5D harmonic superspace and construct the corresponding quadratic action for the spin-2 supermultiplet.\footnote{In this paper, a supermultiplet which contains bosonic fields of integer spins $s$ and $s-1$, as well as one Dirac-type fermionic field of spin $s-1/2$ on shell is referred to as a spin-$s$ supermultiplet.} In Section \ref{Sec:higher-spins}, this formulation will be extended for the description of higher-spin multiplets in 5D harmonic superspace. The hypermultiplet coupling to the 5D Weyl multiplets is discussed in Appendix \ref{AppendixC}.

So far, the harmonic superspace formulation for 4D $\cN=2$ supergravity-matter systems has not yet been extended to five dimensions. Such a generalization may be obtained following the original 4D approach \cite{GIKOS,Galperin:1987em,Galperin:1987ek} (see \cite{GIOS} for a review) or by extending Butter's formalism for curved harmonic superspace \cite{Butter:2015nza}. These developments will be discussed elsewhere. Here we will restrict our analysis to the linearized version of one of the off-shell versions of 5D $\cN=1$ Poincar\'e supergravity. This version proves to provide a simplest uplift to massless higher-spin supermultiplets. 

As mentioned above, off-shell formulations for 5D minimal supergravity are obtained by coupling the Weyl multiplet to two compensating supermultiplets. One of the compensators is a vector supermultiplet. In the harmonic superspace framework, prepotentials for the linearized Weyl multiplet are given by the superfields $h^{++m}$, $h^{++\hat\alpha+}$ and $h^{(+4)}$ with constraints and gauge transformations described in Appendix \ref{AppendixC}. The coupling to the vector multiplet compensator is accounted for by adding the prepotential $h^{++5}$ such that the covariant harmonic derivative becomes
\begin{equation}
    {\cal D}^{++} = D^{++} + h^{++ m}\partial_m + h^{++\hat\alpha +}D^-_{\hat\alpha} + h^{(+4)}D^{--}
    + h^{++5}\partial_5\,,
\label{eq2.2}
\end{equation}
where $D^{++}$ and $D^{--}$ are flat harmonic derivatives and $\partial_5$ is the central charge operator. The functional form of $h^{++5}$ is fixed by the requirement that the operator (\ref{eq2.2}) preserves the analyticity of a massive $q$-hypermultiplet (with a non-vanishing central charge).

The second compensating multiplet will be chosen to be a nonlinear multiplet which was introduced in Ref.~\cite{deWvanHvanP} in the case of 4D $\cN=2$ supergravity.\footnote{Comprehensive discussions of the nonlinear multiplets can be found in \cite{Butter:2015nza, Kuzenko:2025bud}.} Within the 5D conformal superspace approach \cite{Butter:2014xxa}, the nonlinear multiplet is described by a matrix superfield $L_{\mathbf a}^i$, where $i$ is an $SU(2)_R$ index while $\mathbf{a}$ corresponds to an external $SU(2)$ group. It satisfies the reality condition $\overline{L_{\mathbf a}^i} = - L^{\mathbf a}_i$, and is characterized by the nonlinear constraints
\begin{equation}
\det L = 1 \,, \qquad L^{\mathbf{a} (i } \nabla^{j}_{\hat \alpha} L^{ k) }_{ \mathbf{a} } = 0\,,
\label{NLMconstraint}
\end{equation}
which are superconformal provided that $L_{\mathbf a}^i$ is primary and dimensionless. The action for 5D $\cN=1$ Poincar\'e supergravity in this setting is proportional to the vector multiplet action with the wrong sign. Several equivalent expressions for the supergravity action exist in the literature, including the following
\cite{Butter:2014xxa}:
\begin{align}
S_{\rm SG} =
\frac{1}{4}  \int \mathrm{d}^{5|8}z\, E\,   V_{ij} { H }_{\rm VM}^{ij} \,,\qquad
 H_{\rm VM}^{ij} = \frac{\mathrm{i}}{2} W \nabla^{ij}  W + \mathrm{i} (\nabla^{\hat{\alpha} (i } W) \nabla_{\hat \alpha}^{j)} W \,,
 \label{SGaction}
\end{align}
where  $V_{ij}$ is Mezincescu's prepotential\footnote{Within the conformal superspace framework of \cite{Butter:2014xxa}, 
Mezincescu's prepotential \cite{Z2}, $V_{ij} =V_{ji}$, is a real primary superfield of dimension $-2$ defined modulo transformation (6.50) in \cite{Butter:2014xxa}.
} for the vector multiplet,   $W$ is the field strength of the vector multiplet, and $\nabla^{ij} = \nabla^{\hat{\alpha} (i }\nabla_{\hat \alpha}^{j)}$, see \cite{Butter:2014xxa} for the technical details. The nonlinear multiplet appears in the action \eqref{SGaction} implicitly in the sense that the constraints \eqref{NLMconstraint} express some component fields of the Weyl multiplet in terms of those belonging to the nonlinear multiplet.
The curved superspace geometry, which corresponds to the supergravity formulation under consideration, is completely specified by two curvature tensors: a spinor superfield $\Xi_{\hat \alpha}^i$ and a super Weyl tensor 
$W_{ab }= -W_{ba}$.\footnote{This geometry may be obtained from 5D conformal superspace by analogy with the 4D ${\cal N}=2$ analysis of \cite{Kuzenko:2025bud}, and is similar to the 4D $\cN=2$ superspace geometry introduced in \cite{Howe, Castellani:1980cu, Gates}.} The supergravity equation of motion corresponding to \eqref{SGaction} is
\begin{align}
\Xi_{\hat \alpha}^i =0~.
\label{spinorEoM}
\end{align}

In the harmonic superspace framework \cite{GIOS,Galperin:1987ek}, the non-linear multiplet is described by an analytic superfield $N^{++}$ satisfying the nonlinear constraint\footnote{The relationship between $N^{++}$ and $L_{\mathbf a}^i$ is described in \cite{Butter:2015nza}.} 
\begin{equation}
  {\cal D}^{++} N^{++} + (N^{++})^2 - h^{(+4)} = 0\,,
  \label{nonlinear}
\end{equation}
and transforming under the conformal supergravity gauge group by the rule
\begin{equation}
    \delta N^{++} = \lambda^{++}\,,
    \label{deltaN}
\end{equation}
where $\lambda^{++}$ is an analytic superfield parameter described in detail in Appendix \ref{AppendixC}. The second term in (\ref{nonlinear}) should be dropped in the linearized theory. The gauge freedom (\ref{deltaN}) may be used to choose the gauge 
\begin{equation}
  N^{++} = 0\,,
\end{equation}
which implies $\lambda^{++} = 0$. In this gauge, one of the conformal supergravity prepotentials vanishes,
\begin{equation}
  h^{(+4)} = 0 \,,
\end{equation}
as a corollary of the constraint (\ref{nonlinear}). Thus we are left with the prepotentials $h^{++m}$, $h^{++\hat\alpha+}$ and $h^{++5}$, which describe 5D Poincar\'e supergravity in the harmonic superspace. In the next section, we will study this supergravity in detail in the linearized case.


\section{Linearized 5D minimal supergravity in harmonic superspace}
\label{Sec:Supergravity}

We introduce the prepotential formulation for linearized 5D $\cN=1$ Poincar\'e supergravity by analogy with the 4D $\cN=2$ analysis by Zupnik \cite{Zupnik:1998td}.

\subsection{Prepotentials}
\label{Sec:s2-prepotentials}

5D $\cN=1$ harmonic superspace with central charge may be parametrized by the coordinates $(x^m,\theta^\pm_{\hat\alpha},u^\pm_i,x^5)$, where $m=0,1,2,3,4$ is the Lorentz index while $\hat\alpha=1,2,3,4$ is the spinor index, and $x^5$ is the central charge coordinate. The $SU(2)_R$ harmonics $u^\pm_i$ obey the defining relations $u^{+i}u^-_j - u^{-i}u^+_j = \delta^i_j$, $u^{+i}u^+_i = u^{-i}u^-_i=0$. Associated with the harmonic coordinates are the covariant harmonic derivatives $D^{++}$, $D^{--}$ and $D^0$ with commutation relations of the $su(2)$ Lie algebra
\begin{equation}
  [D^{++}, D^{--}] = D^0\,,\quad
  [D^0, D^{++}] = 2D^{++}\,,\quad
  [D^0, D^{--}] = -2D^{--}\,.
\label{harmonic-derivatives-algebra}
\end{equation}
Covariant spinor derivatives $D^\pm_{\hat\alpha}$ obey the commutation relations of the 5D $\cN=1$ superalgebra with central charge:
\begin{equation}
  \{ D^+_{\hat\alpha} , D^-_{\hat\beta} \} = 2\mathrm{i} \Gamma^{m}_{\hat\alpha\hat\beta} \partial_{m} + 2\mathrm{i}\varepsilon_{\hat\alpha\hat\beta}\partial_5\,,\qquad
  \{ D^+_{\hat\alpha} , D^+_{\hat\beta} \} = \{ D^-_{\hat\alpha} , D^-_{\hat\beta} \} = 0\,.
\label{D-algebra}  
\end{equation}
For more details of our superspace notation and conventions see Appendix \ref{AppA}.

By analogy with the 4D case \cite{Galperin:1987em,Galperin:1987ek}, the spin-2 supermultiplet in harmonic superspace may be described by a set of prepotentials $h^{++M} := (h^{++m},h^{++\hat\alpha+},h^{++5})$ which comprise the following differential operator\footnote{Associated with the covariant spinor derivative $D^+_{\hat\alpha}$ there is also a prepotential $h^{++\hat\alpha-}$. This prepotential, however, plays no role in the supergravity model because it can be safely gauged away. We will ignore analogous prepotentials in our construction of actions for massless higher spins in the 5D harmonic superspace.}
\begin{equation}
  H^{++}:= h^{++M} D_M = h^{++m}\partial_m + h^{++\hat\alpha+}D^-_{\hat\alpha} + h^{++5}\partial_5\,,
  \label{H-definition}
\end{equation}
where $D_M = (\partial_m , D^-_{\hat\alpha},\partial_5)$. By construction, the operator (\ref{H-definition}) maps an analytic superfield $\Phi_A$ ($D^+_{\hat\alpha}\Phi_A=0$),  into another one, i.e., $D^+_{\hat\alpha}(H^{++}\Phi_A) = 0$. As a corollary, the prepotentials $h^{++M}$ are constrained by the analyticity conditions
\begin{subequations}
\label{h-analyticity-properties}
\begin{align}
  D^+_{\hat\alpha} h^{++m} = &\,2\mathrm{i}\Gamma^m_{\hat\alpha\hat\beta}h^{++\hat\beta+}\,,\\
  D^{+\hat\alpha} h^{++5} =&\, 2\mathrm{i}h^{++\hat\alpha+}\,,\\
  D^+_{\hat\alpha} h^{++\hat\beta+} =&\, 0 \,.
\end{align}
\end{subequations}
These equations show that the prepotentials $h^{++m}$ and $h^{++5}$ are constrained superfields in the full superspace while $h^{++\hat\alpha+}$ is an analytic superfield. Moreover, by construction, all prepotentials are independent of the central charge variable,
\begin{equation}
  \partial_5 h^{++M} = 0\,.
\end{equation}
Only massive matter (hypermultiplet) superfields considered below may have non-trivial dependence on $x^5$.

By construction, linearized gauge transformations of the prepotentials $h^{++M}$ are
\begin{equation}
  \delta_\lambda h^{++M} = D^{++}\lambda^M\,,
\label{h-gauge-transformations}
\end{equation}
where $\lambda^M := (\lambda^m,\lambda^{\hat\alpha+},\lambda^5)$ is a set of superfield gauge parameters subject to the analyticity constraints:
\begin{subequations}
\label{lambda-analyticity-properties}
\begin{align}
  D^+_{\hat\alpha} \lambda^{m} = &\,2\mathrm{i}\Gamma^m_{\hat\alpha\hat\beta}\lambda^{\hat\beta+}\,,\\
  D^{+\hat\alpha} \lambda_5 =&\, 2\mathrm{i} \lambda^{\hat\alpha+}\,,\\
  D^+_{\hat\alpha} \lambda^{\hat\beta+} =&\, 0 \,.
\end{align}
\end{subequations}
Modulo a term with the derivative $D^+_{\hat\alpha}$, the gauge transformation (\ref{h-gauge-transformations}) may be cast in the form
\begin{equation}
  \delta_\lambda H^{++} = [D^{++},\Lambda]\,,
\label{delta-H++}
\end{equation}
where $\Lambda$ is the following first-order differential operator
\begin{equation}
  \Lambda := \lambda^M D_M = \lambda^m\partial_m + \lambda^{\hat\alpha+}D^-_{\hat\alpha} + \lambda^{++5}\partial_5\,.
\label{spin-2-Lambda}
\end{equation}

Within the harmonic superspace formulation for supergravity, the prepotentials $h^{++M}$ have a clear geometric meaning: they represent vielbeins for a (super)diffeomorphism-covariant harmonic derivative,
\begin{equation}
  {\cal D}^{++} = D^{++} + \kappa H^{++} = D^{++} + \kappa h^{++M}D_M\,, 
\label{D++covariant} 
\end{equation}
where $\kappa$ is a coupling constant. A (super)diffeomorphism-covariant extension of the harmonic derivative $D^{--}$ may be represented in a similar form,
\begin{equation}
  {\cal D}^{--} = D^{--} + \kappa H^{--} \,,
\end{equation}
where the operator $H^{--}$ consists of the following four prepotentials:
\begin{equation}
  H^{--} = h^{--m}\partial_m + h^{--\hat\alpha+}D^-_{\hat\alpha} + h^{--\hat\alpha-}D^+_{\hat\alpha} + h^{--5}\partial_5\,.
\end{equation}
Covariant harmonic derivatives ${\cal D}^{++}$ and ${\cal D}^{--}$ should obey the commutation relations of the $su(2)$ Lie algebra isomorphic to (\ref{harmonic-derivatives-algebra}):
\begin{equation}
  [{\cal D}^{++}, {\cal D}^{--}] = D^0\,,\quad
  [D^0, {\cal D}^{++}] = 2{\cal D}^{++}\,,\quad
  [D^0, {\cal D}^{--}] = -2{\cal D}^{--}\,,
\end{equation}
where $D^0$ remains flat. These equations imply the so-called harmonic zero-curvature condition for the prepotentials,
\begin{equation}
\label{zero-curvature-full}
  [D^{++},H^{--}] - [D^{--},H^{++}] + \kappa[H^{++},H^{--}] = 0\,.
\end{equation}

For studies of linearized supergravity, it is sufficient to consider the zero-curvature condition (\ref{zero-curvature-full}) to zeroth order in the coupling constant $\kappa$. In this case, this equation may written in the unfolded form
\begin{subequations}
\label{zero-curvature-linearized}
\begin{align}
  D^{++} h^{--m,5} - D^{--} h^{++m,5} =&\, 0\,,
  \label{zero-curvature-linearized-a}\\
  D^{++} h^{--\hat\alpha+} - D^{--}h^{++\hat\alpha+} =&\,0\,,\\
  D^{++} h^{--\hat\alpha-} + h^{--\hat\alpha+} =&\,0\,,
\end{align}
\end{subequations}
and explicitly solved in terms of harmonic distributions:
\begin{subequations}
\label{h--solutions}
\begin{align}
  h^{--m,5}(z,u) =& \int \mathrm{d}u' \frac{h^{++m,5}(z,u')}{(u^+ u'^+)^2}\,,\\
  h^{--\hat\alpha+}(z,u) =& \int \mathrm{d}u' \frac{(u^+ u'^-)}{(u^+ u'^+)^2} h^{++\hat\alpha+}(z,u')\,,\\
  h^{--\hat\alpha-}(z,u) =& -\int \mathrm{d}u' \frac{(u^- u'^-)}{(u^+ u'^+)^2}h^{++\hat\alpha+}(z,u')\,.
\end{align}
\end{subequations}
It is straightforward to verify that these expressions solve the equations (\ref{zero-curvature-linearized}) by using the properties of the harmonic distributions described, e.g., in Refs.~\cite{GIOS,Galperin:1985bj}. The solutions (\ref{h--solutions}) represent spin-2 counterparts of analogous solutions to the harmonic zero-curvature condition in the Yang-Mills theory in harmonic superspace \cite{Z,Z2}. It is also possible to show that the gauge transformations (\ref{h-gauge-transformations}) imply the following transformation laws for the prepotentials (\ref{h--solutions}):
\begin{subequations}
\label{h--gauge-transformations}
\begin{align}
  \delta_\lambda h^{--m,5} =&\, D^{--}\lambda^{m,5}\,,\\
  \delta_\lambda h^{--\hat\alpha-} =&\, D^{--}\lambda^{\hat\alpha-}\,,\\
  \delta_\lambda h^{--\hat\alpha+} =&\, D^{--}\lambda^{\hat\alpha+} + \lambda^{\hat\alpha-}\,,
\end{align}
\end{subequations}
where $\lambda^{\hat\alpha-}(z,u) = -\int \mathrm{d}u' \frac{\lambda^{\hat\alpha+}(z,u'))}{(u^+ u'^+)}$ is a general solution of the equation $D^{++}\lambda^{\hat\alpha-} + \lambda^{\hat\alpha+} = 0$. 

\subsection{Unconstrained prepotentials}
The supergravity prepotentials $h^{++M}$ introduced above obey the constraints (\ref{h-analyticity-properties}). These constraint may be solved in terms of two unconstrained prepotentials, a full-superspace superfield $\Psi^{\hat\alpha-}$, and an analytic superfield $v^{++5}$, $D^+_{\hat\alpha}v^{++5}=0$:
\begin{subequations}
\label{pre-prepotentials}
\begin{align}
  h^{++m} =& -\frac{\mathrm{i}}4(\Gamma^m)^{\hat\alpha\hat\beta} (D^+)^2 D^+_{\hat\alpha}\Psi^-_{\hat\beta} \,,\\
  h^{++\hat\alpha+} =& -(D^+)^4 \Psi^{\hat\alpha-}\,,\\
  h^{++5} =& \frac{\mathrm{i}}4 (D^+)^2D^+_{\hat\alpha}\Psi^{\hat\alpha-} + v^{++5}\,.
\end{align}
\end{subequations}
The prepotentials $\Psi^{\hat\alpha-}$ and $v^{++5}$ are defined modulo gauge transformations
\begin{equation}
  \delta_\rho \Psi^{\hat\alpha-} = D^{+\hat\alpha}\rho_1^{--} + D^+_{\hat\beta} \rho_2^{--\hat\alpha\hat\beta}\,,\qquad
  \delta_\rho v^{++5} = -8\mathrm{i}(D^+)^4 \rho_1^{--}\,,
\label{s2-pre-gauge-symmetry}
\end{equation}
with two independent unconstrained superfield parameters $\rho_1^{--}$ and $\rho_2^{--\hat\alpha\hat\beta}=\rho_2^{--(\hat\alpha\hat\beta)}$. The prepotentials $h^{++M}$ remain invariant under (\ref{s2-pre-gauge-symmetry}).

Recall that the gauge parameters $\lambda^M$ obey the  constraints (\ref{lambda-analyticity-properties}) which are analogous to the ones for the prepotentials (\ref{h-analyticity-properties}). Therefore, by analogy with Eqs.~(\ref{pre-prepotentials}), the constraints (\ref{lambda-analyticity-properties}) may be resolved in the following form
\begin{subequations}
\label{pre-lambda}
\begin{align}
  \lambda^{m} =&\, -\frac{\mathrm{i}}4(\Gamma^m)^{\hat\alpha\hat\beta} (D^+)^2 D^+_{\hat\alpha} l^{---}_{\hat\beta} \,,\\
  \lambda^{\hat\alpha+} =&\, -(D^+)^4 l^{--\hat\alpha-}\,,\\
  \lambda^{5} =&\, \frac{\mathrm{i}}4 (D^+)^2D^+_{\hat\alpha} l^{--\hat\alpha-} + l^{5}\,,
\end{align}
\end{subequations}
with an unconstrained general superfield $l^{--\hat\alpha-}$ and an analytic superfield $l^5$, $D^+_{\hat\alpha}l^5 = 0$. In terms of these superfields, the $\lambda$-gauge transformations of the prepotentials $\Psi^{\hat\alpha -}$ and $v^{++5}$ are
\begin{equation}
    \delta_\lambda \Psi^{\hat\alpha-} = D^{++} l^{--\hat\alpha-}\,,\qquad \delta_\lambda v^{++5} = D^{++} l^5\,.
\label{pre-prepotentials-gauge}
\end{equation}
These variations imply the gauge transformations (\ref{h-gauge-transformations}).

As will be shown below, the unconstrained prepotentials will be necessary for deriving superfield equations of motions in the linearized supergravity and for obtaining conserved currents for the $q$-hypermultiplet. The gauge symmetries (\ref{s2-pre-gauge-symmetry}) and (\ref{pre-prepotentials-gauge}) will play crucial role in obtaining corresponding conservation equations.

\subsection{Action for the linearized Poincar\'e supergravity}

It is possible to construct the following two candidate actions quadratic in the superfields $h^{++M}$:
\begin{subequations}
\label{trial-actions}
\begin{align}
  S_1 =& \int \mathrm{d}^{5|8}z \mathrm{d}u\, h^{++m} h^{--}_m =\int \mathrm{d}^{5|8}z \mathrm{d}u \mathrm{d}u'\frac{h^{++m}(z,u) h^{++}_m(z,u')}{(u^+u'^+)^2} \,,\\
  S_2 =& \int \mathrm{d}^{5|8}z \mathrm{d}u\, h^{++5} h^{--5} = \int \mathrm{d}^{5|8}z \mathrm{d}u\mathrm{d}u'\frac{h^{++5}(z,u) h^{++5}(z,u')}{(u^+u'^+)^2}\,.
\end{align}
\end{subequations}
Here $\mathrm{d}^{5|8}z \mathrm{d}u$ is the full superspace measure which is related to the analytic superspace measure $\mathrm{d}\zeta^{(-4)}$ in the standard manner,
\begin{equation}
  \int \mathrm{d}^{5|8}z \mathrm{d}u \,{\cal L} = \int \mathrm{d}\zeta^{(-4)} (D^+)^4 {\cal L}\,,
\label{analytic-measure}
\end{equation}
where
\begin{equation}
  (D^+)^4 := -\frac1{32}(D^+)^2 (D^+)^2 = -\frac1{32}D^{+\hat\alpha}D^+_{\hat\alpha} D^{+\hat\beta}D^+_{\hat\beta}\,.
\end{equation}

Under $\lambda$-gauge transformations $\delta h^{++M} = D^{++}\lambda^M$ the actions (\ref{trial-actions}) transform as follows
\begin{subequations}
\label{variations-of-trial-actions}
\begin{align}
  \delta_\lambda S_1 =& -2 \int \mathrm{d}^{5|8}z \mathrm{d}u\, \lambda_m D^{--} h^{++m}\,,\\
  \delta_\lambda S_2 =& -2 \int \mathrm{d}^{5|8}z \mathrm{d}u\, \lambda^5 D^{--} h^{++5}\,.
\end{align}
\end{subequations}
Here we integrated by parts the harmonic derivative $D^{++}$ and made use of the harmonic zero-curvature condition (\ref{zero-curvature-linearized-a}). Next, in Eqs.~(\ref{variations-of-trial-actions}) we switch to the analytic superspace measure by the rule (\ref{analytic-measure}) and apply the following corollaries of the analyticity properties of prepotentials (\ref{h-analyticity-properties}) and gauge parameters (\ref{lambda-analyticity-properties}):
\begin{subequations}
\label{D4-identities}
\begin{align}
  (D^+)^4(\lambda_m D^{--}h^{++m}) =&\, 6\mathrm{i}\lambda^{\hat\alpha+}\Gamma^m_{\hat\alpha\hat\beta}\partial_m h^{++\hat\beta+}\,,\\
  (D^+)^4(\lambda^5 D^{--}h^{++5}) =&\,2\mathrm{i}\lambda^{\hat\alpha+}\Gamma^m_{\hat\alpha\hat\beta} \partial_m h^{++\hat\beta+}\,.
\end{align}
\end{subequations}
In deriving these relations we have used the anticommutation relations of covariant spinor derivatives (\ref{D-algebra}) and the identity $\Gamma^n \Gamma^m \Gamma_n = 3\Gamma^m$ which follows from Eq.~(\ref{3-Gamma}). As a result, the variations (\ref{variations-of-trial-actions}) reduce to the same expression in the analytic subspace,
\begin{equation}
  \delta_\lambda S_1 = 3\delta_\lambda S_2 = -12\mathrm{i} \int \mathrm{d}\zeta^{(-4)} \lambda^{\hat\alpha+} \Gamma^m_{\hat\alpha\hat\beta} \partial_m h^{++\hat\beta+}\,.
\end{equation}
Thus, modulo an overall normalization coefficient, gauge-invariant quadratic supergravity action reads
\begin{equation}
S = S_1-3S_2 = \int \mathrm{d}^{5|8}z\mathrm{d}u(h^{++m}h^{--}_m - 3h^{++5}h^{--5})\,.
\label{s2-action}
\end{equation}
This is a linearized version of the Poincar\'e supergravity action \eqref{SGaction}.
It would be interesting to develop a non-linear generalization of the supergravity action in the harmonic superspace which reduces to (\ref{s2-action}).


\subsection{Equations of motion}
Although the general variation of the action (\ref{s2-action}) has a simple form,
\begin{equation}
  \delta S = 2\int \mathrm{d}^{5|8}z \mathrm{d}u \left( \delta h^{++m} h^{--}_m - 3\delta h^{++5} h^{--5} \right)\, ,
\label{delta-S-2}
\end{equation}
it is not completely trivial to derive the corresponding equations of motion, since the prepotentials $h^{++M}$ obey the constraints (\ref{h-analyticity-properties}). However, the general solution of these constraints in terms of unconstrained superfields $\Psi^{\hat\alpha-}$ and $v^{++5}$ is given in Eqs.~(\ref{pre-prepotentials}). Substituting Eqs.~(\ref{pre-prepotentials}) into Eq.~(\ref{delta-S-2}) we find the variation of the action (\ref{s2-action}) in terms of the unconstrained prepotentials:
\begin{equation}
\label{deltaS2}
\begin{aligned}
  \delta S =&\,\frac{\mathrm{ i}}2\int \mathrm{d}^{5|8}z \mathrm{d}u\left( (\Gamma^m)^{\hat\alpha}{}_{\hat\beta}\delta\Psi^{\hat\beta-} (D^+)^2 D^+_{\hat\alpha} h^{--}_m  
  - 3 \delta \Psi^{\hat\beta-} (D^+)^2 D^+_{\hat\beta}h^{--5}\right)\\
  &-6\int \mathrm{d}\zeta^{(-4)}\delta v^{++5} (D^+)^4 h^{--5}\,.
\end{aligned}
\end{equation} 
Let us consider the variational derivative of the action
with respect to $\Psi^{\hat\alpha-}$,
\begin{equation}
  \Xi^+_{\hat\alpha} := 2\mathrm{i}\frac{\delta S}{\delta\Psi^{\hat\alpha-}} = (\Gamma^m)_{\hat\alpha}{}^{\hat\beta}(D^{+})^2 D^+_{\hat\beta} h^{--}_m + 3 (D^+)^2D^+_{\hat\alpha} h^{--5} \,.
  \label{Xi}
\end{equation}
This  superfield proves to be gauge-invariant,
$\delta_\lambda \Xi^+_{\hat\alpha}=0$, and obeys 
the equation
\begin{equation}
    D^{++} \Xi^+_{\hat\alpha} = 0\quad \implies \quad
    \Xi^+_{\hat\alpha} (z,u) = \Xi^i_{\hat\alpha} (z)u_i^+~.
\end{equation}
The harmonic-independent superfield $\Xi^i_{\hat\alpha} $
is the linearized version of the curvature spinor introduced in section \ref{Sec:sugra-overview}. 

The equation of motion for $\Psi^{\hat\alpha-}$, which follows from \eqref{deltaS2}, is 
\begin{equation}
    \Xi^+_{\hat\alpha} = 0\, .
    \label{s2-EOM-a}
\end{equation}
This is the linearized version of \eqref{spinorEoM}. It should be pointed out that $\Xi^+_{\hat\alpha}$
may be expressed in the form
\begin{equation}
    \Xi^+_{\hat\alpha} = D^+_{\hat\alpha} T\,,
\end{equation}
where
\begin{equation}
    T := D^+_{\hat\alpha} D^+_{\hat\beta} \Gamma_m^{\hat\alpha\hat\beta} h^{--m} + 3(D^+)^2 h^{--5}\,.
\end{equation}
This superfield gets shifted under the $\lambda$-gauge transformation (\ref{h--gauge-transformations}), $\delta_\lambda T = -8\mathrm{i}\Omega$, where $\Omega = \partial_m\lambda^m - D^-_{\hat\alpha}\lambda^{\hat\alpha+}$ is an analytic superfield.

The equation of motion for $v^{++5}$, which follows from \eqref{deltaS2}, is
\begin{equation}
  0=\frac{\delta S}{\delta v^{++5}} =-6(D^+)^4 h^{--5}\,.
\end{equation}
This equation is, however, not independent. It is obtained  from Eq.~(\ref{s2-EOM-a}) by applying the derivative $D^{+\hat\alpha}$. 

\subsection{Linearized super-Weyl tensor}

The superfield description of the Weyl multiplet for conformal supergravity was developed in Ref. \cite{KT-M08, Butter:2014xxa}. The super Weyl tensor is a primary dimension-1 antisymmetric tensor superfield, $W_{ab}=-W_{ba}$.
In the linearized case, there is a unique (modulo an overall coefficient) gauge-invariant expression for this tensor in terms of the prepotentials (\ref{h--solutions}):
\begin{align}
    W_{ab} =&\, (D^+)^2 \partial_{[a}h_{b]}^{--} + \frac14 \varepsilon_{abcde}(\Gamma^e)^{\hat\alpha\hat\beta} D^+_{\hat\alpha}D^+_{\hat\beta} \partial^c h^{--d} \nonumber\\
    & -(\Sigma_{ab})_{\hat\alpha\hat\beta} (D^+)^2 (D^{+\hat\alpha}h^{--\hat\beta-} + D^{-\hat\alpha}h^{--\hat\beta+})\,.
    \label{superWeyl}
\end{align}
Using the identities (\ref{lambda-analyticity-properties}) and (\ref{3Gamma}) it is possible to prove that (\ref{superWeyl}) is invariant under the $\lambda$-gauge transformations (\ref{h--gauge-transformations}), $\delta_\lambda W_{ab} = 0$. In a similar way, the identities (\ref{h-analyticity-properties}) and (\ref{zero-curvature-linearized}) allow one to verify that this tensor is harmonic-independent, $D^{++} W_{ab} = 0$. 

The expression (\ref{superWeyl}) is a 5D counterpart of the 4D $\cN=2$ super-Weyl tensor constructed in harmonic superspace in Ref.~\cite{Ivanov:2024gjo}.

\subsection{Hypermultiplet coupling}

We now turn to coupling a $q$-hypermultiplet to supergravity. 

\subsubsection{Gauge-invariant cubic vertex}

In the harmonic superspace framework, the hypermultiplet is described by an analytic superfield $q^+$ and its tilde-conjugate, $\tilde q^+$, which comprise an $SU(2)$ doublet $q^{+a} = (\tilde q^+, q^+)$. The $SU(2)$ index $a$ is raised and lowered with the use of the antisymmetric $\varepsilon$-tensor, $q^+_a = \varepsilon_{ab}q^{+b} = (q^+,-\tilde q^+)$. The free hypermultiplet action is given by an analytic subspace integral:
\begin{equation}
  S_q = - \int\mathrm{d}\zeta^{(-4)} \tilde q^+ D^{++} q^+ = -\frac12\int\mathrm{d}\zeta^{(-4)} q^{+a} D^{++} q^+_a\,.
\label{Sq-free}
\end{equation}
The power of this formulation is that it naturally extends to include the interaction with the background spin-2 supermultiplet: it is sufficient to promote the harmonic derivative $D^{++}$ to the covariant derivative ${\cal D}^{++}$ given by Eq.~(\ref{D++covariant}). As a result, the cubic interaction vertex of the hypermultiplet with the background spin-2 supermultiplet reads
\begin{align}
  S_\mathrm{int} =& -\frac{\kappa}2\int\mathrm{d}\zeta^{(-4)}q^{+a} H^{++} q^+_a
  \nonumber\\
  =& -\frac{\kappa}2\int\mathrm{d}\zeta^{(-4)}q^{+a}\left( h^{++m}\partial_m + h^{++\hat\alpha+}D^-_{\hat\alpha} + h^{++5} \partial_5 \right) q^+_a
  \,.
\label{Sint-spin2}
\end{align}

It is possible to show that the action $S_q + S_\mathrm{int}$ is invariant under spin-2 gauge transformations (\ref{delta-H++}) with the hypermultiplet fields transforming by the rule
\begin{equation}
  \delta_\lambda q^{+a} = -\kappa\left( \Lambda + \frac12\Omega \right)q^{+a}\,,
\end{equation}
where the operator $\Lambda$ is given by Eq.~(\ref{spin-2-Lambda}) while $\Omega$ is a compensating superfield of the form
\begin{equation}
  \Omega = (-1)^{\varepsilon_M} D_M \lambda^M = \partial_m \lambda^m - D^-_{\hat\alpha} \lambda^{\hat\alpha+}\,.
\label{Omega-definition}
\end{equation}
Here $\varepsilon_M$ is the Grassmann parity of the operator $D_M$. Using the properties (\ref{lambda-analyticity-properties}) it is easy to verify that $\Omega$ is an analytic superfield, $D^+_{\hat\alpha}\Omega = 0$. 

Integrating by parts the derivatives $D_M$ from the operator $\Lambda = \lambda^M D_M$, the variations of the actions (\ref{Sq-free}) and (\ref{Sint-spin2}) may be brought to the form
\begin{align}
  \delta_\lambda S_q =& \frac\kappa2 \int\mathrm{d}\zeta^{(-4)} q^{+a}[D^{++},\Lambda] q^+_a\,,\\
  \delta_\lambda S_\mathrm{int} =& \frac\kappa2 \int\mathrm{d}\zeta^{(-4)}q^{+a}\left( \kappa [H^{++},\Lambda] - \delta_\lambda H^{++} \right)q^+_a\,.
\end{align}
These two variations annihilate each other provided the prepotentials transform by the rule
\begin{equation}
  \delta_\lambda h^{++M} = {\cal D}^{++}\lambda^M - \kappa \Lambda h^{++M}\,.
\end{equation}
This variation is a nonlinear extension of the gauge transformation (\ref{h-gauge-transformations}). In what follows we will restrict ourselves to the linear gauge transformations of the prepotentials, as the construction of non-linear supergravity models leads us too far from the goals of this paper.

\subsubsection{Equations of motion and conserved currents}
\label{Sec:supercurrent}

Unlike the prepotentials $h^{++M}$, the hypermultiplet may have non-trivial dependence on the central charge coordinate $x^5$ if it has a non-vanishing mass $m$. This dependence, however, may only be cyclic,
\begin{equation}
  q^+(x,\theta^+, u, x^5) = e^{\mathrm{i}m x^5} \boldsymbol{q}^+(x,\theta^+,u)\,,\qquad
  \tilde q^+(x,\theta^+, u, x^5) = e^{-\mathrm{i}m x^5} \tilde{\boldsymbol{q}}^+(x,\theta^+,u)\,.
\end{equation}
This implies the equations $\partial_5 q^+ = \mathrm{i}m q^{+}$ and $\partial_5 \tilde q^+ = -\mathrm{i}m\tilde q^{+}$, which may be combined as follows:
\begin{equation}
  \partial_5 q^{+a} = -m J q^{+a}\,,\qquad
  J := \left( 
    \begin{array}{cc}
      \mathrm{i} & 0 \\ 0 & -\mathrm{i}
    \end{array}
   \right).
\label{J-matrix}
\end{equation}
Here and below $Jq^{+a}$ stands for $Jq^{+a} := J^a{}_b q^{+b}$.

The free hypermultiplet equation of motion 
\begin{equation}
  D^{++}q^{+a} = 0
\label{free-q-EOM}
\end{equation}
implies $D^{--}D^{--}q^{+a} = 0$. By acting with the derivatives $D^+_{\hat\alpha}$ on the latter equation it is possible to extract its differential consequences:
\begin{equation}
  (\Gamma^m)_{\hat\alpha}{}^{\hat\beta} D^-_{\hat\beta} \partial_m q^{+a} = m J D^-_{\hat\alpha} q^{+a}\,,\qquad
  (\partial^m\partial_m - m^2) q^{+a} = 0\,.
\label{extra-q-EOM}
\end{equation}

When the interaction with the background supergravity is included, the variation of action $S_q + S_\mathrm{int}$ with respect to $q^{+a}$ yields the following generalization of the free hypermultiplet equation of motion (\ref{free-q-EOM})
\begin{equation}
  \left({\cal D}^{++} + \frac{\kappa}2\Gamma^{++} \right) q^{+a} =0\,,
\end{equation}
where $\Gamma^{++}$ is the connection (which should not be confused with the $SO(4,1)$ $\Gamma$-matrices $\Gamma^m$),
\begin{equation}
  \Gamma^{++} = (-1)^{\varepsilon_M} D_M h^{++M} = \partial_m h^{++m} - D^-_{\hat\alpha} h^{++\hat\alpha+}\,.
\end{equation}
Under the gauge transformations (\ref{h-gauge-transformations}) this superfield varies as 
\begin{equation}
    \delta_\lambda \Gamma^{++} = D^{++}\Omega\,,
\end{equation}
where $\Omega$ is defined in Eq.~(\ref{Omega-definition}).

To find conserved matter currents, it is necessary to express the action (\ref{Sint-spin2}) in terms of unconstrained prepotentials $\Psi^{\hat\alpha-}$. The relations (\ref{pre-prepotentials}) imply the identity
\begin{equation}
  H^{++}q^+_a = -(D^+)^4(\Psi^{\hat\alpha-}D^-_{\hat\alpha} q^+_a) - m v^{++5} J q^+_a\,,
\end{equation}
which allows us to represent the action (\ref{Sint-spin2}) in the required form:
\begin{equation}
  S_\mathrm{int} = \frac\kappa2\int\mathrm{d}^{5|8}z \mathrm{d}u\, q^{+a}\Psi^{\hat\alpha-}D^-_{\hat\alpha}q^+_a + \frac m2 \kappa\int \mathrm{d}\zeta^{(-4)}q^{+a}v^{++5} Jq^+_a\,. 
\end{equation}
The corresponding conserved currents are:
\begin{align}
  {\cal J}^+_{\hat\alpha}:=&\,\kappa^{-1}\frac{\delta S_\mathrm{int}}{\delta\Psi^{\hat\alpha-}} = \frac12 q^{+a}D^-_{\hat\alpha} q^+_a = -\frac12 D^+_{\hat\alpha} (q^{+a}D^{--}q^+_a)\,,\\
  {\cal T}^{++}:=&\,\kappa^{-1}\frac{\delta S_\mathrm{int}}{\delta v^{++5}} = \frac m2q^{+a} J q^+_a\,.
\end{align}
These currents are conserved on the free hypermultiplet equation of motion:
\begin{equation}
  D^{++}q^{+a} = 0\quad\Rightarrow\quad
  D^{++}{\cal J}^+_{\hat\alpha} =0\,,\quad
  D^{++}{\cal T}^{++}=0\,.
\end{equation}
These conservation conditions follow from the $\lambda$-gauge symmetry of the prepotentials (\ref{pre-prepotentials-gauge}). $\rho$-gauge symmetry (\ref{s2-pre-gauge-symmetry}) dictates the following conservation laws for the currents:
\begin{align}
    D^+_{(\hat\alpha} {\cal J}^+_{\hat\beta)} = &\,0\,,
    \label{DJ}\\
  D^{+\hat\alpha}{\cal J}^+_{\hat\alpha} + 8\mathrm{i} {\cal T}^{++} =&\,0\,.
  \label{DJT}
\end{align}
This equations hold on the hypermultiplet equations of motion (\ref{extra-q-EOM}). 

Equation (\ref{DJ}) has a general solution in the form ${\cal J}^+_{\hat\alpha} = D^+_{\hat\alpha}{\cal J}$, where $\cal J$ is an unconstrained scalar superfield (supercurrent). In terms of this superfield, the second conservation law (\ref{DJT}) reads
\begin{equation}
    (D^{+})^2 {\cal J} + 8\mathrm{i}{\cal T}^{++} = 0\,.
    \label{J-T-conservation}
\end{equation}
In the case of the free hypermultiplet model, the scalar supercurrent has the simple form
\begin{equation}
    {\cal J} = \frac12 q^{+a}D^{--}q^+_a\,.
    \label{J}
\end{equation}

In 4D field theories, the scalar supercurrent $\cal J$ and the multiplet of anomalies ${\cal T}^{++}$ were found in Ref.~\cite{KT,Butter:2010sc}. These superfields are related to the supercurrent constructed by Sohnius \cite{Sohnius-supercurrent}. In Appendix \ref{AppendixC4} we show that the supercurrent (\ref{J}) originates from the $q$-hypermultiplet model coupled to the conformal supergravity background.


\subsection{Component structure of prepotentials}
\label{Sec:s=2-component-structure}

The component structure of the prepotentials $h^{++M}$ can be studied in the analytic basis with coordinates $(x^m_A,\theta^+_{\hat\alpha},u^\pm_i,x^5_A)$, which are related to the central basis $(x^m,\theta^\pm_{\hat\alpha},u^\pm_i,x^5)$ through the identities
\begin{equation}
  x^m_A = x^m + \mathrm{i}\theta^{+\hat\alpha}\Gamma^m_{\hat\alpha\hat\beta} \theta^{-\hat\beta}\,,\qquad
  x^5_A = x^5 + \mathrm{i}\theta^{+\hat\alpha}\theta^-_{\hat\alpha}\,.
\label{analytic-coordinates}
\end{equation}
In this basis, the analyticity constraint is satisfied  because the corresponding covariant spinor derivative is short, $D^+_{\hat\alpha} =\frac\partial{\partial\theta^{-\hat\alpha}}$. Explicit expressions for the other spinor and harmonic derivatives are given by Eqs.~(\ref{analytic-derivatives1}) and (\ref{analytic-derivatives2}).

\subsubsection{Bosonic fields}
The constraints (\ref{h-analyticity-properties}) and (\ref{lambda-analyticity-properties}) may be explicitly solved in the analytic coordinates:
\begin{align}
  h^{++m} =&\, h^{++m}_A(x_A,\theta^+,u) + 2\mathrm{i}\theta^{-\hat\alpha}\Gamma^m_{\hat\alpha\hat\beta} h^{++\hat\beta+}(x_A,\theta^+,u)\,,\\
  h^{++5} =&\, h^{++5}_{A}(x_A,\theta^+,u) - 2\mathrm{i}\theta^-_{\hat\alpha} h^{++\hat\alpha+}(x_A,\theta^+,u) \,,\\
  \lambda^{m} =&\, \lambda^{m}_A(x_A,\theta^+,u) + 2\mathrm{i}\theta^{-\hat\alpha}\Gamma^m_{\hat\alpha\hat\beta} \lambda^{\hat\beta+}(x_A,\theta^+,u)\,,\\
  \lambda^5 =&\, \lambda^5_{A}(x_A,\theta^+,u) - 2\mathrm{i}\theta^-_{\hat\alpha} \lambda^{\hat\alpha+}(x_A,\theta^+,u) \,.
\end{align}
The gauge transformations
\begin{subequations}
\label{analytic-gauge-transformation}
\begin{align}
  \delta h^{++m}_A =&\, D^{++}\lambda^m_A + 2\mathrm{i}\theta^{+\hat\alpha}\Gamma^m_{\hat\alpha\hat\beta}\lambda^{\hat\beta+}\,,\\
  \delta h^{++5}_{A} =&\, D^{++}\lambda^5_{A} - 2\mathrm{i}\theta^+_{\hat\alpha}\lambda^{\hat\alpha+}\,,\\
  \delta h^{++\hat\alpha+} =&\,D^{++}\lambda^{\hat\alpha+}
\end{align}
\end{subequations}
allow us to impose the Wess-Zumino gauge and remove infinite tails of auxiliary fields. In this gauge, the following bosonic component fields survive in the prepotential and gauge superfield parameters:
\begin{subequations}
\label{h-lambda-components}
\begin{align}
  h^{++m}_A =&\, \mathrm{i}(\theta^+\Gamma_n\theta^+) B^{nm}(x_A) + (\theta^+)^2 f^m(x_A) + (\theta^+)^4 V^{m\,ij}(x_A) u^-_i u^-_j\,,\\
  h^{++5}_{A} =&\, \mathrm{i}(\theta^+\Gamma_m \theta^+) A^m(x_A) +(\theta^+)^2\varphi(x_A) + (\theta^+)^4 D^{ij}(x_A)u^-_i u^-_j\,,\\
  h^{++\hat\alpha+} =&\, (\theta^+)^2\theta^+_{\hat\beta} P^{\hat\alpha\hat\beta}(x_A) \,,\\
  \lambda^m_A =&\, b^m(x_A)\,,\\
  \lambda_5 =&\, a(x_A)\,,\\
  \lambda^{\hat\alpha+} =& \,\theta^+_{\hat\beta}l^{\hat\alpha\hat\beta}(x_A)\,,
\end{align}
\end{subequations}
where $(\theta^+ \Gamma_m \theta^+) := \theta^+_{\hat\alpha} \Gamma_m^{\hat\alpha\hat\beta}\theta^+_{\hat\beta}$.

A general spin-tensor $l^{\hat\alpha\hat\beta}$ may be decomposed into three irreducible components:
\begin{equation}
  l^{\hat\alpha\hat\beta} = \varepsilon^{\hat\alpha\hat\beta} l_1 + (\Gamma_m)^{\hat\alpha\hat\beta} l^m_2 + \frac12 (\Sigma_{mn})^{\hat\alpha\hat\beta} l^{mn}_3\,,
\end{equation}
with
\begin{equation}
  l_1=-\frac14 \varepsilon_{\hat\alpha\hat\beta}l^{\hat\alpha\hat\beta}\,,\quad
  l^m_2 =-\frac14 \Gamma^m_{\hat\alpha\hat\beta} l^{\hat\alpha\hat\beta}\,,\quad
  l^{mn}_3 = l_3^{[mn]} = \Sigma^{mn}_{\hat\alpha\hat\beta} l^{\hat\alpha\hat\beta}\,.
\end{equation}
Making use of this decomposition, we substitute Eqs.~(\ref{h-lambda-components}) into the gauge transformations (\ref{analytic-gauge-transformation}) and find the corresponding variations of the component fields:
\begin{align}
  \delta h_A^{++m} =&\, \mathrm{i}(\theta^+\Gamma_n\theta^+)\delta B^{nm} + (\theta^+)^2\delta f^{m} + (\theta^+)^4\delta V^{m\,ij}u^-_i u^-_j\,,\nonumber\\
  =&\,\mathrm{i}(\theta^+\Gamma_n\theta^+)\partial^n b^{m} + 2\mathrm{i}(\theta^+\Gamma^{m}\theta^+)l_1 + 2\mathrm{i}(\theta^+)^2 l_2^{m} -\mathrm{i}(\theta^+\Gamma_n\theta^+) l_3^{nm}\,,\\
  \delta h^{++5}_{A} =&\, \mathrm{i}(\theta^+\Gamma_n\theta^+)\delta A^{n} + (\theta^+)^2\delta\varphi + (\theta^+)^4\delta D^{ij}u^-_iu^-_j\nonumber\\
  =&\,\mathrm{i}(\theta^+\Gamma_n\theta^+)\partial^n a +2\mathrm{i}(\theta^+)^2l_1 -2\mathrm{i} (\theta^+\Gamma_n\theta^+) l_2^{n}\,,\\
  \delta h^{++\hat\alpha} =&\, (\theta^+)^2 \theta^+_{\hat\beta} \delta P^{\hat\alpha\hat\beta} = \mathrm{i}(\theta^+)^2 \theta^+_{\hat\beta} (\Gamma_n)_{\hat\gamma}{}^{\hat\beta} \partial^n l^{\hat\alpha\hat\gamma}\,.
\end{align}
By analyzing these relations it is easy to see that the gauge freedom in $l_1$, $l_2^m$ and $l_3^{mn}$ may be used to gauge away some of the remaining auxiliary fields and achieve the standard gauge variations for the physical ones:
\begin{itemize}
    \item The scalar $l_1$ allows us to gauge away the component $\varphi$ from $h^{++5}_A$. In the gauge $\varphi=0$, we have $l_1=0$.
    \item The vector $f^m$ in $h^{++m}_A$ may be completely gauged away by $l_2^m$. In the gauge $f^m=0$, we have $l_2^m =0$.
    \item The antisymmetric tensor $l_3^{mn}$ allows us to gauge away the antisymmetric component of the tensor $B^{mn}$. In the gauge $B^{[mn]}=0$, the tensor $l_3^{mn}$ reads 
    \begin{equation}
        l_3^{mn} = \partial^{[m}b^{n]}\,.
    \end{equation}
    \item The physical components $A_m$ and $B_{mn} = B_{(mn)}$ have standard gauge variations in terms of the unconstrained gauge parameters $b_m$ and $a$:
    \begin{equation}
        \delta B_{mn} = \partial_{(m} b_{n)}\,,\qquad
        \delta A_m = \partial_m a\,.
    \end{equation}
    \item The auxiliary field $P^{\hat\alpha\hat\beta}$ has the following residual gauge transformation:
    \begin{equation}
        \delta P^{\hat\alpha\hat\beta} =\frac{\mathrm{i}}{2} \Gamma_m^{\hat\alpha\hat\beta} \partial_n \partial^{[m}b^{n]}\,.
    \end{equation}
\end{itemize}
As a result, we conclude that, modulo the gauge freedom associated with $b_m$ and $a$, the physical components $B_{(mn)}$ and $A_m$ as well as the auxiliary ones $V^{m\,ij}$, $D^{ij}$ and $P^{\hat\alpha\hat\beta}$ comprise 48 bosonic degrees of freedom (d.o.f.) off shell. The quadratic action for these fields is given in Appendix \ref{sec:s=2-component-action}, see Eq.~(\ref{s=2-component-action}).

\subsubsection{Fermionic fields}
\label{Sec:s=2-fermions}
Upon imposing the Wess-Zumino gauge, the following fermionic components survive in the prepotentials and gauge parameters:
\begin{align}
  h^{++m}_A =& \, -2 (\theta^+)^2\theta^+_{\hat\alpha}\psi^{\hat\alpha\, m\,i}u^-_i + (\theta^+)^2\theta^{+\hat\alpha}\Gamma^m_{\hat\alpha\hat\beta} \chi^{\hat\beta\,i}u^-_i\,,\\
  h^{++5}_{A} =& \,(\theta^+)^2\theta^+_{\hat\alpha}\chi^{\hat\alpha\,i}u^-_i\,,\\
  h^{++\hat\alpha+} =&\, \frac14(\theta^+)^4 \rho^{\hat\alpha\,i}u^-_i \,,\\
  \lambda^{m}_A =&\, -\frac{\mathrm{i}}{2}\theta^{+\hat\alpha}\Gamma^{m}_{\hat\alpha\hat\beta}\epsilon^{\hat\beta i}u^-_i\,, \\
  \lambda^5_{A} =&\,\frac{\mathrm{i}}{2} \theta^+_{\hat\alpha}\epsilon^{\hat\alpha i}u^-_i\,,\\
  \lambda^{\hat\alpha+} =&\, \frac14\epsilon^{\hat\alpha i} u^+_i -\frac{\mathrm{i}}{4} (\theta^+\Gamma^n\theta^+)\partial_n \epsilon^{\hat\alpha i}u^-_i\,.
\end{align}
Substituting these decompositions into Eqs.~(\ref{analytic-gauge-transformation}), we find the residual gauge transformations of the fermionic components:
\begin{equation}
  \delta\psi^{\hat\alpha m i} = \partial^m \epsilon^{\hat\alpha i}\,,\quad
  \delta \chi^{\hat\alpha i} = (\Gamma^m)^{\hat\alpha}{}_{\hat\beta} \partial_m \epsilon^{\hat\beta i}\,,\quad
  \delta \rho^{\hat\alpha i} = \Box \epsilon^{\hat\alpha i}\,.
\end{equation}
There are 40 d.o.f.\ in the gravitino field $\psi^{\hat\alpha\, m\, i}$ and $2\times8$ d.o.f.\ in the auxiliary spinors $\chi^{\hat\alpha\,i}$ and $\rho^{\hat\alpha i}$, but it is possible to gauge away 8 d.o.f.\ using $\epsilon^{\hat\alpha i}$. Thus, there are 48 fermionic d.o.f.\ off shell.

We conclude that the supergravity prepotentials $h^{++M}$ contain 48+48 degrees of freedom off shell. In the component field formulation, this off-shell version of supergravity was studied in Ref.~\cite{Zucker:1999ej}.


\section{Massless higher-spin supermultiplets}
\label{Sec:higher-spins}

In this section we construct massless higher-spin supermultiplets in harmonic superspace and analyze several related aspects such as their component structure.  

\subsection{Prepotentials and gauge transformations}
\label{Sec:prepotentials}
The prepotentials $h^{++M}$ describing the massless spin-2 supermultiplet may be naturally extended to arbitrary integer spin $s\geq3$ by attaching $s-2$ Lorentz indices,
\begin{equation}
  h^{++M m_1\ldots m_{s-2}} = (h^{++m m_1\ldots m_{s-2}}, h^{++\hat\alpha+m_1\ldots m_{s-2}}, h^{++5m_1\ldots m_{s-2}})\,.
\label{spin-s-prepotentials}
\end{equation}
By construction, these prepotentials are independent of the central charge variable,
\begin{equation}
  \partial_5 h^{++Mm_1\ldots m_{s-2}} = 0\,,
\end{equation}
and are required to obey the following algebraic constraints:
\begin{itemize}
\item[(i)] All Lorentz indices are symmetrized,
\begin{subequations}
\begin{align}
  h^{++ m_1\ldots m_{s-1}} =& \,h^{++(m_1\ldots m_{s-1})}\,,\\
  h^{++\hat\alpha+ m_1\ldots m_{s-2}} =&\, h^{++\hat\alpha+( m_1\ldots m_{s-2})}\,,\\
  h^{++5 m_1\ldots m_{s-2}} =&\, h^{++5(m_1\ldots m_{s-2})}\,;
\end{align}
\end{subequations}
\item[(ii)] The prepotentials are traceless in Lorentz indices,
\begin{equation}
  \eta_{m_1m_2}h^{++M m_1m_2\ldots m_{s-2}} = 0\,;
\label{h-tracelessness}
\end{equation}
\item[(iii)] The Grassmann-odd prepotential is $\Gamma$-traceless:
\begin{equation}
  (\Gamma_{m_1})_{\hat\alpha\hat\beta} h^{++\hat\beta+m_1\ldots m_{s-2}} =0\,.
\label{Gamma-tracelessness}
\end{equation}
\end{itemize}
These constraints eliminate reducible (lower-spin) components from this multiplet and retain the irreducible spin-$s$ massless supermultiplet. As will be shown below, these constraints are consistent with gauge invariance and supersymmetry. Moreover, the constraint (\ref{Gamma-tracelessness}) will be crucial in constructing a gauge-invariant quadratic action for higher spin supermultiplets.

Associated with the prepotentials (\ref{spin-s-prepotentials}) is the differential operator
\begin{align}
  H^{++} = &\,h^{++M m_1\ldots m_{s-2}}D_M \partial_{m_1}\ldots \partial_{m_{s-2}} \nonumber\\
  =&\,(h^{++m m_1\ldots m_{s-2}}\partial_m + h^{++\hat\alpha+m_1\ldots m_{s-2}}D^-_{\hat\alpha} + h^{++5m_1\ldots m_{s-2}}\partial_5)\partial_{m_1}\ldots \partial_{m_{s-2}}\,.
  \label{H++-general}
\end{align}
We require this operator to preserve analyticity, i.e., it obeys $D^+_{\hat\alpha}(H^{++}\Phi_A) = 0$, for an analytic superfield $\Phi_A$. This condition implies the following analyticity constraints on the prepotentials:
\begin{subequations}
\label{h-analyticity-properties-general}
\begin{align}
  D^+_{\hat\alpha} h^{++mm_1\ldots m_{s-2}} = &\,2\mathrm{i}\Gamma^{(m}_{\hat\alpha\hat\beta} h^{++\hat\beta+m_1\ldots m_{s-2})}\,,\\
  D^+_{\hat\alpha} h^{++5m_1\ldots m_{s-2}} =&\, 2\mathrm{i}\varepsilon_{\hat\alpha\hat\beta} h^{++\hat\beta+ m_1\ldots m_{s-2}}\,,\\
  D^+_{\hat\alpha}h^{++\hat\beta+ m_1\ldots m_{s-2}} =&\,0\,.
\end{align}
\end{subequations}
These equations show that $h^{++\hat\beta+ m_1\ldots m_{s-2}}$ is analytic while $h^{++mm_1\ldots m_{s-2}}$ and $h^{++5m_1\ldots m_{s-2}}$ are linear,
\begin{equation}
  D^+_{\hat\alpha} D^+_{\hat\beta}h^{++mm_1\ldots m_{s-2}} = D^+_{\hat\alpha} D^+_{\hat\beta} h^{++5m_1\ldots m_{s-2}} =0\,.
\end{equation}

We postulate the gauge transformations of the prepotentials in the following form:
\begin{equation}
  \delta_\lambda h^{++Mm_1\ldots m_{s-2}} = D^{++} \lambda^{Mm_1\ldots m_{s-2}}\,,
\label{h-gauge-transformation}
\end{equation}
where the superfield gauge parameters
\begin{equation}
  \lambda^{Mm_1\ldots m_{s-2}} = (\lambda^{mm_1\ldots m_{s-2}},\lambda^{\hat\alpha+m_1\ldots m_{s-2}},\lambda^{5m_1\ldots m_{s-2}})
\label{lambda-gauge-parameters}
\end{equation}
respect the same symmetry properties as the prepotentials: they are symmetric and traceless in all vector indices,
\begin{subequations}
\begin{align}
  \lambda^{m_1\ldots m_{s-1}} =&\, \lambda^{(m_1\ldots m_{s-1})}\,,\\
  \lambda^{\hat\alpha+m_1\ldots m_{s-2}} =&\, \lambda^{\hat\alpha+(m_1\ldots m_{s-2})}\,,\\
  \lambda^{5m_1\ldots m_{s-2}} =&\, \lambda^{5(m_1\ldots m_{s-2})}\,,\\
  \eta_{m_1m_2}\lambda^{Mm_1m_2\ldots m_{s-2}} =&\,0\,,
\end{align}
\end{subequations}
and the Grassmann-odd gauge parameter is $\Gamma$-traceless,
\begin{equation}
  (\Gamma_{m_1})_{\hat\alpha\hat\beta}\lambda^{\hat\beta+m_1\ldots m_{s-2}} = 0\,.
\label{lambda-Gamma-tracelessness}
\end{equation}
Moreover, since the gauge transformation (\ref{h-gauge-transformation}) should preserve the analyticity properties of the prepotentials (\ref{h-analyticity-properties-general}), the gauge parameters must obey analogous constraints:
\begin{subequations}
\label{lambda-analyticity-general}
\begin{align}
  D^+_{\hat\alpha} \lambda^{mm_1\ldots m_{s-2}} = &\,2\mathrm{i}\Gamma^{(m}_{\hat\alpha\hat\beta} \lambda^{\hat\beta+m_1\ldots m_{s-2})}\,,\\
  D^+_{\hat\alpha} \lambda^{5m_1\ldots m_{s-2}} =&\, 2\mathrm{i}\varepsilon_{\hat\alpha\hat\beta} \lambda^{\hat\beta+ m_1\ldots m_{s-2}}\,,\\
  D^+_{\hat\alpha}\lambda^{\hat\beta+ m_1\ldots m_{s-2}} =&\,0\,.
\end{align}
\end{subequations}

By analogy with the spin-2 supermultiplet considered in Sec.~\ref{Sec:s2-prepotentials}, we introduce a set of four prepotentials with negative $U(1)$ charges,
\begin{equation}
  h^{--m_1\ldots m_{s-1}}\,,\quad
  h^{--5m_1\ldots m_{s-2}}\,,\quad
  h^{--\hat\alpha+m_1\ldots m_{s-2}}\,,\quad
  h^{--\hat\alpha-m_1\ldots m_{s-2}}\,,
\label{h--prepotentials-general}
\end{equation}
which are defined by the following equations
\begin{subequations}
\label{zero-curvature-general}
\begin{align}
  D^{++} h^{--m_1\ldots m_{s-1}} - D^{--} h^{++m_1\ldots m_{s-1}} =&\, 0\,,\\
  D^{++} h^{--5m_1\ldots m_{s-2}} - D^{--} h^{++5m_1\ldots m_{s-2}} =&\, 0\,,\\
  D^{++} h^{--\hat\alpha+m_1\ldots m_{s-2}} - D^{--}h^{++\hat\alpha+m_1\ldots m_{s-2}}  =&\,0\,,\\
  D^{++} h^{--\hat\alpha-m_1\ldots m_{s-2}} + h^{--\hat\alpha+m_1\ldots m_{s-2}} =&\,0\,.
\end{align}
\end{subequations}
Explicit solutions of these equations in terms of harmonic distributions are analogous to the ones for the spin-2 supermultiplet, Eqs.~(\ref{h--solutions}). 

Gauge transformations of the prepotentials (\ref{h--prepotentials-general}) are
\begin{subequations}
\label{h--gauge-transformations-general}
\begin{align}
  \delta_\lambda h^{--m_1\ldots m_{s-1}} =&\, D^{--}\lambda^{m_1\ldots m_{s-1}}\,,\\
  \delta_\lambda h^{--5m_1\ldots m_{s-2}} =&\, D^{--}\lambda^{5m_1\ldots m_{s-2}}\,,\\
  \delta_\lambda h^{--\hat\alpha-m_1\ldots m_{s-2}} =&\, D^{--}\lambda^{\hat\alpha-m_1\ldots m_{s-2}}\,,\\
  \delta_\lambda h^{--\hat\alpha+m_1\ldots m_{s-2}} =&\, D^{--}\lambda^{\hat\alpha+m_1\ldots m_{s-2}} + \lambda^{\hat\alpha-m_1\ldots m_{s-2}}\,,
\end{align}
\end{subequations}
where $\lambda^{\hat\alpha-m_1\ldots m_{s-2}}$ is defined by the equation $D^{++}\lambda^{\hat\alpha-m_1\ldots m_{s-2}} + \lambda^{\hat\alpha+m_1\ldots m_{s-2}} = 0$. 

\subsection{Unconstrained prepotentials}

The prepotentials $h^{++M m_1\ldots m_{s-2}}$ and their gauge parameters $\lambda^{M m_1\ldots m_{s-1}}$ obey the constraints (\ref{h-analyticity-properties-general}) and (\ref{lambda-analyticity-general}), respectively. These constraints may be solved in terms of unconstrained full superspace and analytic superfields. 

Starting with the prepotentials, it is possible to show that the general solution to (\ref{h-analyticity-properties-general}) may be represented in the form
\begin{subequations}
  \label{h-Psi-solution}
\begin{align}
  h^{++m_1\ldots m_{s-1}} = &\,-\frac{\mathrm{i}}4\Gamma^{(m_1}_{\hat\alpha\hat\beta} (D^+)^2 D^{+\hat\alpha} \Psi^{\hat\beta- m_2\ldots m_{s-1})}\,,\\
  h^{++\hat\alpha+m_1\ldots m_{s-2}} =&\,-(D^+)^4 \Psi^{\hat\alpha- m_1\ldots m_{s-2}}\,,\\
  h^{++5m_1\ldots m_{s-2}} =&\, \frac{\mathrm{i}}4(D^+)^2 D^+_{\hat\alpha}\Psi^{\hat\alpha-m_1\ldots m_{s-2}} + v^{++5m_1\ldots m_{s-2}}\,,
\end{align}
\end{subequations}
where $\Psi^{\hat\alpha-m_1\ldots m_{s-2}}$ is an unconstrained superfield in the full superspace while $v^{++5m_1\ldots m_{s-2}}$ is an analytic superfield, 
\begin{equation}
    D^+_{\hat\alpha} v^{++5m_1\ldots m_{s-2}} = 0\,.
\end{equation}
To preserve all symmetry properties of $h^{++Mm_1\ldots m_{s-2}}$, the tensors $\Psi^{\hat\alpha-m_1\ldots m_{s-2}}$ and \\ $v^{++5m_1\ldots m_{s-2}}$ must be symmetric and ($\Gamma$-)traceless,
\begin{subequations}
\label{Psi-v-symmetries}
\begin{align}
  \Psi^{\hat\alpha-m_1\ldots m_{s-2}} =& \,\Psi^{\hat\alpha-(m_1\ldots m_{s-2})}\,, & (\Gamma_{m_1})_{\hat\alpha\hat\beta}\Psi^{-\hat\beta m_1\ldots m_{s-2}} =&\,0\,,\\
  v^{++5m_1\ldots m_{s-2}} =&\, v^{++5(m_1\ldots m_{s-2})}\,,&
  \eta_{m_1 m_2} v^{++5m_1 m_2\ldots m_{s-2}}=&\,0\,.
\end{align}
\end{subequations}
These prepotentials are defined modulo gauge transformations 
\begin{subequations}
\label{pre-gauge-symmetry}
  \begin{align}
  \delta_\rho\Psi^{\hat\alpha-m_1\ldots m_{s-2}} = &\, D^{+\hat\alpha} \rho_1^{-- m_1\ldots m_{s-2}}
  +\frac{s-2}{s+2}(\Gamma^{(m_1})^{\hat\alpha}{}_{\hat\beta} (\Gamma_n)^{\hat\beta}{}_{\hat\gamma} D^{+\hat\gamma} \rho_1^{-- m_2\ldots m_{s-2})n} \nonumber\\&
  + D^+_{\hat\beta}\rho_2^{--\hat\alpha\hat\beta m_1\ldots m_{s-2}} \,,\\
  \delta_\rho v^{++5m_1\ldots m_{s-2}} =&\, -8\mathrm{i} (D^+)^4 \rho_1^{--m_1\ldots m_{s-2}}\,,
\end{align}
\end{subequations}
with two independent unconstrained superfields $\rho_1^{--m_1\ldots m_{s-2}}$ and $\rho_2^{--\hat\alpha\hat\beta m_1\ldots m_{s-2}}$ subject to the symmetry properties
\begin{align}
    \rho_1^{--m_1\ldots m_{s-2}}=&\,\rho_1^{--(m_1\ldots m_{s-2})}\,, & \rho_1^{--m_1 m_2\ldots m_{s-2}}\eta_{m_1 m_2}=&\,0\,,\\
    \rho_2^{--\hat\alpha\hat\beta m_1\ldots m_{s-2}} =&\,\rho_2^{--(\hat\alpha\hat\beta)(m_1\ldots m_{s-2})}\,, & (\Gamma_{m_1})_{\hat\alpha\hat\beta}\rho_2^{--\hat\beta\hat\gamma m_1\ldots m_{s-2}} =&\,0\,.
\end{align}
The prepotentials $h^{++Mm_1\ldots m_{s-2}}$ are inert under the $\rho$-gauge transformations (\ref{pre-gauge-symmetry}). 

Gauge symmetry (\ref{pre-gauge-symmetry}) should not be confused with the $\lambda$-gauge symmetry (\ref{h-gauge-transformation}) with the parameters $\lambda^{Mm_1\ldots m_{s-2}}$ subject to the constraints (\ref{lambda-analyticity-general}). By analogy with Eqs.~(\ref{h-Psi-solution}), the constraints (\ref{lambda-analyticity-general}) may be solved in terms of an unconstrained superfield $l^{--\hat\alpha- m_1\ldots m_{s-2}}$ and an analytic superfield $l^{5m_1\ldots m_{s-2}}$ ($D^+_{\hat\alpha}l^{5m_1\ldots m_{s-2}}=0$):
\begin{subequations}
\begin{align}
  \lambda^{m_1\ldots m_{s-1}} = &\,-\frac{\mathrm{i}}4\Gamma^{(m_1}_{\hat\alpha\hat\beta} (D^+)^2 D^{+\hat\alpha} l^{--\hat\beta- m_2\ldots m_{s-1})}\,,\\
  \lambda^{\hat\alpha+m_1\ldots m_{s-2}} =&\,-(D^+)^4 l^{--\hat\alpha- m_1\ldots m_{s-2}}\,,\\
  \lambda^{5m_1\ldots m_{s-2}} =&\, \frac{\mathrm{i}}4(D^+)^2 D^+_{\hat\alpha}l^{--\hat\alpha-m_1\ldots m_{s-2}} + l^{5m_1\ldots m_{s-2}}\,.
\end{align}
\end{subequations}
These superfields obey the same algebraic constraints as the prepotentials (\ref{Psi-v-symmetries}),
\begin{subequations}
\label{lambda-symmetries}
\begin{align}
  l^{--\hat\alpha- m_1\ldots m_{s-2}} =&\, l^{--\hat\alpha-(m_1\ldots m_{s-2})}\,, & (\Gamma_{m_1})_{\hat\alpha\hat\beta}l^{--\hat\beta- m_1\ldots m_{s-2}} =&\,0\,,\\
  l^{5m_1\ldots m_{s-2}} =&\, l^{5(m_1\ldots m_{s-2})}\,, &
  \eta_{m_1 m_2} l^{5m_1 m_2\ldots m_{s-2}}=&\,0\,.
\end{align}
\end{subequations}
In terms of these superfield parameters, we find $\lambda$-gauge transformations of the unconstrained prepotentials which follow from Eq.~(\ref{h-gauge-transformation}):
\begin{subequations}
\label{l-Psi-gauge-transformations}
\begin{align}
\delta_\lambda \Psi^{\hat\alpha-m_1\ldots m_{s-2}} = &\,D^{++}l^{--\hat\alpha-m_1\ldots m_{s-2}}\,,\\
\delta_\lambda v^{++5m_1\ldots m_{s-2}} = &\,D^{++}l^{5m_1\ldots m_{s-2}}\,.
\end{align}
\end{subequations}

As will be shown below, both $\lambda$- and $\rho$-gauge symmetries (\ref{l-Psi-gauge-transformations}) and (\ref{pre-gauge-symmetry}) play a fundamental role in the formulation of conserved higher currents.

\subsection{Example: component structure of spin-3 supermultiplet}

The component structure of the prepotentials (\ref{spin-s-prepotentials}) for arbitrary higher spin $s$ is given in Appendix \ref{App:component-structure}. In this subsection, for simplicity, we will present the component structure of these prepotentials for the $s=3$ supermultiplet and will show that the constraints and gauge symmetries described above are sufficient to eliminate all non-physical components from these superfields. On the one hand, the study of the component field structure in this section follows the same logic as in Sec.~\ref{Sec:s=2-component-structure} for the $s=2$ supermultiplet. On the other hand, the analysis presented here demonstrates some complications arising for higher spins which are absent in the $s=2$ case.

The spin-3 supermultiplet is described by the prepotentials $h^{++mn}$, $h^{++5m}$ and $h^{++\hat\alpha+m}$ subject to the constraints (\ref{h-analyticity-properties-general}) and gauge transformations (\ref{h-gauge-transformation}). The analyticity constraints may be solved explicitly in the analytic coordinates (\ref{analytic-coordinates}):
\begin{align}
  h^{++mn} =&\, h^{++mn}_A(x_A,\theta^+,u) + 2\mathrm{i}\theta^{-\hat\alpha}\Gamma^{(m}_{\hat\alpha\hat\beta} h^{++\hat\beta+n)}(x_A,\theta^+,u)\,,\\
  h^{++5m} =&\, h^{++5m}_{A}(x_A,\theta^+,u) - 2\mathrm{i}\theta^-_{\hat\alpha} h^{++\hat\alpha+m}(x_A,\theta^+,u) \,,\\
  \lambda^{mn} =&\, \lambda^{mn}_A(x_A,\theta^+,u) + 2\mathrm{i}\theta^{-\hat\alpha}\Gamma^{(m}_{\hat\alpha\hat\beta} \lambda^{\hat\beta+n)}(x_A,\theta^+,u)\,,\\
  \lambda^{5m} =&\, \lambda^{5m}_{A}(x_A,\theta^+,u) - 2\mathrm{i}\theta^-_{\hat\alpha} \lambda^{\hat\alpha+m}(x_A,\theta^+,u) \,,
\end{align}
where $h^{++mn}_A$, $h^{++5m}_A$, $\lambda^{mn}_A$ and $\lambda^{5m}_A$ are analytic superfields. These analytic prepotentials have the following $\lambda$-gauge transformation laws:
\begin{subequations}
\label{s=3-gauge-transformations}
\begin{align}
  \delta_\lambda h^{++mn}_A =&\, D^{++}\lambda^{mn}_A + 2\mathrm{i}\theta^{+\hat\alpha}\Gamma^{(m}_{\hat\alpha\hat\beta}\lambda^{\hat\beta+n)}\,,\\
  \delta_\lambda h^{++5m}_{A} =&\, D^{++}\lambda^{5m}_{A} - 2\mathrm{i}\theta^+_{\hat\alpha}\lambda^{\hat\alpha+m}\,,\\
  \delta_\lambda h^{++\hat\alpha+m} =& \,D^{++}\lambda^{\hat\alpha+m}\,.
\end{align}
\end{subequations}
In addition, there are the algebraic ($\Gamma$-)tracelessness constraints:
\begin{align}
  h^{++mn}\eta_{mn}=h^{++mn}_A\eta_{mn}=&\,0\,,\\
  \lambda^{mn}\eta_{mn} = \lambda^{mn}_A \eta_{mn} =&\,0\,,\\
  \Gamma^{m}_{\hat\alpha\hat\beta}h^{++\hat\beta+n}\eta_{mn}=\Gamma^{m}_{\hat\alpha\hat\beta}\lambda^{\hat\beta+n}\eta_{mn}=&\,0\,.\label{eq3.24}
\end{align}

\subsubsection{Bosonic fields}
\label{Sec:s=3-bosons}
In the Wess-Zumino gauge, the following bosonic components survive in the prepotentials
\begin{subequations}
\label{s=3-prepotentials}
\begin{align}
  h^{++mn}_A =&\, \mathrm{i}(\theta^+\Gamma_p\theta^+) B^{p\,mn}(x_A) + (\theta^+)^2 f^{mn}(x_A)+ (\theta^+)^4 V^{mn\,ij}(x_A)u^-_i u^-_j\,,\\
  h^{++5m}_{A} =&\, \mathrm{i}(\theta^+\Gamma_p \theta^+) A^{p\,m}(x_A) +(\theta^+)^2\varphi^m(x_A) + (\theta^+)^4 D^{m\,ij}(x_A)u^-_i u^-_j\,,\\
  h^{++\hat\alpha+m} =&\, (\theta^+)^2\theta^+_{\hat\beta} P^{\hat\alpha\hat\beta m}(x_A) \,,
\end{align}
\end{subequations}
and in the gauge superfield parameters
\begin{equation}
\label{s=3-gauge-parameters}
  \lambda^{mn}_A = b^{mn}(x_A)\,,\quad
  \lambda^{5m}_A = a^m(x_A)\,,\quad
  \lambda^{\hat\alpha+m} = \theta^+_{\hat\beta}l^{\hat\alpha\hat\beta m}(x_A)\,.
\end{equation}
A general spin-tensor $l^{\hat\alpha\hat\beta p}$ is reducible, with the following components:
\begin{equation}
  l^{\hat\alpha\hat\beta p} = \varepsilon^{\hat\alpha\hat\beta} l_1^p + (\Gamma_m)^{\hat\alpha\hat\beta} l_2^{mp} + \frac12 (\Sigma_{mn})^{\hat\alpha\hat\beta} l_3^{mn\,p}\,,
\label{s=3-l-decomposition}
\end{equation}
where
\begin{equation}
  l_1^p=-\frac14 \varepsilon_{\hat\alpha\hat\beta}l^{\hat\alpha\hat\beta p}\,,\quad
  l_2^{mp} =-\frac14 \Gamma^m_{\hat\alpha\hat\beta} l^{\hat\alpha\hat\beta p}\,,\quad
  l_3^{mn\,p} = \Sigma^{mn}_{\hat\alpha\hat\beta} l^{\hat\alpha\hat\beta p}\,.
\end{equation}
Note that $l_3^{mn\,p}$ is antisymmetric in its first two indices, $l_3^{mn\,p}=l_3^{[mn]p}$ while $l_2^{mp}$ is a general tensor with no symmetry constraints. Additional relations between these components arise from the constraint $(\Gamma_m)_{\hat\alpha\hat\beta}l^{\hat\beta\hat\gamma m} =0$: 
\begin{equation}
    l_1^m + \frac12\eta_{pq} l_3^{mp\,q} = 0\,,\quad
    l_2^{[mn]} + \frac18 \varepsilon^{mn}{}_{pqr} l_3^{pq\,r} = 0\,,\quad
    \eta_{mn} l_2^{mn} = 0\,.
\label{s=3-l-constraints}
\end{equation}
The same constraints hold for the components of the auxiliary field $P^{\hat\alpha\hat\beta m}$ which contains 64 d.o.f.

Taking into account the decomposition (\ref{s=3-l-decomposition}), we substitute the gauge superfield parameters (\ref{s=3-gauge-parameters}) into Eqs.~(\ref{s=3-gauge-transformations}) and find gauge variations of the prepotentials (\ref{s=3-prepotentials}):
\begin{align}
  \delta_\lambda h_A^{++mn} =&\, \mathrm{i}(\theta^+\Gamma_p\theta^+)\delta B^{p\,mn} + (\theta^+)^2\delta f^{mn} + (\theta^+)^4\delta V^{mn\,ij}u^-_i u^-_j\,,\nonumber\\
  =&\,\mathrm{i}(\theta^+\Gamma_p\theta^+)\partial^p b^{mn} + 2\mathrm{i}(\theta^+\Gamma^{(m}\theta^+)l_1^{n)} + 2\mathrm{i}(\theta^+)^2 l_2^{(mn)} \nonumber\\&
  -\mathrm{i}(\theta^+\Gamma_p\theta^+) l_3^{[p(m]n)}\,,\label{eq3.30}\\
  \delta_\lambda h^{++5m}_{A} =&\, \mathrm{i}(\theta^+\Gamma_n\theta^+)\delta A^{n\,m} + (\theta^+)^2\delta\varphi^m + (\theta^+)^4\delta D^{m\,ij}u^-_iu^-_j\nonumber\\
  =&\,\mathrm{i}(\theta^+\Gamma_n\theta^+)\partial^n a^m +2\mathrm{i}(\theta^+)^2 l_1^m -2\mathrm{i} (\theta^+\Gamma_n\theta^+) l_2^{nm}\,,\label{eq3.31}\\
  \delta_\lambda h^{++\hat\alpha+m} =&\,(\theta^+)^2\theta^+_{\hat\beta}\delta P^{\hat\alpha\hat\beta m} = \mathrm{i}(\theta^+)^2 \theta^+_{\hat\beta} (\Gamma_n)_{\hat\gamma}{}^{\hat\beta}\partial^n l^{\hat\alpha\hat\gamma m}\,.\label{eq3.32}
\end{align}
Using the gauge freedom of the tensors $l_1^m$, $l_2^{mn}$ and $l_3^{mn\,p}$ we can gauge away any remaining non-physical degrees of freedom in the prepotentials:
\begin{itemize}
\item The gauge parameter $l_1^m$ allows us to gauge away the vector component $\varphi^m$ from $h^{++5m}_A$. In the gauge $\varphi^m = 0$ we have $\delta\varphi^m = 2\mathrm{i} l_1^m = 0$. 
\item As follows from Eq.~(\ref{s=3-l-constraints}), the tensor $l_2^{(mn)}$ is traceless. Therefore, it allows us to gauge away the component $f^{mn}$ from $h^{++mn}_A$. In the gauge $f^{mn} = 0$, we have $\delta f^{mn} = 2\mathrm{i} l_2^{(mn)} = 0$.
\item In the gauge $l_2^{(mn)} = 0$, the field $A^{n\,m}$ in Eq.~(\ref{eq3.31}) has the following transformation law:
\begin{equation}
    \delta A^{n\,m} = \partial^n a^m - 2 l_2^{[nm]}\,. 
\end{equation}
Thus, we can impose the gauge $A^{[nm]} = 0$ in which $l_2^{[nm]} = \frac12 \partial^{[n} a^{m]}$. As a result, the component $A^{n\,m}$ is given by a symmetric tensor, $A^{n\,m} = A^{(nm)}$ describing a spin-2 field with standard gauge freedom
\begin{equation}
  \delta A_{nm} =\partial_{(n}a_{m)}\,.  \quad \mbox{(10 d.o.f.)}
\end{equation}
\item Eq.~(\ref{eq3.30}) shows that the component $B^{p\,mn}$ has the following transformation law:
\begin{equation}
    \delta B^{p\,mn} = \partial^p b^{mn} - l_3^{[p(m]n)}\,.
\end{equation}
One could try to impose the gauge $B^{[p(m]n)} = 0$ in which $l_3^{[p(m]n)} = \partial^{[p} b^{(m]n)}$. In this gauge, the field $B^{p\,mn}$ would be totally symmetric, $B^{p\,mn} = B^{(pmn)}$, with standard gauge freedom $\delta B_{pmn} = \partial_{(p} b_{mn)}$. This gauge, however, is not consistent because it is not compatible with the constraint $B^{p\,mn}\eta_{mn} = 0$. In the consistent gauge, the tensor $B^{p\,mn}$ is expressed via a totally symmetric and traceless tensor $B_1^{pmn}=B_1^{(pmn)}$, $B_1^{pmn}\eta_{mn}=0$, and a vector $B_2^m$ such that
\begin{equation}
    B^{p\,mn} = B_1^{pmn} + \eta^{p(m}B_2^{n)} - \frac15 \eta^{mn}B_2^p\,.
    \label{eq3.36}
\end{equation}
In this gauge, the remaining components of the gauge parameter $l_3$ are fixed, $l_3^{[p(m]n)} = \partial^{[p} b^{(m]n)} - \frac16 \eta^{p(m}\partial_q b^{n)q} + \frac16 \eta^{mn}\partial_q b^{pq}$, and the tensors $B_1^{pmn}$ and $B_2^m$ have the following gauge transformation laws:
\begin{equation}
    \delta B_1^{pmn} = \partial^{(p}b^{mn)} - \frac27 \eta^{(mn}\partial_q b^{p)q}\,,\qquad
    \delta B_2^m = \frac5{14}\partial_n b^{nm}\,.
\end{equation}
Finally, it is possible to combine $B_1^{pmn}$ and $B_2^m$ into a totally symmetric tensor ${\boldsymbol B}^{pmn}$ with no tracelessness condition imposed and a standard spin-3 field gauge transformation
\begin{equation}
    {\boldsymbol B}^{pmn} = B_1^{pmn} + \frac45 \eta^{(pm}B_2^{n)}\,,\qquad
    \delta {\boldsymbol B}^{pmn} = \partial^{(p}b^{mn)}\,.
\end{equation}
This is a Fronsdal's spin-3 field with 21 d.o.f. However, ${\boldsymbol B}^{pmn}$ cannot appear as a component in the superfield prepotential $h_A^{++mn}$, whereas the correct component field containing spin-3 degrees of freedom is given by Eq.~(\ref{eq3.36}).

\item The auxiliary component $V^{mn\,ij}$ is symmetric in both Lorentz and $SU(2)_R$ indices. It contains 42 d.o.f. The auxiliary field $D^{m\, ij}$ adds up 15 d.o.f.
\item Remembering that the auxiliary component $P^{\hat\alpha\hat\beta m}$ is $\Gamma$-traceless, $(\Gamma_m)_{\hat\alpha\hat\beta} P^{\hat\beta\hat\gamma m} = 0$, we conclude that it contains 64 d.o.f. As follows from Eq.~(\ref{eq3.32}), this field has the following transformation law
\begin{equation}
    \delta P^{\hat\alpha\hat\beta m} = \mathrm{i}(\Gamma_n)_{\hat\gamma}{}^{\hat\beta} \partial^n l^{\hat\alpha\hat\gamma m}\,.
\end{equation}
This gauge transformation, however, does not decrease the number of degrees of freedom since all components of the gauge parameter $l^{\hat\alpha\hat\gamma m}$ have been fully fixed above and expressed via gauge parameters $a^m$ and $b^{mn}$.
\end{itemize}
As a result, there are 152 bosonic d.o.f.\ in this supermultiplet.

\subsubsection{Fermionic fields}

The fermionic fields in the spin-3 supermultiplet may be found in a similar way to the spin-2 case considered in Sec.~\ref{Sec:s=2-fermions}. In the Wess-Zumino gauge, the following fermionic components survive in the analytic prepotentials:
\begin{subequations}
\label{s=3-fermionic-components}
\begin{align}
  h^{++mn}_A =&\, - (\theta^+)^2\theta^+_{\hat\alpha}\left[ 2\psi^{\hat\alpha\, mn\,i} +(\Gamma^{(m})^{\hat\alpha}{}_{\hat\beta} \chi^{\hat\beta\, n) i} -\frac15 \eta^{mn} 
  (\Gamma_{p})^{\hat\alpha}{}_{\hat\beta} \chi^{\hat\beta\, p\, i}\right]u^-_i\,,\label{s=3-fermionic-components-a}\\
  h^{++5m}_{A} =&\, (\theta^+)^2\theta^+_{\hat\alpha}\chi^{\hat\alpha\, m\,i}u^-_i\,,\\
  h^{++\hat\alpha+m} =&\, \frac14 (\theta^+)^4 \rho^{\hat\alpha\, m\,i}u^-_i \,.
\end{align}
\end{subequations}
Here $\psi^{\hat\alpha \,mn\, i}$ is symmetric and traceless in the Lorentz indices with 112 independent components while the unconstrained field $\chi^{\hat\alpha \, m \,i}$ has 40 components. The auxiliary component field $\rho^{\hat\alpha \, m \, i}$ contains 32 d.o.f.\ since it is $\Gamma$-traceless, $(\Gamma_m)_{\hat\alpha\hat\beta}\rho^{\hat\beta \, m \, i} = 0$, as a corollary of Eq.~(\ref{eq3.24}). The last two terms in Eq.~(\ref{s=3-fermionic-components-a}) may be eliminated by a redefinition of $\psi^{\hat\alpha\,mn\,i}$, but we keep them to simplify the gauge variation of $\psi^{\hat\alpha\,mn\,i}$ below.

The residual gauge transformations of the prepotentials (\ref{s=3-fermionic-components}) have the form (\ref{s=3-gauge-transformations}) with the gauge superfield parameters given by
\begin{align}
  \lambda^{mn}_A =&\, -\frac{\mathrm{i}}{2}\theta^{+\hat\alpha}\Gamma^{(m}_{\hat\alpha\hat\beta}\epsilon^{\hat\beta\, n) i}u^-_i\,, \\
  \lambda^{5m}_{A} =&\, \frac{\mathrm{i}}{2} \theta^+_{\hat\alpha}\epsilon^{\hat\alpha\, m\, i}u^-_i\,,\\
  \lambda^{\hat\alpha+m} =&\, \frac14 \epsilon^{\hat\alpha\, m\, i} u^+_i -\frac{\mathrm{i}}{4} (\theta^+\Gamma^n\theta^+)\partial_n \epsilon^{\hat\alpha\, m\, i}u^-_i\,,
\end{align}
where the gauge parameter $\epsilon^{\hat\beta\, m\, i }$ is $\Gamma$-traceless, $(\Gamma_m)_{\hat\alpha\hat\beta}\epsilon^{\hat\beta\, m\, i }=0$ because of the constraint (\ref{eq3.24}). Hence, this field has 32 independent components. We find residual gauge transformations of the fermionic component field in the prepotentials (\ref{s=3-fermionic-components}):
\begin{equation}
    \delta \psi^{\hat\alpha\, mn\,i} = \partial^{(m}\epsilon^{\hat\alpha\, n)i} -\frac15 \eta^{mn} \partial_p \epsilon^{\hat\alpha \, p \,i} \,,\quad
    \delta\chi^{\hat\alpha\,m\,i} = (\Gamma^n)^{\hat\alpha}{}_{\hat\beta}\partial_n\epsilon^{\hat\beta\,m\,i}\,,\quad
    \delta \rho^{\hat\alpha\,m\,i} = \Box\epsilon^{\hat\alpha\, m \,i}\,.
    \label{s=3-component-gauge-transformations}
\end{equation}
Thus, we have $152$ fermionic d.o.f.\ off shell which exactly balances the number of bosonic d.o.f.\ in the spin-3 supermultiplet found in Sec.~\ref{Sec:s=3-bosons}. 

This demonstrates that the Ansatz for the prepotentials $h^{++M m_1\ldots m_{s-2}}$ introduced in Sec.~\ref{Sec:prepotentials} is necessary and sufficient to describe an irreducible massless spin-3 supermultiplet. A similar analysis for arbitrary integer $s$ is given in Appendix~\ref{App:component-structure}.

Gauge transformations (\ref{s=3-component-gauge-transformations}) hint that the field $\psi^{\hat\alpha\,mn\,i}$ can be combined with some components of $\chi^{\hat\alpha\, m\,i}$ to form one Fang-Fronsdal field ${\boldsymbol\psi}^{\hat\alpha\,mn\,i}$. This construction goes as follows. The general field $\chi^{\hat\alpha\, m\,i}$ with no symmetry constraints may be decomposed into a $\Gamma$-traceless part $\chi'^{\hat\alpha\,m\,i}$ and a spinor $\phi^{\hat\alpha\,i}$:
\begin{equation}
    \chi^{\hat\alpha\, m\,i} = \chi'^{\hat\alpha\, m\,i} + (\Gamma^m)^{\hat\alpha}{}_{\hat\beta} \phi^{\beta\,i}\,,\qquad
    (\Gamma_m)^{\hat\alpha}{}_{\hat\beta}\chi'^{\hat\beta\,m\,i} = 0\,,
\end{equation}
where
\begin{equation}
    \phi^{\hat\alpha\,i} = -\frac15 (\Gamma_m)^{\hat\alpha}{}_{\hat\beta} \chi^{\hat\beta\,m\,i}\,.
\end{equation}
For these components, the gauge transformations (\ref{s=3-component-gauge-transformations}) imply
\begin{align}
    \delta\chi'^{\hat\alpha\,m\,i} =&\, (\Gamma^n)^{\hat\alpha}{}_{\hat\beta} \partial_n \epsilon^{\hat\beta\,m\,i} - \frac25 (\Gamma^m)^{\hat\alpha}{}_{\hat\beta}\partial_n\epsilon^{\hat\beta\,n\,i}\,,\\
    \delta\phi^{\hat\alpha\,i} =&
    ,\frac25 \partial_m \epsilon^{\hat\alpha\,m\,i}\,.
\end{align}
As a result, the Fang-Fronsdal field may be composed of $\psi^{\hat\alpha\,mn\,i}$ and $\phi^{\hat\alpha\,i}$:
\begin{equation}
    {\boldsymbol\psi}^{\hat\alpha\,mn\,i} = \psi^{\hat\alpha\,mn\,i} + \frac12 \eta^{mn}\phi^{\hat\alpha\,i}\,,\qquad
    \delta{\boldsymbol\psi}^{\hat\alpha\,mn\,i} = \partial^{(m}\epsilon^{\hat\alpha\,n)i}\,.
\end{equation}
The remaining $\Gamma$-traceless components $\chi'^{\hat\alpha\,m\,i}$ and $\rho^{\hat\alpha\,m\,i}$ play the role of auxiliary fields.

\subsection{Gauge-invariant action}
With the prepotentials introduced in Sec.~\ref{Sec:prepotentials}, it is possible to construct the following two candidate actions in the full superspace:
\begin{subequations}
\label{S1andS2}
\begin{align}
  S_1 =& \int \mathrm{d}^{5|8}z \mathrm{d}u\, h^{++m_1\ldots m_{s-1}} h^{--}_{m_1\ldots m_{s-1}} \nonumber\\
  =&\int \mathrm{d}^{5|8}z \mathrm{d}u \mathrm{d}u'\frac{h^{++m_1\ldots m_{s-1}}(z,u) h^{++}_{m_1\ldots m_{s-1}}(z,u')}{(u^+u'^+)^2}  \,,\\
  S_2 =& \int \mathrm{d}^{5|8}z \mathrm{d}u\, h^{++5m_1\ldots m_{s-2}} h^{--5}_{m_1\ldots m_{s-2}} \nonumber\\
  =&\int \mathrm{d}^{5|8}z \mathrm{d}u \mathrm{d}u' \frac{h^{++5m_1\ldots m_{s-2}}(z,u) h^{++5}_{m_1\ldots m_{s-2}}(z,u')}{(u^+u'^+)^2} \,.
\end{align}
\end{subequations}
Using the harmonic zero-curvature equations (\ref{zero-curvature-general}) we find the transformations of the above actions under gauge variations of the prepotentials (\ref{h-gauge-transformation}) 
\begin{align}
  \delta_\lambda S_1 =&-2\int \mathrm{d}^{5|8}z\mathrm{d}u \,\lambda_{m_1\ldots m_{s-1}} D^{--} h^{++m_1\ldots m_{s-1}}\,,\\
  \delta_\lambda S_2 =&-2\int \mathrm{d}^{5|8}z\mathrm{d}u\, \lambda^5_{m_1\ldots m_{s-2}}D^{--}h^{++5m_1\ldots m_{s-2}}\,.
\end{align}

Next, we switch to the analytic superspace measure by the rule (\ref{analytic-measure}) and apply the following corollaries of the analyticity properties of the prepotentials (\ref{h-analyticity-properties-general}) and gauge parameters (\ref{lambda-analyticity-general})
\begin{align}
  (D^+)^4 (\lambda_{m_1\ldots m_{s-1}}D^{--}h^{++m_1\ldots m_{s-1}}) =&\,
  2\mathrm{i}\lambda^{\hat\rho+}_{(m_2\ldots m_{s-1}} (\Gamma_{m_1)})_{\hat\rho}{}^{\hat\alpha} (\Gamma^n)_{\hat\alpha}{}^{\hat\beta} (\Gamma^{m_1})_{\hat\beta\hat\sigma}\nonumber\\&
  \times \partial_n h^{++\hat\sigma+m_2\ldots m_{s-1}}\,,\label{D4-id1}\\
  (D^+)^4(\lambda^5_{m_1\ldots m_{s-2}}D^{--}h^{++5m_1\ldots m_{s-2}})
  =&\,2\mathrm{i}\lambda^{\hat\alpha+}_{m_1\ldots m_{s-2}}\Gamma^n_{\hat\alpha\hat\beta} \partial_n h^{++\hat\beta+m_1\ldots m_{s-2}}\,.
  \label{D4-id2}
\end{align}
Unlike the spin-2 case considered in the previous section, the expressions (\ref{D4-id1}) and (\ref{D4-id2}) have a different structure. The former expression, however, may be brought to the form of the latter upon commuting $\Gamma$-matrices, e.g.,
\begin{equation}
  \Gamma_{m_2} \Gamma^n \Gamma^{m_1} = -2\eta^{nm_1}\Gamma_{m_2} + 2\delta_{m_2}^{m_1}\Gamma^n + \Gamma^{m_1} \Gamma_{m_2}\Gamma^n \,,
\end{equation}
and considering the $\Gamma$-tracelessness properties of the prepotential (\ref{Gamma-tracelessness}) and gauge parameter (\ref{lambda-Gamma-tracelessness}):
\begin{equation}
  (D^+)^4 (\lambda_{m_1\ldots m_{s-1}}D^{--}h^{++m_1\ldots m_{s-1}}) =
  2\mathrm{i}\left(2+\frac1{s-1}\right)\lambda^{\hat\alpha+}_{m_1\ldots m_{s-2}}\Gamma^n_{\hat\alpha\hat\beta}\partial_n h^{++\hat\beta+m_1\ldots m_{s-2}}\,.
\end{equation}
As a result, gauge variations of the actions (\ref{S1andS2}) have the same form
\begin{equation}
  \delta_\lambda S_2 = \left(2+\frac1{s-1}\right)^{-1} \delta_\lambda S_1
  =-4\mathrm{i}\int \mathrm{d}\zeta^{(-4)} \lambda^{\hat\alpha+}_{m_1\ldots m_{s-2}}\Gamma^n_{\hat\alpha\hat\beta} \partial_n h^{++\hat\beta+m_1\ldots m_{s-2}}\,.
\end{equation}
Thus, modulo an overall normalization coefficient, we obtain the gauge-invariant quadratic action for an integer higher spin-$s$ supermultiplet:
\begin{align}
    S =& S_1 -\left(2+\frac1{s-1}\right)S_2 \nonumber\\
    =& \int \mathrm{d}^{5|8}z \mathrm{d}u\left[h^{++m_1\ldots m_{s-1}} h^{--}_{m_1\ldots m_{s-1}} - \frac{2s-1}{s-1} h^{++5m_1\ldots m_{s-2}} h^{--5}_{m_1\ldots m_{s-2}} \right].
\label{spin-s-action}
\end{align}

We stress that the only conditions required for gauge invariance of the action (\ref{spin-s-action}) are the $\Gamma$-tracelessness of the Grassmann-odd prepotential (\ref{Gamma-tracelessness}) and gauge parameter (\ref{lambda-Gamma-tracelessness}). Although not required for gauge invariance, the tracelessness conditions of other prepotentials (\ref{lambda-Gamma-tracelessness}) may be consistently imposed and preserved by gauge symmetry. These conditions single out an irreducible spin-$s$ supermultiplet.


\subsection{Equations of motion}
Although the general variation of the action (\ref{spin-s-action}) has a moderately simple form,
\begin{equation}
  \delta S = 2\int \mathrm{d}^{5|8}z\mathrm{d}u \left(\delta h^{++m_1\ldots m_{s-1}} h^{--}_{m_1\ldots m_{s-1}} - \frac{2s-1}{s-1} \delta h^{++5m_1\ldots m_{s-2}} h^{--5}_{m_1\ldots m_{s-2}} \right),
\label{spin-s-action-variation}
\end{equation}
it is not suitable for obtaining equations of motion because the prepotentials $h^{++Mm_1\ldots m_{s-2}}$ obey the constraints (\ref{h-analyticity-properties-general}). In Eqs.~(\ref{h-Psi-solution}), these constraints are resolved in terms of an unconstrained prepotential $\Psi^{\hat\alpha-m_1\ldots m_{s-2}}$ in the full superspace and an analytic prepotential $v^{++5m_1\ldots m_{s-2}}$ ($D^+_{\hat\alpha}v^{++5m_1\ldots m_{s-2}}=0$). Therefore, the correct gauge-invariant equations of motions for massless higher-spin supermultiplets in 5D harmonic superspace arise from variations of the action (\ref{spin-s-action}) with respect to the unconstrained prepotentials:
\begin{align}
  0=\frac{\delta S}{\delta \Psi^{\hat\alpha-m_1\ldots m_{s-2}}} =& \,\frac{\mathrm{i}}2 \Gamma^{m_{s-1}}_{\hat\alpha\hat\beta} D^{+\hat\beta} (D^+)^2 h^{--}_{m_1\ldots m_{s-1}} 
  -\frac{\mathrm{i}}2 \frac{2s-1}{s-1}D^+_{\hat\alpha} (D^+)^2 h^{--5}_{m_1\ldots m_{s-2}} \nonumber\\
   &-\frac{\mathrm{i}}2 \frac{s-2}{s-1}(\Gamma_{(m_1})_{\hat\alpha}{}^{\hat\beta}(\Gamma^n)_{\hat\beta}{}^{\hat\gamma}D^+_{\hat\gamma}(D^+)^2 h^{--5}_{m_2\ldots m_{s-2})n} \,,\label{EOM1}\\
  0=\frac{\delta S}{\delta v^{++5m_1\ldots m_{s-2}}} =&\, -2\frac{2s-1}{s-1}(D^+)^4 h^{--5}_{m_1\ldots m_{s-2}}\,.
\end{align}
Note that the latter equation is a differential consequence of the former one. The term in the second line of Eq.~(\ref{EOM1}) is required for $\Gamma$-tracelessness of this equation and ensures its $\lambda$-gauge invariance.

\subsection{Hypermultiplet coupling and conserved higher currents}

A procedure for the construction of cubic hypermultiplet vertices with higher-spin supermultiplets in 4D harmonic superspace was developed in Refs.~\cite{Buchbinder:2022kzl,Buchbinder:2022vra}. The structure of these vertices is essentially fixed by the higher-spin gauge symmetry. Following similar ideas, we find the cubic interaction vertex of the hypermultiplet with the higher spin prepotentials in the 5D harmonic superspace in the following general form:
\begin{equation}
  S_\mathrm{int} = -\frac{\kappa}2 \int \mathrm{d}\zeta^{(-4)}q^{+a} \hat H^{++} J^{P(s)} q^+_a\,.
  \label{Sint-general}
\end{equation}
The ingredients of this action are as follows:
\begin{itemize}
\item $\kappa$ is a coupling constant;
\item $J$ is the constant matrix acting on the $SU(2)$ index $a$ of the hypermultiplet $q^{+a}$ as given by Eq.~(\ref{J-matrix}). As was noticed in Refs.~\cite{Buchbinder:2022vra,Buchbinder:2022kzl}, this matrix is required for odd spins $s$ and is absent for interactions with even spin supermultiplets. This condition is taken into account by the parity function $P(s)$:
\begin{equation}
  P(s) := \frac{1+(-1)^{s+1}}{2} = \begin{cases}
    1\,, & s \mbox{ odd}\,,\\
    0\,, & s \mbox{ even}\,.
  \end{cases}
\end{equation}
\item By definition, the operator $\hat H^{++}$ involves the prepotential $h^{++Mm_1\ldots m_{s-2}}$, and higher derivatives,
\begin{align}
  \hat H^{++}q^{+a} :=&\, \frac12\{ h^{++M m_1\ldots m_{s-2}}D_M,\partial_{m_1}\ldots\partial_{m_{s-2}} \}q^{+a} \label{H-hat}\\
  =&\, \frac12 h^{++Mm_1\ldots m_{s-2}}D_M\partial_{m_1}\ldots\partial_{m_{s-2}}q^{+a}
  \nonumber\\&
  +\frac12 \partial_{m_1}\ldots\partial_{m_{s-2}} (h^{++Mm_1\ldots m_{s-2}}D_Mq^{+a})\,.\nonumber
\end{align}
The curly braces stand for the anticommutator of differential operators. The operator $\hat H^{++}$ is similar to $H^{++}$ defined by Eq.~(\ref{H++-general}), but the derivatives $\partial_m$ in the last term in Eq.~(\ref{H-hat}) act on both the prepotential $h^{++Mm_1\ldots m_{s-2}}$ and the hypermultiplet $q^{+a}$. Integrating by parts these derivatives, it is possible to represent the action (\ref{Sint-general}) in the form
\begin{align}
  S_\mathrm{int} =& -\frac{\kappa}4 \int \mathrm{d}\zeta^{(-4)} q^{+a} \{ h^{++Mm_1\ldots m_{s-2}}D_M,\partial_{m_1}\ldots\partial_{m_{s-2}} \} J^{P(s)}q^+_a  \label{Sint-explicit}\\  
  =&-\frac\kappa4 \int \mathrm{d}\zeta^{(-4)} h^{++Mm_1\ldots m_{s-2}}
  [q^{+a}\stackrel{\longleftrightarrow}{D_M} \partial_{m_1}\ldots \partial_{m_{s-2}}J^{P(s)}q^+_a]\,,
\nonumber
\end{align}
where 
\begin{equation}
\begin{aligned}
  q^{+a}\stackrel{\longleftrightarrow}{D_M}\partial_{m_1}\ldots \partial_{m_{s-2}}J^{P(s)} q^+_a := &\,q^{+a}D_M \partial_{m_1}\ldots \partial_{m_{s-2}}J^{P(s)}q^+_a \\& -(D_M q^{+a}) \partial_{m_1}\ldots \partial_{m_{s-2}}J^{P(s)}q^+_a\,.
\end{aligned}
\end{equation}
\end{itemize}

Following Refs.~\cite{Buchbinder:2022vra,Buchbinder:2022kzl} for the 4D higher-spin theory in the harmonic superspace, we define the $\lambda$-gauge transformation of the hypermultiplet in the following way:
\begin{equation}
  \delta_\lambda q^{+a} = -\frac\kappa2\left(\Lambda + \frac12\Omega \right)J^{P(s)}q^{+a}\,,
\label{hyper-gauge-transformation}
\end{equation}
where $\Lambda$ and $\Omega$ are differential operators of the form
\begin{align}
  \Lambda = &\{\lambda^{Mm_1\ldots m_{s-2}}D_M, \partial_{m_1}\ldots \partial_{m_{s-2}} \}\,,\\
  \Omega = &\{ (-1)^{\varepsilon_M} (D_M \lambda^{Mm_1 \ldots m_{s-2}}), \partial_{m_1}\ldots \partial_{m_{s-2}} \}\,.
\end{align}
Here $\lambda^{Mm_1\ldots m_{s-2}}$ are the gauge parameters (\ref{lambda-gauge-parameters}) subject to the constraints (\ref{lambda-analyticity-general}) which ensure that the operators $\Lambda$ and $\Omega$ preserve analyticity. The derivatives in these operators are distributed so that upon integration by parts, the variation of the free hypermultiplet action (\ref{Sq-free}) under (\ref{hyper-gauge-transformation}) takes the form
\begin{align}
  \delta_\lambda S_q = &\,\frac\kappa4 \int \mathrm{d}\zeta^{(-4)} q^{+a}[D^{++},\Lambda] J^{P(s)}q^+_a \nonumber \\
  =&\,\frac\kappa4 \int \mathrm{d}\zeta^{(-4)} q^{+a}\{(D^{++}\lambda^{Mm_1\ldots m_{s-2}})D_M,\partial_{m_1}\ldots \partial_{m_{s-2}} \}J^{P(s)}q^+_a\,.
\end{align}
By comparing this expression with the variation of the action (\ref{Sint-explicit}) under gauge transformations (\ref{h-gauge-transformation}),
\begin{align}
  \delta_\lambda S_\mathrm{int} =& -\frac\kappa4 \int \mathrm{d}\zeta^{(-4)} q^{+a} \{ \delta h^{++Mm_1\ldots m_{s-2}}D_M,\partial_{m_1}\ldots\partial_{m_{s-2}} \} J^{P(s)}q^+_a + O(\kappa^2) 
  \nonumber\\
  =& -\frac\kappa4 \int \mathrm{d}\zeta^{(-4)} q^{+a} \{ D^{++} \lambda^{Mm_1\ldots m_{s-2}}D_M,\partial_{m_1}\ldots\partial_{m_{s-2}} \} J^{P(s)}q^+_a + O(\kappa^2)\,,
\end{align}
we conclude that $S_q + S_\mathrm{int}$ is gauge invariant at order $O(\kappa)$.

Recall that the prepotentials $h^{++Mm_1\ldots m_{s-2}}$ are expressed via the unconstrained prepotentials as in Eq.~(\ref{h-Psi-solution}). These equations imply the following identity
\begin{align}
  \hat H^{++}q^+_a = &-\frac12(D^+)^4\left(\{ \Psi^{\hat\alpha-m_1\ldots m_{s-2}}D^-_{\hat\alpha},\partial_{m_1}\ldots \partial_{m_{s-2}}\} q^+_a\right) 
  \nonumber\\&
  - \frac m2 \{v^{++5m_1\ldots m_{s-2}}, \partial_{m_1}\ldots \partial_{m_{s-2}} \} J q^+_a\,,
\end{align}
which allows us to represent the action (\ref{Sint-explicit}) via unconstrained prepotentials:
\begin{align}
  S_\mathrm{int} =& \frac\kappa4\int \mathrm{d}^{5|8}z \mathrm{d}u\, \Psi^{\hat\alpha-m_1\ldots m_{s-2}}\left( q^{+a} \stackrel{\longleftrightarrow}{D^-_{\hat\alpha}}\partial_{m_1}\ldots\partial_{m_{s-2}}J^{P(s)}q^+_a \right) \nonumber\\
  &+ \kappa\frac m2\int \mathrm{d}\zeta^{(-4)}v^{++5m_1\ldots m_{s-2}} q^{+a}\partial_{m_1}\ldots \partial_{m_{s-2}}  J^{P(s)+1}q^+_a\,. 
\end{align}
The corresponding conserved currents are:
\begin{subequations}
  \label{conserved-currents}
\begin{align}
  {\cal J}^+_{\hat\alpha m_1\ldots m_{s-2}}:=&\,\frac{\kappa^{-1}\delta S_\mathrm{int}}{\delta\Psi^{\hat\alpha-m_1\ldots m_{s-2}}} = \frac14 q^{+a} \stackrel{\longleftrightarrow}{D^-_{\hat\alpha}}\partial_{m_1}\ldots \partial_{m_{s-2}}J^{P(s)} q^+_a -\Gamma\mbox{-traces}\,,\label{eq3.75a}\\
  {\cal T}^{++}_{m_1\ldots m_{s-2}}:=&\,\frac{\kappa^{-1}\delta S_\mathrm{int}}{\delta v^{++5m_1\ldots m_{s-2}}} = \frac m2q^{+a} J^{P(s)+1}\partial_{m_1}\ldots \partial_{m_{s-2}}q^+_a-\mbox{traces}\,.
\end{align}
\end{subequations}

The $\lambda$-gauge transformations of the prepotentials (\ref{l-Psi-gauge-transformations}) imply the conservation conditions of the currents (\ref{conserved-currents}) on the free hypermultiplet equation of motion:
\begin{equation}
  D^{++}q^{+a} = 0\quad\Rightarrow\quad
  D^{++}{\cal J}^+_{\hat\alpha m_1\ldots m_{s-2}} =0\,,\quad
  D^{++}{\cal T}^{++}_{m_1\ldots m_{s-2}}=0\,.
\end{equation}
The $\rho$-gauge symmetry (\ref{pre-gauge-symmetry}) leads to the following conservation laws:
\begin{align}
  D^{+\hat\alpha} {\cal J}^+_{\hat\alpha m_1\ldots m_{s-2}} + 8\mathrm{i} {\cal T}^{++}_{m_1\ldots m_{s-2}} =&\,0\,,\label{DJT-general-s}\\
  D^{+}_{(\hat\alpha} {\cal J}^+_{\hat\beta) m_1\ldots m_{s-2}} =&\,0\,.
  \label{DJ-general-s}
\end{align}
These equations hold on the free hypermultiplet equation of motion (\ref{extra-q-EOM}).

The general solution of Eq.~(\ref{DJ-general-s}) may be written in terms of a symmetric tensor with vanishing $U(1)$ charge, ${\cal J}^+_{\hat\alpha m_1\ldots m_{s-2}} = D^+_{\hat\alpha} {\cal J}_{m_1\ldots m_{s-2}}$. Equation (\ref{DJT-general-s}) implies the following conservation law for this superfield:
\begin{equation}
    (D^{+})^2 {\cal J}_{m_1\ldots m_{s-2}} + 8\mathrm{i}{\cal T}^{++}_{m_1\ldots m_{s-2}} = 0\,.
\end{equation}
Superfields ${\cal J}_{m_1\ldots m_{s-2}}$ and ${\cal T}^{++}_{m_1\ldots m_{s-2}}$ represent higher-spin generalizations of the supercurrent $\cal J$ and multiplet of anomalies ${\cal T}^{++}$ constructed in Sec.~\ref{Sec:supercurrent}. Higher-spin supercurrents in 4D $\cN=2$ field theories in projective and harmonic superspaces were constructed in Refs.~\cite{Kuzenko:2021pqm,Kuzenko:2024vms,Buchbinder:2022vra}.

In the \emph{massless} hypermultiplet model, the supercurrent ${\cal J}_{m_1\ldots m_{s-2}}$ may be found from Eq.~(\ref{eq3.75a}):
\begin{equation}
    {\cal J}_{m_1\ldots m_{s-2}} = \frac14 q^{+a} \stackrel{\longleftrightarrow}{D^{--}}\partial_{m_1}\ldots \partial_{m_{s-2}}J^{P(s)} q^+_a\,,\qquad
    (D^+)^2 {\cal J}_{m_1\ldots m_{s-2}} =0\,.
\end{equation}
This supercurrent is harmonic independent, $D^{++} {\cal J}_{m_1\ldots m_{s-2}}=0$.
This definition, however, is not unique, as there is some freedom in the distribution of the partial derivatives between the two hypermultiplet fields, 
as was demonstrated in \cite{Kuzenko:2023vgf} in the 4D $\cN=2$ case.
The unique higher-spin hypermultiplet supercurrents, compatible with the superconformal symmetry off the mass-shell, were constructed 
in Ref.~\cite{Kuzenko:2024vms} both in the projective and harmonic superspace approaches.
A similar analysis for the hypermultiplet model in 5D harmonic superspace will be carried out elsewhere.


\section{Summary and outlook}
\label{Sec:Summary}

In this paper, we constructed quadratic actions for massless higher-spin supermultiplets in 5D harmonic superspace. Within this framework, the higher-spin supermultiplets are described by the prepotentials $h^{++M m_1\ldots m_{s-2}}=(h^{++m m_1\ldots m_{s-2}}, h^{++\hat\alpha+m_1\ldots m_{s-2}}, h^{++5m_1\ldots m_{s-2}})$ subject to the constraints (\ref{h-analyticity-properties-general}) and gauge transformations (\ref{h-gauge-transformation}). These superfields are symmetric and traceless in Lorentz indices while the Grassmann-odd prepotential $h^{++\hat\alpha+m_1\ldots m_{s-2}}$ is also $\Gamma$-traceless. The latter constraint appears crucial for constructing gauge-invariant actions for $s>2$. These actions are similar to the ones  found in Refs.~\cite{Buchbinder:2021ite,Buchbinder:2022kzl,Buchbinder:2022vra} for higher-spin supermultiplets in 4D harmonic superspace. In our formulation, two of the three prepotentials (\ref{spin-s-prepotentials}) are not analytic while the manifest analyticity of the prepotentials was enforced in the works \cite{BIZ24,Buchbinder:2021ite,Buchbinder:2022kzl,Buchbinder:2022vra}. The advantage of our formulation is the manifest supersymmetry of all constructed superfield actions. 

We showed that the analyticity constraints (\ref{h-analyticity-properties-general}) are resolved via an unconstrained full superspace prepotential $\Psi^{\hat\alpha- m_1\ldots m_{s-2}}$ and an analytic superfield $v^{++5m_1\ldots m_{s-2}}$. Superfield equations of motion arise upon variation of the action with respect to these unconstrained prepotentials. 

We constructed the gauge-invariant cubic interaction vertex of $q$-hypermultiplet with higher-spin supermultiplets. In the harmonic superspace framework, this interaction naturally arises from covariantization of the harmonic superspace derivative $D^{++}$, by analogy with the vector multiplet minimal coupling. This cubic vertex allows us to derive higher conserved currents in the free hypermultiplet model. 

In the particular case of $s=2$, our superfield action describes the 5D linearized supergravity model with 48+48 component degrees of freedom off shell. In the component field form, this supergravity model was studied in Ref.~\cite{Zucker:1999ej}. A non-linear formulation for this variant of supergravity in the harmonic superspace framework remains an open problem. This supergravity theory originates from the conformal supergravity coupled to vector and non-linear multiplets playing the role of conformal compensators. It would be interesting to develop superfield formulations of other variants of 5D supergravity models corresponding to different choices of conformal compensators. We expect that each of these supergravity models can be generalized to a variant formulation of the higher-spin supermultiplets.

In conclusion, we would like to briefly discuss some open problems. First, we note that massless off-shell higher-spin supermultiplets in 5D harmonic superspace can be used to obtain off-shell {\it massive} higher-spin supersymmetric theories in four dimensions by performing dimensional reduction. This was our main motivation for the present work. Second, let us point out that our approach can be extended to study conformal higher-spin supermultiplets. In particular, one can build cubic 5D $q$-hypermultiplet interaction vertices with conformal higher-spin supermultiplets, by analogy with the works in the 4D case \cite{Kuzenko:2021pqm,BIZ24}. Another important problem is to extend the present formulation to the AdS$_5$ harmonic superspace which was introduced in Ref.~\cite{Kuzenko:2007aj}.\footnote{An analytic superspace measure in AdS$_4$ harmonic superspace was explicitly constructed in the recent work \cite{Ivanov:2025jdp}.} Finally, the approach developed in the present paper provides a direction for constructing higher-spin supermultiplets in 6D harmonic superspace because of its similarity to the 5D case. 

The linearized supergravity action \eqref{s2-action} may be rewritten in terms of ordinary superfields. The constrained prepotentials in \eqref{s2-action} can be expressed in terms of $\Psi^{\hat\alpha-}$ and $v^{++5}$ according to the relations \eqref{pre-prepotentials}. Next, the gauge freedom allows us to bring  $\Psi^{\hat\alpha-}$ to the form
\eqref{spinor gauge}, thus resulting in the unconstrained harmonic-independent superfield $\psi^{\hat\alpha i}(z)$. 
Instead of dealing with the analytic prepotential 
$v^{++5}$, it suffices to work with a Mezincescu-type prepotential $V^{ij}(z)$ 
\cite{Mezincescu} which originates as a gauge-fixed version 
of $v^{++5}$, 
\begin{equation}
v^{++5} (z,u) = (D^+)^4 V^{--}(z,u)~,\qquad 
V^{--}(z,u) = V^{ij}(z) u^-_i u^-_j~.
\label{Mez}
\end{equation}
Plugging the expressions \eqref{spinor gauge} and \eqref{Mez} into the action \eqref{s2-action}, one may evaluate the $u$-integral to obtain an integral over the Minkowski  superspace.\footnote{In the 4D $\cN=2$ case, there exist different linearized supergravity actions \cite{Butter:2010sc,Butter:2011ym}
which correspond to different choices of the second compensating supermultiplet. It is plausible that there exist alternative off-shell formulations for massless higher-spin supermultiplets.} The actions for higher-spin supermultiplets may also be recast in terms of ordinary superfields. This will be discussed elsewhere.


\acknowledgments
SMK acknowledges the kind hospitality of the INFN, Sezione di Padova, the Department of Physics of the University of Mons, and the Department of Physics of the University of Munich during his visits in June--July 2025. We are grateful to Jake Stirling for comments on the manuscript. 
This work is supported in part by the Australian Research Council, project No. DP230101629.

\appendix

\section{Notation and conventions}
\label{AppA}

\subsection{5D \texorpdfstring{$\Gamma$}{Gamma}-matrices and spinor notation}

In 5D Minkowski space, spinors carry $USp(2,2)$ indices which we denote by hatted Greek letters, e.g., $\hat\alpha=1,2,3,4$ to distinguish them from $SL(2,\mathbb{C})$ spinor indices $\alpha,\dot\alpha$ in 4D Minkowski space. The corresponding 5D $\Gamma$-matrices $(\Gamma_m)_{\hat\alpha}{}^{\hat\beta}$ obey the Clifford algebra,
\begin{equation}
\{ \Gamma_{m} \, , \,\Gamma_{n} \} 
= - 2 \eta_{ m n} \,\mathds{1}_{4\times4}\,, \qquad
\eta_{mn} = \mathrm{diag}(-1,1,1,1,1)\,.
\end{equation}
The group $USp(2,2)$ possesses invariant antisymmetric tensors $\varepsilon^{\hat \alpha \hat \beta}=-\varepsilon^{\hat\beta\hat\alpha}$ and $\varepsilon_{\hat \alpha \hat \beta}=-\varepsilon_{\hat\beta\hat\alpha}$ which may be chosen in the following $2\times2$ block form 
\begin{equation}
\varepsilon^{\hat \alpha \hat \beta}=
\left(
\begin{array}{cc}
 \varepsilon^{\alpha \beta} &0 \\
0& -\varepsilon_{\dot\alpha \dot \beta}    
\end{array}
\right), \qquad
\varepsilon_{\hat \alpha \hat \beta}=
\left(
\begin{array}{cc}
 \varepsilon_{\alpha \beta} &0 \\
0& -\varepsilon^{\dot \alpha \dot \beta}    
\end{array}
\right),\qquad
\varepsilon_{\hat\alpha\hat\beta}\varepsilon^{\hat\beta\hat\gamma} = \delta_{\hat\alpha}^{\hat\gamma}\,,
\label{epsilon-tensor}
\end{equation}
where $\varepsilon^{\alpha\beta}$ and $\varepsilon^{\dot\alpha\dot\beta}$ are antisymmetric $SL(2,\mathbb{C})$-invariant tensors with $\varepsilon^{12}=\varepsilon_{21}=1$. The tensors (\ref{epsilon-tensor}) may be used to raise and lower spinor indices, e.g.,
\begin{equation}
  (\Gamma_m)^{\hat\alpha\hat\beta} = \varepsilon^{\hat\alpha\hat\gamma} (\Gamma_m)_{\hat\gamma}{}^{\hat\beta}\,,\qquad
  (\Gamma_m)_{\hat\alpha\hat\beta} = \varepsilon_{\hat\beta\hat\gamma} (\Gamma_m)_{\hat\alpha}{}^{\hat\gamma}\,.
\end{equation}
The $\Gamma$-matrices with the spinor indices on the same level are antisymmetric and $\varepsilon$-traceless:
\begin{equation}
  (\Gamma_m)_{\hat\alpha\hat\beta} = -(\Gamma_m)_{\hat\beta\hat\alpha}\,, \qquad
  \varepsilon^{\hat\alpha\hat\beta} (\Gamma_m)_{\hat\alpha\hat\beta} = 0\,.
\end{equation}

The matrices 
$\{ \mathds{1}, \Gamma_{m} , \Sigma_{mn} \} $ form a basis in the space of  $4 \times 4$ matrices, where 
\begin{equation}
\Sigma_{mn} =-\Sigma_{nm} := -\frac14 [\Gamma_{m} , \Gamma_{n} ]\,.
\end{equation}
$\Gamma$ and $\Sigma$ matrices allow us to convert the Lorentz indices into the spinor ones for vectors $V_m$ and antisymmetric tensors $F_{mn}$:
\begin{align} 
  V_{\hat \alpha \hat \beta} &= V^{m} ( \Gamma_{m})_{\hat \alpha \hat \beta}\,,
  & V_{m}  &= -\frac14 \,( \Gamma_{m})^{\hat \alpha \hat \beta} V_{\hat \alpha \hat \beta}\,, \\
  F_{\hat \alpha \hat \beta} &= \frac12 F^{mn} (\Sigma_{mn})_{\hat \alpha \hat \beta}\,,  & 
  F_{mn}  &= (\Sigma_{mn})^{\hat \alpha \hat \beta} F_{\hat \alpha \hat \beta}\,.
\end{align} 
These results can be easily checked using the identities (see e.g. \cite{Z}):
\begin{align}
 \varepsilon _{\hat \alpha \hat \beta \hat \gamma \hat \delta} 
=&\, \varepsilon_{\hat \alpha \hat \beta} \, \varepsilon_{\hat \gamma \hat \delta}
+ \varepsilon_{\hat \alpha \hat \gamma} \, \varepsilon_{\hat \delta \hat \beta}
+\varepsilon_{\hat \alpha \hat \delta} \, \varepsilon_{\hat \beta \hat \gamma}\,,  \\
 \varepsilon_{\hat \alpha \hat \gamma} \, \varepsilon_{\hat \beta \hat \delta}
-\varepsilon_{\hat \alpha \hat \delta} \, \varepsilon_{\hat \beta \hat \gamma}
=&\,-\frac12 ( \Gamma^{\hat m})_{\hat \alpha \hat \beta}
( \Gamma_{\hat m})_{\hat \gamma \hat \delta}
+\frac12 \varepsilon_{\hat \alpha \hat \beta} \varepsilon_{ \hat \gamma \hat \delta}\,,\\
 \varepsilon_{\hat \alpha \hat \beta \hat \gamma \hat \delta} 
=&\,\frac12( \Gamma^{\hat m})_{\hat \alpha \hat \beta}( \Gamma_{\hat m})_{\hat \gamma \hat \delta}
+\frac12 \varepsilon_{\hat \alpha \hat \beta} \varepsilon_{ \hat \gamma \hat \delta}\,,
\end{align}
with $\varepsilon _{\hat \alpha \hat \beta \hat \gamma \hat \delta} $ the completely antisymmetric rank-four tensor. Complex conjugation gives 
\begin{equation}
(\varepsilon_{\hat \alpha \hat \beta})^* = - \varepsilon^{\hat \alpha \hat \beta}\,,
\qquad 
(V_{\hat \alpha \hat \beta})^* = V^{\hat \alpha \hat \beta}\,,
\qquad 
(F_{\hat \alpha \hat \beta})^* = F^{\hat \alpha \hat \beta}\,,
\end{equation}
provided $V^{m}$ and  $F^{mn} $ are real.

Identities with products of $\Gamma$-matrices:
\begin{subequations}
\label{3Gamma}
\begin{align}
  \Gamma^m\Gamma^n = &\, -\eta^{mn}\mathds{1} - 2\Sigma^{mn}\,,\\
  \Gamma^m\Gamma^n\Gamma^p = &\, (-\eta^{mn}\eta^{pk} + \eta^{pm}\eta^{nk} - \eta^{np}\eta^{mk})\Gamma_k + \varepsilon^{mnpkl}\Sigma_{kl}\,,\label{3-Gamma}\\
  \Gamma^m\Gamma^n\Gamma^r\Gamma^s =&\, (\eta^{mn}\eta^{rs}-\eta^{mr}\eta^{ns}+\eta^{ms}\eta^{nr})\mathds{1} - \varepsilon^{mnrsp}\Gamma_p + 2\eta^{mn}\Sigma^{rs} \nonumber\\
  &-2\eta^{mr}\Sigma^{ns}+2\eta^{nr}\Sigma^{ms}+2\eta^{rs}\Sigma^{mn} -2\eta^{ns}\Sigma^{mr} +2\eta^{ms}\Sigma^{nr}\,,\\
  \Gamma^m\Gamma^n\Gamma^r\Gamma^s\Gamma^t =&\,\varepsilon^{mnrst}\mathds{1} + \Gamma^m(\eta^{nr}\eta^{st}-\eta^{ns}\eta^{rt}+\eta^{rs}\eta^{nt}) \nonumber\\&
  +\Gamma^n(-\eta^{rs}\eta^{mt}+\eta^{rt}\eta^{ms}-\eta^{st}\eta^{mr}) + \Gamma^r(\eta^{st}\eta^{mn}-\eta^{ms}\eta^{nt}+\eta^{mt}\eta^{ns})\nonumber\\&
  +\Gamma^s(-\eta^{mt}\eta^{nr}+\eta^{nt}\eta^{mr}-\eta^{mn}\eta^{rt}) + \Gamma^t(\eta^{mn}\eta^{rs}-\eta^{mr}\eta^{ns}+\eta^{nr}\eta^{ms})\nonumber\\&
  +2\varepsilon^{mnrsp}\Sigma_p{}^t - \eta^{mn}\varepsilon^{rstpq}\Sigma_{pq} + \eta^{mr}\varepsilon^{nstpq}\Sigma_{pq} - \eta^{nr}\varepsilon^{mstpq}\Sigma_{pq} \nonumber\\&
  -\eta^{ms}\varepsilon^{nrtpq}\Sigma_{pq} + \eta^{ns}\varepsilon^{nrtpq}\Sigma_{pq} - \eta^{rs}\varepsilon^{mstpq}\Sigma_{pq}\,,
\end{align}
\end{subequations}
where $\varepsilon^{mnrsp}$ is the completely antisymmetric tensor with $\varepsilon^{01235} = -1$.

\subsection{5D \texorpdfstring{$\cN=1$}{N=1} harmonic superspace with central charge}
The flat 5D superspace with central charge is parametrized by  coordinates $(x^m,  \theta^{\hat \alpha}_i, x^5 )$, where the Grassmann variable obeys the pseudo-Majorana reality condition, $(\theta_i^{\hat\alpha})^* = \theta^i_{\hat\alpha}$, and $x^5$ stands for the central charge variable. Here $i=1,2$ is the $SU(2)_R$ index. Covariant spinor derivatives in this superspace may be chosen in the form
\begin{equation}
  D^i_{\hat \alpha} = \frac{\partial}{\partial \theta^{\hat \alpha}_i} - \mathrm{i} \Gamma^m_{\hat \alpha \hat \beta} \theta^{\hat \beta i} \partial_m - \mathrm{i}\theta^i_{\hat\alpha}\partial_5\,,
\end{equation}
with $\partial_5$ the central charge operator. These derivatives obey the algebra
\begin{equation}
  \{D^i_{\hat \alpha}, D^j_{\hat \beta} \} = -2\mathrm{i}
\varepsilon^{ij}\Gamma^m_{\hat \alpha \hat \beta} \partial_m - 2\mathrm{i}\varepsilon_{\hat \alpha \hat \beta}\varepsilon^{ij} \partial_5 \,,
\qquad [ D^i_{\hat \alpha}, \partial_m ] = [ D^i_{\hat \alpha} ,\partial_5 ] = 0\,.
\end{equation}

Harmonic variables $u^{\pm}_i$ obey the orthogonality and completeness conditions,
\begin{equation}
  u^{+i}u^{-}_i = 1\,,\quad
  u^{+i}u^{+}_i = u^{-i}u^{-}_i = 0\,,\quad
  u^{+i}u^{-}_j - u^{-i}u^{+}_j = \delta^i_j\,.
\end{equation}
These may be used to define harmonic projections of Grassmann coordinates $\theta^{\hat\alpha\pm} = u^\pm_i \theta^{\hat\alpha i}$ and covariant spinor derivatives, 
\begin{equation}
  D^\pm_{\hat\alpha} := u^\pm_i D^i_{\hat\alpha} = \pm \frac{\partial}{\partial \theta^{\mp\hat\alpha}} - \mathrm{i}\Gamma^m_{\hat\alpha\hat\beta} \theta^{\pm\hat\beta}\partial_m - \mathrm{i}\theta^\pm_{\hat\alpha}\partial_5\,.
  \label{D-central}
\end{equation}
They obey the following anticommutation relations:
\begin{equation}
  \{ D^+_{\hat\alpha} , D^-_{\hat\beta} \} = 2\mathrm{i} \Gamma^{m}_{\hat\alpha\hat\beta} \partial_{m} + 2\mathrm{i}\varepsilon_{\hat\alpha\hat\beta}\partial_5\,,\qquad
  \{ D^+_{\hat\alpha} , D^+_{\hat\beta} \} = \{ D^-_{\hat\alpha} , D^-_{\hat\beta} \} = 0\,.
\end{equation}

The harmonic derivatives in the central basis are defined as
\begin{equation}
\begin{aligned}
  D^{++}  = &\,u^{+i} \frac{\partial}{\partial u^{-i}} \equiv \partial^{++}\,,\qquad
  D^{--}  = u^{-i} \frac{\partial}{\partial u^{+i}} \equiv \partial^{--} \,,\\
  D^0  = &\,u^{+i} \frac{\partial}{\partial u^{+i}} - u^{-i} \frac{\partial}{\partial u^{-i}} \equiv \partial^0\,.
  \label{D-harmonic-central}
\end{aligned}
\end{equation}
They obey the commutation relations of the $su(2)$ algebra:
\begin{equation}
  [D^{++}, D^{--}] = D^0\,,\quad
  [D^0, D^{++}] = 2D^{++}\,,\quad
  [D^0, D^{--}] = -2D^{--}\,.
\end{equation}

The crucial feature of the harmonic superspace is the existence of an invariant analytic subspace $(x^m_A,\theta^+_{\hat\alpha},u^\pm_i,x^5_A)$ in the full superspace. Here the coordinates $x^m_A$ and $x^5_A$ are defined as
\begin{equation}
  x^m_A = x^m + \mathrm{i}\theta^{+\hat\alpha}\Gamma^m_{\hat\alpha\hat\beta} \theta^{-\hat\beta}\,,\qquad
  x^5_A = x^5 + \mathrm{i}\theta^{+\hat\alpha}\theta^-_{\hat\alpha}\,.
\label{analytic-x}
\end{equation}
In the analytic basis, the covariant spinor derivatives (\ref{D-central}) have the following form:
\begin{subequations}
\label{analytic-derivatives1}
\begin{align}
  D^+_{\hat\alpha} =&\frac\partial{\partial\theta^{-\hat\alpha}}\,,\label{D+analytic}\\
  D^-_{\hat\alpha} =&-\frac\partial{\partial\theta^{+\hat\alpha}} - 2\mathrm{i}\Gamma^m_{\hat\alpha\hat\beta}\theta^{-\hat\beta}\partial_{Am} - 2\mathrm{i}\theta^-_{\hat\alpha}\partial_{A5}\,,
\end{align}
\end{subequations}
where $\partial_{Am} = \partial/\partial x^m_A$ and $\partial_{A5} = \partial/\partial x^5_A$. Harmonic derivatives (\ref{D-harmonic-central}) are now given by the following expressions:
\begin{subequations}
\label{analytic-derivatives2}
\begin{align}
  D^{++} =&\, \partial^{++} + \theta^{+\hat\alpha} \frac\partial{\partial\theta^{-\hat\alpha}} + \mathrm{i}\theta^{+\hat\alpha}\Gamma^m_{\hat\alpha\hat\beta}\theta^{+\hat\beta}\partial_{Am} + \mathrm{i}(\theta^+)^2\partial_{A5}\,,\\
  D^{--} =&\, \partial^{--} +\theta^{-\hat\alpha}\frac\partial{\partial\theta^{+\hat\alpha}} + \mathrm{i}\theta^{-\hat\alpha}\Gamma^m_{\hat\alpha\hat\beta}\theta^{-\hat\beta}\partial_{Am} +\mathrm{i}(\theta^-)^2\partial_{A5}\,,\\
  D^0 =&\, \partial^0 + \theta^{+\hat\alpha}\frac\partial{\partial\theta^{+\hat\alpha}} - \theta^{-\hat\alpha}\frac\partial{\partial\theta^{-\hat\alpha}}\,.
\end{align}
\end{subequations}


\section{Component structure}
\label{App:component-structure}
It is convenient to study the component structure of prepotentials in the analytic basis with coordinates $(x^m_A,\theta^+_{\hat\alpha},u^\pm_i,x^5_A)$ which are related to the central basis $(x^m,\theta^\pm_{\hat\alpha},u^\pm_i,x^5)$ through the identities (\ref{analytic-x}). In this basis, the analyticity constraint is satisfied automatically because the corresponding covariant spinor derivative is short, see Eq.~(\ref{D+analytic}).

\subsection{Prepotentials in the Wess-Zumino gauge: bosonic fields}
\label{AppB1}
In harmonic superspace, 5D higher spin supermultiplets are embedded into three prepotentials $h^{++m_1\ldots m_{s-1}}$, $h^{++5m_1\ldots m_{s-2}}$ and $h^{++\hat\alpha+ m_1\ldots m_{s-2}}$ subject to the analyticity constraints (\ref{h-analyticity-properties-general}). These constraints may be solved explicitly in the analytic coordinates:
\begin{align}
  h^{++m_1\ldots m_{s-1}} =&\, h^{++m_1\ldots m_{s-1}}_A(x_A,\theta^+,u) + 2\mathrm{i}\theta^{-\hat\alpha}\Gamma^{(m_1}_{\hat\alpha\hat\beta} h^{++\hat\beta+m_2\ldots m_{s-1})}(x_A,\theta^+,u)\,,\\
  h^{++5m_1\ldots m_{s-2}} =&\, h^{++5m_1\ldots m_{s-2}}_{A}(x_A,\theta^+,u) - 2\mathrm{i}\theta^-_{\hat\alpha} h^{++\hat\alpha+m_1\ldots m_{s-2}}(x_A,\theta^+,u) \,,
\end{align}
and similar relations hold for the corresponding gauge parameters:
\begin{align}
  \lambda^{m_1\ldots m_{s-1}} =&\, \lambda^{m_1\ldots m_{s-1}}_A(x_A,\theta^+,u) + 2\mathrm{i}\theta^{-\hat\alpha}\Gamma^{(m_1}_{\hat\alpha\hat\beta} \lambda^{\hat\beta+m_2\ldots m_{s-2})}(x_A,\theta^+,u)\,,\\
  \lambda^{5m_1\ldots m_{s-2}} =&\, \lambda^{5m_1\ldots m_{s-2}}_{A}(x_A,\theta^+,u) - 2\mathrm{i}\theta^-_{\hat\alpha} \lambda^{\hat\alpha+m_1\ldots m_{s-2}}(x_A,\theta^+,u) \,.
\end{align}
For these components, the $\lambda$-gauge transformations (\ref{h-gauge-transformation}) take the form
\begin{subequations}
\label{analytic-gauge-transformations}
\begin{align}
  \delta h^{++m_1\ldots m_{s-1}}_A =&\, D^{++}\lambda^{m_1\ldots m_{s-1}}_A + 2\mathrm{i}\theta^{+\hat\alpha}\Gamma^{(m_1}_{\hat\alpha\hat\beta}\lambda^{\hat\beta+m_2\ldots m_{s-1})}\,,\\
  \delta h^{++5m_1\ldots m_{s-2}}_{A} =&\, D^{++}\lambda^{5m_1\ldots m_{s-2}}_{A} - 2\mathrm{i}\theta^+_{\hat\alpha}\lambda^{\hat\alpha+m_1\ldots m_{s-2}}\,,\\
  \delta h^{++\hat\alpha+m_1\ldots m_{s-2}}=&\, D^{++}\lambda^{\hat\alpha+m_1\ldots m_{s-2}}\,.
\end{align}
\end{subequations}
This gauge freedom allows us to gauge away infinite tails of auxiliary fields with $SU(2)_R$ indices, leaving only the following bosonic components in $\theta$-decompositions of the prepotentials:
\begin{subequations}
\label{components-in-prepotentials}
\begin{align}
  h^{++m_1\ldots m_{s-1}}_A =&\, \mathrm{i}(\theta^+\Gamma_n\theta^+) B^{n\,m_1\ldots m_{s-1}}(x_A) + (\theta^+)^2 f^{m_1\ldots m_{s-1}}(x_A)
  \nonumber\\&+ (\theta^+)^4 V^{m_1\ldots m_{s-1}ij}(x_A)u^-_i u^-_j\,,\\
  h^{++5m_1\ldots m_{s-2}}_{A} =&\, \mathrm{i}(\theta^+\Gamma_n \theta^+) A^{n\, m_1\ldots m_{s-2}}(x_A) +(\theta^+)^2\varphi^{m_1\ldots m_{s-2}}(x_A) 
  \nonumber\\&+ (\theta^+)^4 D^{m_1\ldots m_{s-2}ij}(x_A)u^-_i u^-_j\,,\\
  h^{++\hat\alpha+m_1\ldots m_{s-2}} =&\, (\theta^+)^2\theta^+_{\hat\beta} P^{\hat\alpha\hat\beta m_1\ldots m_{s-2}}(x_A) \label{B10}\,.
\end{align}
\end{subequations}
As follows from the constraints (\ref{h-tracelessness}), all of these tensors are symmetric and traceless in Lorentz indices $m_1,m_2,\ldots ,m_{s-1}$. In addition, the component $P^{\hat\alpha\hat\beta m_1\ldots m_{s-2}}$ is $\Gamma$-traceless,
\begin{equation}
    (\Gamma_{m_1})_{\hat\alpha\hat\beta}P^{\hat\beta\hat\gamma m_1\ldots m_{s-2}}=0\,, 
\label{P-Gamma-traceless}
\end{equation}
which follows from Eq.~(\ref{Gamma-tracelessness}).

In the Wess-Zumino gauge, residual gauge transformations are given by the following parameters
\begin{align}
  \lambda^{m_1\ldots m_{s-1}}_A =&\, b^{m_1\ldots m_{s-1}}(x_A)\,,\\
  \lambda^{5m_1\ldots m_{s-2}}_A =&\, a^{m_1\ldots m_{s-2}}(x_A)\,,\\
  \lambda^{\hat\alpha+m_1\ldots m_{s-2}} =&\, \theta^+_{\hat\beta} l^{\hat\alpha\hat\beta m_1\ldots m_{s-2}}(x_A)\,.
\end{align}
where all tensors are symmetric and traceless in vector indices, and $l^{\hat\alpha\hat\beta m_1\ldots m_{s-2}}$ is $\Gamma$-traceless:
\begin{equation}
  (\Gamma_{m_1})_{\hat\alpha\hat\beta}l^{\hat\beta\hat\gamma m_1\ldots m_{s-2}} =0 \,.
\label{l-Gamma-traceless}
\end{equation}

The spinor indices in the tensor $l^{\hat\alpha\hat\beta m_1\ldots m_{s-2}}$ may be converted into the vector ones by the rule
\begin{equation}
  l^{\hat\alpha\hat\beta m_1\ldots m_{s-2}} = \varepsilon^{\hat\alpha\hat\beta} l_1^{m_1\ldots m_{s-2}} + (\Gamma_n)^{\hat\alpha\hat\beta} l_2^{n (m_1\ldots m_{s-2})} + \frac12 (\Sigma_{rs})^{\hat\alpha\hat\beta} l_3^{[rs](m_1\ldots m_{s-2})}\,,
\label{l-decomposition}
\end{equation}
where
\begin{align}
  l_1^{m_1\ldots m_{s-2}}=&-\frac14 \varepsilon_{\hat\alpha\hat\beta}l^{\hat\alpha\hat\beta m_1\ldots m_{s-2}} = \begin{array}{|c|}\hline
   s-2\\\hline
  \end{array}\,'
    \,,\\
  l_2^{n (m_1\ldots m_{s-2})} =&-\frac14 \Gamma^n_{\hat\alpha\hat\beta} l^{\hat\alpha\hat\beta m_1\ldots m_{s-2}} = \begin{array}{|c|}\hline
   \phantom{m}\\\hline
  \end{array}\times \begin{array}{|c|}\hline
   s-2\\\hline
  \end{array}\,'\,,\\
  l_3^{[rs](m_1\ldots m_{s-2})} =&\, \Sigma^{rs}_{\hat\alpha\hat\beta} l^{\hat\alpha\hat\beta m_1\ldots m_{s-2}}= \begin{array}{|c|}\hline
   \phantom{m}\\\hline\phantom{m}\\\hline
  \end{array}\times \begin{array}{|c|}\hline
   s-2\\\hline
  \end{array}\,'\,.
\end{align}
Here the rectangles represent Young's tableau with a prime for the tracelessness constraint. The constraint (\ref{l-Gamma-traceless}) implies the following relations between these components
\begin{align}
  \eta_{nm}l_2^{n(mm_2\ldots m_{s-2})} =&\,0\label{EqB20}\,,\\
  l_1^{m_1\ldots m_{s-2}} - \frac12 \eta_{rs}l_3^{[r (m_1]m_2\ldots m_{s-2})s} =&\,0\,,\label{l1-via-l3}\\
  l_2^{[p(k]m_2\ldots m_{s-2})} +\frac1{8} \varepsilon^{pk}{}_{rsn}l_3^{[rs](nm_2\ldots m_{s-2})} =&\,0 \,.
  \label{l2-via-l3}
\end{align}
The same constraints hold for the components of the auxiliary field $P^{\hat\alpha\hat\beta m_1\ldots m_{s-1}}$ in Eq.~(\ref{P-Gamma-traceless}).

Making use of the identities
\begin{align}
  \theta^{+}_{\hat\alpha}\lambda^{\hat\alpha+m_1\ldots m_{s-2}} =&\, -(\theta^+)^2 l_1^{m_1\ldots m_{s-2}} + (\theta^+\Gamma_n\theta^+) l_2^{n(m_{m_1\ldots m_{s-2}})}\,,\\
  \theta^{+\hat\alpha}\Gamma^{(m_1}_{\hat\alpha\hat\beta} \lambda^{\hat\beta+m_2\ldots m_{s-2})} =&\, (\theta^+)^2 l_2^{(m_1\ldots m_{s-1})} + (\theta^+\Gamma^{(m_1}\theta^+)l_1^{m_2\ldots m_{s-1})} \nonumber\\& - \frac12(\theta^+\Gamma_n\theta^+)l_3^{[n(m_1]\ldots m_{s-1})}\,,
\end{align}
we find residual gauge symmetry of the prepotentials,
\begin{align}
  \delta h_A^{++m_1\ldots m_{s-1}} =&\, \mathrm{i}(\theta^+\Gamma_n\theta^+)\delta B^{n\,m_1\ldots m_{s-1}} + (\theta^+)^2\delta f^{m_1\ldots m_{s-1}} + (\theta^+)^4\delta V^{m_1\ldots m_{s-1}\,ij}u^-_i u^-_j\,,\nonumber\\
  =&\,\mathrm{i}(\theta^+\Gamma_n\theta^+)\partial^n b^{m_1\ldots m_{s-1}} + 2\mathrm{i}(\theta^+\Gamma^{(m_1}\theta^+)l_1^{m_2\ldots m_{s-1})} + 2\mathrm{i}(\theta^+)^2 l_2^{(m_1\ldots m_{s-1})}
  \nonumber\\&\, -\mathrm{i}(\theta^+\Gamma_n\theta^+) l_3^{[n(m_1]m_2\ldots m_{s-1})}\,,\label{EqB25}\\
  \delta h^{++5m_1\ldots m_{s-2}}_{A} =&\, \mathrm{i}(\theta^+\Gamma_n\theta^+)\delta A^{n\,m_1\ldots m_{s-2}} + (\theta^+)^2\delta\varphi^{m_1\ldots m_{s-2}} + (\theta^+)^4\delta D^{m_1\ldots m_{s-2}\,ij}u^-_iu^-_j\nonumber\\
  =&\,\mathrm{i}(\theta^+\Gamma_n\theta^+)\partial^n a^{m_1\ldots m_{s-2}} +2\mathrm{i}(\theta^+)^2l_1^{m_1\ldots m_{s-2}} \nonumber\\ &
  -2\mathrm{i} (\theta^+\Gamma_n\theta^+) l_2^{n(m_1\ldots m_{s-2})}\,,\label{EqB26}\\
  \delta h^{++\hat\alpha m_1\ldots m_{s-2}} =& \,(\theta^+)^2 \theta^+_{\hat\beta} \delta P^{\hat\alpha\hat\beta m_1\ldots m_{s-2}} = \mathrm{i}(\theta^+)^2 \theta^+_{\hat\beta} (\Gamma_n)_{\hat\gamma}{}^{\hat\beta} \partial^n l^{\hat\alpha\hat\gamma m_1\ldots m_{s-2}}\,.
  \label{EqB27}
\end{align}

Next, we will use the gauge freedom represented by the tensors $l_1$, $l_2$ and $l_3$ to gauge away any remaining non-physical degrees of freedom and achieve the conventional gauge transformations for the physical ones. We do this in the following steps:
\begin{itemize}
\item The field $l_1^{m_1\ldots m_{s-2}}$ is symmetric and traceless. It may be used to fully gauge away $\varphi^{m_1\ldots m_{s-1}}$ from $h_A^{++5m_1\ldots m_{s-2}}$. In the gauge $\varphi^{m_1\ldots m_{s-2}} = 0$ we have
$\delta\varphi^{m_1\ldots m_{s-2}} = 2\mathrm{i} l_1^{m_1\ldots m_{s-2}} =0$.
\item Eq.~(\ref{EqB20}) shows that the tensor $l_2^{n(m_1\ldots m_{s-2})}$ is traceless in all indices. Thus, its totally symmetric component $l_2^{(m_1\ldots m_{s-1})}$ may be used to gauge away $f^{{m_1\ldots m_{s-1}}}$ from $h_A^{++m_1\ldots m_{s-1}}$. In the gauge $f^{{m_1\ldots m_{s-1}}} = 0$ we have $\delta f^{{m_1\ldots m_{s-1}}} = 2\mathrm{i} l_2^{(m_1\ldots m_{s-1})} =0$.
\item Eq.~(\ref{EqB26}) gives the following gauge transformation law for the tensor $A$:
\begin{equation}
  \delta A^{n\, m_1\ldots m_{s-2}} = \partial^n a^{m_1\ldots m_{s-2}} - 2 l_2^{n(m_1\ldots m_{s-2})}\,.
  \label{B.24}
\end{equation}
Recall that $l_2^{n(m_1\ldots m_{s-2})}$ is traceless with vanishing totally symmetric part because of the gauge fixing imposed in the previous step. Hence, it may be used to gauge away hook-type mixed symmetry components from the tensor $A$, leaving only the symmetric part. The number of independent components in $l_2$ is
\begin{equation}
5\left(
\begin{array}{c}
  s+2 \\ s-2
\end{array}
\right) 
-5\left(
\begin{array}{c}
  s \\ s-4
\end{array}
\right)
-\left(
\begin{array}{c}
  s+3 \\ s-1
\end{array}
\right)
+\left(
\begin{array}{c}
  s-1 \\ s-5
\end{array}
\right),
\end{equation}
while the tensor $A^{n\,m_1\ldots m_{s-2}}$ contains 
\begin{equation}
5\left(
\begin{array}{c}
  s+2 \\ s-2
\end{array}
\right) 
-5\left(
\begin{array}{c}
  s \\ s-4
\end{array}
\right)
\end{equation}
independent components before gauge fixing. Making use of the gauge freedom in $l_2$ we are left with 
\begin{equation}
\left(
\begin{array}{c}
  s+3 \\ s-1
\end{array}
\right)
-\left(
\begin{array}{c}
  s-1 \\ s-5
\end{array}
\right)
\end{equation}
independent components in the tensor $A$, which is exactly the number of components in a double-traceless tensor of rank $(s-1)$. Such a double-traceless tensor may be decomposed into a pair of totally symmetric and traceless tensors with $(s-1)$ and $(s-3)$ Lorentz indices, respectively. This suggests the following decomposition of the tensor $A^{n\,m_1\ldots m_{s-1}}$ provided that its mixed-symmetry components are gauged away using the gauge freedom in $l_2$:
\begin{equation}
    A^{n\, m_1\ldots m_{s-2}} = A_1^{nm_1\ldots m_{s-2}} + \eta^{n(m_1}A_2^{m_2\ldots m_{s-2})} - \frac{s-3}{2s-3}\eta^{(m_1m_2}A_2^{m_3\ldots m_{s-2
    })n} \,.
    \label{A1A2-decomposition}
\end{equation}
Here both $A_1$ and $A_2$ are totally symmetric and traceless,
\begin{align}
    A_1^{m_1\ldots m_{s-1}} =&\, A_1^{(m_1\ldots m_{s-1})}\,,&
    \eta_{m_1m_2} A_1^{m_1m_2\ldots m_{s-1}}=&\,0\,,\\
    A_2^{m_1\ldots m_{s-3}} =&\, A_2^{(m_1\ldots m_{s-3})}\,,&
    \eta_{m_1m_2} A_2^{m_1m_2\ldots m_{s-3}}=&\,0\,.
\end{align}
In this gauge, the hook-type components of the tensor $l_2$ obey
\begin{equation}
    l_2^{[n(m_1]m_2\ldots m_{s-2})} = \frac12 \partial^{[n}a^{(m_1]m_2\ldots m_{s-2})} - \mbox{traces}.
\label{l2-via-a}
\end{equation}

Equation (\ref{A1A2-decomposition}) implies
\begin{equation}
    \eta_{nm_1} A^{n\,m_1m_2\ldots m_{s-2}} = \frac{s(2s-1)}{(s-2)(2s-3)} A_2^{m_2\ldots m_{s-2}}\,.
\end{equation}
Remembering that $l_2^{n(m_1\ldots m_{s-2})}$ is traceless, from Eq.~(\ref{B.24}) we find the gauge transformation of the tensor $A_2$:
\begin{equation}
    \delta A_2^{m_1\ldots m_{s-3}} = \frac{(s-2)(2s-3)}{s(2s-1)}\partial_n a^{nm_1\ldots m_{s-3}}\,.
\end{equation}
The same equation (\ref{B.24}) dictates the gauge transformation law for $A_1$:
\begin{equation}
    \delta A_1^{m_1\ldots m_{s-1}} = \partial^{(m_1} a^{m_2\ldots m_{s-1})} - \frac{s-2}{2s-1}\eta^{(m_1m_2} \partial_n a^{m_3\ldots m_{s-1})n}\,.
\end{equation}

The two tensors $A_1$ and $A_2$ may be combined into one totally symmetric and double-traceless tensor $\boldsymbol{A}^{m_1\ldots m_{s-1}}=\boldsymbol{A}^{(m_1\ldots m_{s-1})}$
\begin{equation}
    {\boldsymbol A}^{m_1\ldots m_{s-1}} = A_1^{m_1\ldots m_{s-1}} + \frac{s}{2s-3}\eta^{(m_1m_2}A_2^{m_3\ldots m_{s-1})}\,,\qquad
    {\boldsymbol A}''=0\,,
\end{equation}
with gauge freedom
\begin{equation}
  \delta{\boldsymbol A}^{m_1\ldots m_{s-1}} = \partial^{(m_1} a^{m_2\ldots m_{s-1})}\,.
\end{equation}
Note that for $s\leq4$ the condition of double-tracelessness of the tensor $\boldsymbol A$ is not required. This tensor $\boldsymbol A$, however, cannot be embedded as a component into the superfield $h^{++5m_1\ldots m_{s-2}}_{A}$. Rather, the two tensors $A_1$ and $A_2$ in the combination (\ref{A1A2-decomposition}) enter this superfield.

\item 
In a similar way we single out physical degrees of freedom in $B^{n\,m_1\ldots m_{s-1}} = 
\begin{array}{|c|}
  \hline
  \phantom{m}\\
  \hline
\end{array}\times \begin{array}{|c|}
  \hline
  s-1\\
  \hline
\end{array}\,'$ which contains 
\begin{equation}
  5\left(\begin{array}{c}
    s+3 \\s-1
  \end{array}\right) - 5\left(\begin{array}{c}
    s+1 \\s-3
  \end{array}\right)
  \label{N-comp-before-gf}
\end{equation}
independent components before gauge fixing. Eq.~(\ref{EqB25}) shows that the tensor $B$ has the following transformation law:
\begin{equation}
    \delta B^{n\,m_1\ldots m_{s-1}} = \partial^n b^{m_1\ldots m_{s-1}} - l_3^{[n(m_1]m_2\ldots m_{s-1})}\,.
    \label{B.38}
\end{equation}
We will show that the gauge freedom remaining in $l_3$ is sufficient to gauge away all mixed symmetry components from the tensor $B$. Some of the components of this gauge parameter have already been used for fixing other degrees of freedom owing to Eqs.~(\ref{l1-via-l3}) and (\ref{l2-via-l3}). Let us count the remaining independent components in $l_3$. 

The general tensor of the form (\ref{l-decomposition}) subject to the constraint (\ref{l-Gamma-traceless}) contains
\begin{equation}
  16 \left(\begin{array}{c}
    s+2 \\ s-2
  \end{array} \right) - 16 \left(\begin{array}{c}
    s+1 \\ s-3
  \end{array} \right)
\end{equation}
components because it may be considered as a symmetric, $\Gamma$-traceless tensor with two spinor indices. From this number we have to subtract 
\begin{equation}
  \left(\begin{array}{c}
    s+2 \\ s-2
  \end{array}\right) - \left(\begin{array}{c}
    s \\ s-4
  \end{array}\right),
\end{equation}
which corresponds to the number of components of the symmetric traceless tensor $l_1 = \begin{array}{|c|}
  \hline
  s-2\\
  \hline
\end{array}\,'$. We also have to subtract 
\begin{equation}
  5\left(\begin{array}{c}
    s+2 \\ s-2
  \end{array} \right) 
  - 5\left(\begin{array}{c}
    s \\ s-4
  \end{array} \right) 
  - \left(\begin{array}{c}
    s+1 \\ s-3
  \end{array} \right)
  +\left(\begin{array}{c}
    s-1 \\ s-5
  \end{array} \right),
\end{equation}
which accounts for the components of the tensor $l_2 =\left(\begin{array}{|c|}
  \hline \phantom{m} \\ \hline
\end{array}\times \begin{array}{|c|}
  \hline
  s-2\\
  \hline
\end{array}\,'\right)'$. Hence, after using tensors $l_1$ and $l_2$ for eliminating other components considered above, we are left with 
\begin{equation}
  10\left( 
  \begin{array}{c}
    s+2 \\ s-2
  \end{array}\right) 
  -15\left( 
  \begin{array}{c}
    s+1 \\ s-3
  \end{array}\right)
  +6\left( 
  \begin{array}{c}
    s \\ s-4
  \end{array}\right)
  -\left( 
  \begin{array}{c}
    s-1 \\ s-5
  \end{array}\right)
\end{equation}
independent components in the tensor $l_3$. Subtracting this number from (\ref{N-comp-before-gf}), we find the number of degrees of freedom in the tensor $B^{n\,m_1\ldots m_{s-1}}$ after partial gauge fixing using the gauge freedom from $l_3$:
\begin{align}
    5\left( 
  \begin{array}{c}
    s+3 \\ s-1
  \end{array}\right)
  -10\left( 
  \begin{array}{c}
    s+2 \\ s-2
  \end{array}\right) 
  +10\left( 
  \begin{array}{c}
    s+1 \\ s-3
  \end{array}\right)
  -6\left( 
  \begin{array}{c}
    s \\ s-4
  \end{array}\right)
  +\left( 
  \begin{array}{c}
    s-1 \\ s-5
  \end{array}\right) \nonumber\\
  = \left( 
  \begin{array}{c}
    s+4 \\ s
  \end{array}\right)
  -\left( 
  \begin{array}{c}
    s \\ s-4
  \end{array}\right),
\end{align}
which coincides with the number of components of a symmetric double-traceless tensor of rank $s$. As a result, making use of the remaining gauge freedom in $l_3$, the tensor $B$ may be brought to the form similar to (\ref{A1A2-decomposition}):
\begin{equation}
    B^{n\, m_1\ldots m_{s-1}} = B_1^{nm_1\ldots m_{s-1}} + \eta^{n(m_1}B_2^{m_2\ldots m_{s-2})} - \frac{s-2}{2s-1}\eta^{(m_1m_2}B_2^{m_3\ldots m_{s-1
    })n} \,,
    \label{B1B2-decomposition}
\end{equation}
where $B_1$ and $B_2$ are totally symmetric and traceless tensors,
\begin{align}
    B_1^{m_1\ldots m_{s}} =&\, B_1^{(m_1\ldots m_{s})}\,,&
    \eta_{m_1m_2} B_1^{m_1m_2\ldots m_{s}}=&\,0\,,\\
    B_2^{m_1\ldots m_{s-2}} =&\, B_2^{(m_1\ldots m_{s-2})}\,,&
    \eta_{m_1m_2} B_2^{m_1m_2\ldots m_{s-2}}=&\,0\,.
\end{align}
In this gauge, the components of the tensor $l_3$ obey
\begin{equation}
    l_3^{[n(m_1]m_2\ldots m_{s-1})} = \partial^{[n} b^{(m_1]\ldots m_{s-1})} - \mbox{traces}.
\label{l3-via-b}
\end{equation}
Eq.~(\ref{B.38}) implies the following gauge variations of $B_1$ and $B_2$:
\begin{align}
    \delta B_1^{m_1\ldots m_{s}} =&\, \partial^{(m_1} b^{m_2\ldots m_{s})} - \frac{s-1}{2s+1}\eta^{(m_1m_2} \partial_n b^{m_3\ldots m_{s})n}\,,\\
    \delta B_2^{m_1\ldots m_{s-2}} =&\, \frac{(s-1)(2s-1)}{(s+1)(2s+1)}\partial_n b^{nm_1\ldots m_{s-2}}\,.
\end{align}
These two symmetric and traceless tensors may be combined into one totally symmetric and double-traceless tensor ${\boldsymbol B}^{m_1\ldots m_{s}} = {\boldsymbol B}^{(m_1\ldots m_{s})}$:
\begin{equation}
    {\boldsymbol B}^{m_1\ldots m_{s}} = B_1^{m_1\ldots m_{s}} + \frac{s+1}{2s-1}\eta^{(m_1m_2}B_2^{m_3\ldots m_{s})}\,,\qquad
    {\boldsymbol B}''=0\,,
\end{equation}
with standard gauge transformation
\begin{equation}
  \delta {\boldsymbol B}^{m_1\ldots m_s} = \partial^{(m_1} b^{m_2\ldots m_s)}\,.
\end{equation}
This double-traceless superfield cannot be embedded as a component into the superfield prepotential $h_A^{++m_1\ldots m_{s-1}}$ because the correct component of this superfield is given by Eq.~(\ref{B1B2-decomposition}).

\item Eq.~(\ref{EqB27}) dictates the following gauge transformation law for $P^{\hat\alpha\hat\beta m_1\ldots m_{s-2}}$:
\begin{equation}
    \delta P^{\hat\alpha\hat\beta m_1\ldots m_{s-2}} = \mathrm{i}(\Gamma_n)_{\hat\gamma}{}^{\hat\beta} \partial^n l^{\hat\alpha\hat\gamma m_1\ldots m_{s-2}}\,.
\end{equation}
However, all gauge freedom in the tensor $l$ has already been used in the above steps, and the non-vanishing components of this tensor are given by Eqs.~(\ref{l2-via-a}) and (\ref{l3-via-b}). 
\end{itemize}

\begin{table}[t]
    \centering
\begin{tabular}{l|l}\hline\hline
Bosonic field & Number of independent components after gauge fixing\\\hline
  $B$ & $\left(
\begin{array}{c}
  s+4 \\ s
\end{array}
\right) - \left(
\begin{array}{c}
  s \\ s-4
\end{array}
\right) - \left(
\begin{array}{c}
  s+3 \\ s-1
\end{array}
\right) + \left(
\begin{array}{c}
  s+1 \\ s-3
\end{array}
\right)$  \\
$A$ & $\left(
\begin{array}{c}
  s+3 \\ s-1
\end{array}
\right) - \left(
\begin{array}{c}
  s-1 \\ s-5
\end{array}
\right) - \left(
\begin{array}{c}
  s+2 \\ s-2
\end{array}
\right) + \left(
\begin{array}{c}
  s \\ s-4
\end{array}
\right) $\\
$V$& $3\left(
\begin{array}{c}
  s+3 \\ s-1
\end{array}
\right) - 3\left(
\begin{array}{c}
  s+1 \\ s-3
\end{array}
\right)$ \\
$D$ & $3\left(
\begin{array}{c}
  s+2 \\ s-2
\end{array}
\right) - 3\left(
\begin{array}{c}
  s \\ s-4
\end{array}
\right) $\\ 
$P$& $16\left(
\begin{array}{c}
  s+2 \\ s-2
\end{array}
\right) - 16\left(
\begin{array}{c}
  s+1 \\ s-3
\end{array}
\right) $\\ \hline
Total: & $N_\mathrm{b}=8\left( \begin{array}{c} s+3 \\ s-1 \end{array} \right) + 8\left( \begin{array}{c} s+2 \\ s-2 \end{array} \right) - 8\left( \begin{array}{c} s+1 \\ s-3 \end{array} \right) - 8\left( \begin{array}{c} s \\ s-4 \end{array} \right)$ \\\hline\hline
Fermionic field & \\\hline
$\psi$ & 
$8\left( \begin{array}{c} s+3 \\ s-1 \end{array} \right) - 8\left( \begin{array}{c} s+2 \\ s-2 \end{array} \right)$
\\
$\chi$ & $8\left( \begin{array}{c} s+2 \\ s-2 \end{array} \right) - 8\left( \begin{array}{c} s \\ s-4 \end{array} \right)$ 
\\
$\rho$ & $8\left( \begin{array}{c} s+2 \\ s-2 \end{array} \right) - 8\left( \begin{array}{c} s+1 \\ s-3 \end{array} \right)$
\\\hline
Total: & $N_\mathrm{f}=8\left( \begin{array}{c} s+3 \\ s-1 \end{array} \right) + 8\left( \begin{array}{c} s+2 \\ s-2 \end{array} \right) - 8\left( \begin{array}{c} s+1 \\ s-3 \end{array} \right) - 8\left( \begin{array}{c} s \\ s-4 \end{array} \right)$ \\\hline\hline
\end{tabular}
    \caption{Number of bosonic and fermionic degrees of freedom contained in the prepotentials describing a spin-$s$ supermultiplet off shell. All gauge degrees of freedom are eliminated.}
    \label{tab:DOF}
\end{table}
Table \ref{tab:DOF} summarizes the number of bosonic degrees of freedom left in the prepotentials (\ref{components-in-prepotentials}) after elimination of all unphysical and pure gauge degrees of freedom,
\begin{align}
    N_\mathrm{b} = &\left(
\begin{array}{c}
  s+4 \\ s
\end{array}
\right) + 3\left(
\begin{array}{c}
  s+3 \\ s-1
\end{array}
\right) +18 \left(
\begin{array}{c}
  s+2 \\ s-2
\end{array}
\right) -18 \left(
\begin{array}{c}
  s+1 \\ s-3
\end{array}
\right) - 3\left(
\begin{array}{c}
  s \\ s-4
\end{array}
\right) -\left(
\begin{array}{c}
  s-1 \\ s-5
\end{array}
\right)\nonumber\\
=&\,8 \left[ \left(
\begin{array}{c}
s+3 \\ s-1
\end{array}
\right)
+\left(
\begin{array}{c}
s+2 \\ s-2
\end{array}
\right)
-\left(
\begin{array}{c}
s+1 \\ s-3
\end{array}
\right)
-\left(
\begin{array}{c}
s \\ s-4
\end{array}
\right)
\right].
\label{Nb}
\end{align}
The physical components enter the prepotentials in the following way
\begin{subequations}
\label{h++components}
\begin{align}
h^{++m_1\ldots m_{s-1}} =&\, \mathrm{i}(\theta^+\Gamma_n\theta^+)B^{n m_1\ldots m_{s-1}} + 2\mathrm{i}(\theta^+)^2\theta^{-\hat\alpha}\Gamma^{(m_1}_{\hat\alpha\hat\beta}\theta^+_{\hat\gamma}P^{\hat\beta\hat\gamma m_2\ldots m_{s-1})} \nonumber\\& + (\theta^+)^4 V^{m_1\ldots m_{s-1}ij}u^-_i u^-_j\,,\\
h^{++5m_1\ldots m_{s-2}} =&\, \mathrm{i}(\theta^+\Gamma_n\theta^+)A^{n m_1\ldots m_{s-2}} - 2\mathrm{i}(\theta^+)^2\theta^-_{\hat\alpha}\theta^+_{\hat\beta}P^{\hat\alpha\hat\beta m_1\ldots m_{s-2}} \nonumber\\& + (\theta^+)^4 D^{m_1\ldots m_{s-2}ij}u^-_i u^-_j\,,\\
h^{++\hat\alpha+m_1\ldots m_{s-2}} =&\,(\theta^+)^2 \theta^+_{\hat\beta} P^{\hat\alpha\hat\beta m_1\ldots m_{s-2}}\,,
\end{align}
\end{subequations}
with tensors $A$ and $B$ given by Eqs.~(\ref{A1A2-decomposition}) and (\ref{B1B2-decomposition}), respectively.

\subsection{Prepotentials in the Wess-Zumino gauge: fermionic fields}
\label{sec:prepferm}

Upon imposing the Wess-Zumino gauge, the following fermionic components survive in the $\theta$-decompositions of the prepotentials:
\begin{subequations}
\label{fermionic-component-structure}
\begin{align}
    h_A^{++m_1\ldots m_{s-1}} =& -(\theta^+)^2 \theta^+_{\hat\alpha} \bigg[2\psi^{m_1\ldots m_{s-1}\hat\alpha i}
    + (\Gamma^{(m_1})^{\hat\alpha}{}_{\hat\beta} \chi^{m_2\ldots m_{s-1})\hat\beta i} \nonumber\\&
    -\frac{s-2}{2s-1}\eta^{(m_1m_2} (\Gamma_{n})^{\hat\alpha}{}_{\hat\beta} \chi^{m_3\ldots m_{s-1})n\hat\beta i}
    \bigg]u^-_i\,,\label{fermionic-component-structure-a}\\
    h_A^{++5m_1\ldots m_{s-2}} =&\, (\theta^+)^2 \theta^+_{\hat\alpha} \chi^{m_1\ldots m_{s-2}\hat\alpha i}u^-_i\,,  
    \label{fermionic-component-structure-b} \\
    h^{++\hat\alpha+ m_1\ldots m_{s-2}} = &\,\frac14(\theta^+)^4 \rho^{m_1\ldots m_{s-2}\hat\alpha i}u^-_i\,.
     \label{fermionic-component-structure-c}
\end{align}
\end{subequations}
Here $\psi^{m_1\ldots m_{s-1}\hat\alpha i}$, $\chi^{m_1\ldots m_{s-2}\hat\alpha i}$ and $\rho^{m_1\ldots m_{s-2}\hat\alpha i}$ are symmetric and traceless tensors in all vector indices. The latter field also obeys the $\Gamma$-tracelessness constraint, $(\Gamma_{m_1})_{\hat\alpha\hat\beta}\rho^{m_1\ldots m_{s-2}\hat\beta i} =0$, which follows from Eq.~(\ref{Gamma-tracelessness}). 

Note that the last two terms in Eq.~(\ref{fermionic-component-structure-a}) may be eliminated by a redefinition of $\psi^{m_1\ldots m_{s-1}\hat\alpha i}$. However, we keep these terms because they help represent the gauge variation of $\psi^{m_1\ldots m_{s-1}\hat\alpha i}$ in a suitable form (\ref{B.48}) below.

The prepotentials (\ref{fermionic-component-structure}) are defined modulo residual gauge transformations (\ref{analytic-gauge-transformations}) with the superfield gauge parameters given by
\begin{align}
    \lambda_A^{m_1\ldots m_{s-1}} = & -\frac{\mathrm{i}}{2}\theta^{+\hat\alpha} \Gamma^{(m_1}_{\hat\alpha\hat\beta} \epsilon^{m_2\ldots m_{s-1})\hat\beta i} u^-_i\,,\\
    \lambda_A^{5m_1\ldots m_{s-2}} = &\,\frac{\mathrm{i}}{2}\theta^+_{\hat\alpha} \epsilon^{m_1\ldots m_{s-2}\hat\alpha i}u^-_i\,,\\
    \lambda^{\hat\alpha+m_1\ldots m_{s-2}} =&\, \frac14 \epsilon^{m_1\ldots m_{s-2}\hat\alpha i}u^+_i - \frac{\mathrm{i}}{4} (\theta^+\Gamma^n\theta^+) \partial_n \epsilon^{m_1\ldots m_{s-2}\hat\alpha i}u^-_i\,,
\end{align}
where the tensor $\epsilon^{m_1\ldots m_{s-2}\hat\alpha i}$ is symmetric and traceless in all vector indices, and obeys the $\Gamma$-tracelessness constraint, $(\Gamma_{m_1})_{\hat\alpha\hat\beta}\epsilon^{m_1\ldots m_{s-2}\hat\beta i} =0$, which follows from Eq.~(\ref{lambda-Gamma-tracelessness}). Equations (\ref{analytic-gauge-transformations}) imply the following gauge transformations of the fermionic component fields in Eq.~(\ref{fermionic-component-structure}):
\begin{align}
    \delta \psi^{m_1\ldots m_{s-1}\hat\alpha i} =&\, \partial^{(m_1}\epsilon^{m_2\ldots m_{s-1})\hat\alpha i} - \frac{s-2}{2s-1}\eta^{(m_1m_2}\partial_n \epsilon^{m_3\ldots m_{s-1})n\hat\alpha i}\,,\label{B.48}\\
    \delta \chi^{m_1\ldots m_{s-2}\hat\alpha i} =&\, (\Gamma^n)^{\hat\alpha}{}_{\hat\beta}\partial_n \epsilon^{m_1\ldots m_{s-2}\hat\beta i}\,,\label{B.60}\\
    \delta \rho^{m_1\ldots m_{s-2} \hat\alpha i} = &\, \Box \epsilon^{m_1\ldots m_{s-2}\hat\alpha i}\,.
\end{align}

The auxiliary field $\rho^{m_1\ldots m_{s-2} \hat\alpha i}$ and the gauge parameter $\epsilon^{m_1\ldots m_{s-2}\hat\alpha i}$ contain the same number of degrees of freedom since they respect the same symmetry and constraints. Therefore, when counting the fermionic degrees of freedom it is sufficient to consider only the fields $\psi^{m_1\ldots m_{s-1}\hat\alpha i}$ and $\chi^{m_1\ldots m_{s-2}\hat\alpha i}$ which are both symmetric and traceless in vector indices, but the $\Gamma$-tracelessness is not imposed. Recalling that a symmetric traceless tensor of rank-$s$ contains 
$\left(
\begin{array}{c}
s+4 \\ s
\end{array}
\right) - \left(
\begin{array}{c}
s+2 \\ s-2
\end{array}
\right)
$ independent components, we conclude that the prepotentials $h^{++Mm_1\ldots m_{s-2}}$ contain 
\begin{equation}
    N_\mathrm{f} = 8 \left[ \left(
\begin{array}{c}
s+3 \\ s-1
\end{array}
\right)
+\left(
\begin{array}{c}
s+2 \\ s-2
\end{array}
\right)
-\left(
\begin{array}{c}
s+1 \\ s-3
\end{array}
\right)
-\left(
\begin{array}{c}
s \\ s-4
\end{array}
\right)
\right]
\end{equation}
fermionic degrees of freedom off shell. This number exactly coincides with the number of bosonic d.o.f.\ in Eq.~(\ref{Nb}), 
\begin{equation}
N_\mathrm{b}=N_\mathrm{f}\,.
\end{equation}
Summary of the fermionic degrees of freedom remaining after gauge fixing is given in Table~\ref{tab:DOF}.

It is possible to construct one Fang-Fronsdal field ${\boldsymbol\psi}^{m_1\ldots m_{s-1}\hat\alpha i}$ using $\psi^{m_1\ldots m_{s-1}\hat\alpha i}$ and some components of $\chi^{m_1\ldots m_{s-2}\hat\alpha i}$. For this purpose, we decompose the traceless field $\chi^{m_1\ldots m_{s-2}\hat\alpha i}$ into two $\Gamma$-traceless fields $\chi'^{m_1\ldots m_{s-2}\hat\alpha i}$ and $\phi^{m_1\ldots m_{s-3}\hat\alpha i}$:
\begin{equation}
    \chi^{m_1\ldots m_{s-2}\hat\alpha i} = \chi'^{m_1\ldots m_{s-2}\hat\alpha i} + (\Gamma^{(m_1})^{\hat\alpha}{}_{\hat\beta} \phi^{m_2\ldots m_{s-2})\hat\beta i}\,,
\end{equation}
where
\begin{equation}
    \phi^{m_1\ldots m_{s-3}\hat\alpha i} = -\frac{s-2}{2s-1}(\Gamma_n)^{\hat\alpha}{}_{\hat\beta}\chi^{m_1\ldots m_{s-3}n\hat\beta i}\,.
\end{equation}
Equation (\ref{B.48}) implies the following gauge transformations for these components:
\begin{align}
    \delta\chi'^{m_1\ldots m_{s-2}\hat\alpha i} =&\, (\Gamma_n)^{\hat\alpha}{}_{\hat\beta}\partial^n\epsilon^{m_1\ldots m_{s-2}\hat\beta i} - \frac{2s-4}{2s-1} (\Gamma^{(m_1})^{\hat\alpha}{}_{\hat\beta}\partial_n \epsilon^{m_2\ldots m_{s-2})n\hat\beta i}\,,\\
    \delta\phi^{m_1\ldots m_{s-3}\hat\alpha i} = &\,\frac{2s-4}{2s-1}\partial_n\epsilon^{m_1\ldots m_{s-3}n\hat\alpha i}\,.\label{B.66}
\end{align}
Thus, the Fang-Fronsdal field $\boldsymbol\psi^{m_1\ldots m_{s-1}\hat\alpha i}$ reads
\begin{equation}
    \boldsymbol\psi^{m_1\ldots m_{s-1}\hat\alpha i} = \psi^{m_1\ldots m_{s-1}\hat\alpha i} + \frac12 \eta^{(m_1m_2}\phi^{m_3\ldots m_{s-1})\hat\alpha i}\,.
     \label{e000}
\end{equation}
By construction, this field obeys the algebraic constraint $\eta_{m_1 m_2}(\Gamma_{m_3})^{\hat\alpha}{}_{\hat\beta}\boldsymbol\psi^{m_1\ldots m_{s-1}\hat\beta i} = 0$, and possesses the gauge transformation law 
\begin{equation}
    \delta\boldsymbol\psi^{m_1\ldots m_{s-1}\hat\alpha i} = \partial^{(m_1}\epsilon^{m_2\ldots m_{s-1})\hat\alpha i}\,,
\end{equation}
which follows from Eqs.~(\ref{B.48}) and (\ref{B.66}).

\subsection{Quadratic action for bosonic components}
\label{sec:s=2-component-action}

In addition to the prepotentials $h^{++ m(s-1)}$ and $h^{++5m(s-2)}$, the action (\ref{spin-s-action}) also involves the superfields $h^{-- m(s-1)}$ and $h^{--5m(s-2)}$, which are related to the former ones by the zero-curvature equations (\ref{zero-curvature-general}). By substituting the expressions (\ref{h++components}) into the zero-curvature equations (\ref{zero-curvature-general}) it is possible to find the component structure of the superfields $h^{-- m(s-1)}$ and $h^{--5m(s-2)}$. Then, using the following conventions for Grassmann and harmonic integrations
\begin{align}
  \int \mathrm{d}^{5|8}z \,f(x)(\theta^+)^4(\theta^-)^4 = &\int \mathrm{d}^5x \,f(x)\,,\label{e00}\\
  \int \mathrm{d}u\, f^{ij}u^+_i u^+_j g^{kl}u^-_k u^-_l = &\, f^{ij}g^{kl} \frac16(\varepsilon_{ik}\varepsilon_{jl}+\varepsilon_{il}\varepsilon_{jk}) = \frac13 f^{ij}g_{ij}\,,
\end{align}
we find the quadratic actions for the bosonic components:
\begin{subequations}
\label{S-in-components}
\begin{align}
S_\mathrm{bos}= &\,S_1 - \frac{2s-1}{s-1} S_2\,,\\
S_1 =& \int \mathrm{d}^{5|8}z \mathrm{d}u\, h^{++m(s-1)}h^{--}_{m(s-1)} \nonumber\\ =& \int \mathrm{d}^5x\bigg\{
 -\frac12 B^{p\,m(s-1)}(\partial_p\partial^n - \delta_p^n \Box)B_{n\,m(s-1)}
  +\mathrm{i} \Gamma^p_{\hat\beta\hat\gamma}P^{\hat\beta\hat\gamma\,m_2\ldots m_{s-1}}\partial_{[p} B^n{}_{n]m_2\ldots m_{s-1}}\nonumber\\&
  +\frac18 \Gamma^{(m_1}_{\hat\beta\hat\gamma}P^{\hat\beta\hat\gamma\,m_2\ldots m_{s-1})}(\Gamma_{(m_1})_{\hat\rho\hat\sigma}P^{\hat\rho\hat\sigma}_{m_2\ldots m_{s-1})} 
  -\frac14 \Gamma^{(m_1}_{\hat\alpha\hat\beta}P^{\hat\beta\hat\gamma\,m_2\ldots m_{s-1})}(\Gamma_{(m_1})_{\hat\gamma\hat\delta}P^{\hat\delta\hat\alpha}_{m_2\ldots m_{s-1})} \nonumber\\&
  + \frac{\mathrm{i}}{2} \varepsilon^{klpnm}(\Sigma_{kl})_{\hat\alpha\hat\beta} P^{\hat\alpha\hat\beta m_2\ldots m_{s-1}} \partial_p B_{nmm_2\ldots m_{s-1}} 
  +  \frac19V^{ij\,m(s-1)}V_{ij\,m(s-1)}
  \bigg\},\\
S_2=&\int \mathrm{d}^{5|8}z\mathrm{d}u\, h^{++5m(s-2)}h^{--5}_{m(s-2)} \nonumber\\ =& \int \mathrm{d}^5x \bigg\{ 
    -\frac12 A^{p\, m(s-2)}(\partial_p \partial^n - \delta_p^n \Box)A_{n\,m(s-2)} 
    +\mathrm{i} P^{\hat\alpha\hat\beta\, m(s-2)} (\Sigma^{pn})_{\hat\alpha\hat\beta}\partial_p A_{n\,m(s-2)} \nonumber\\&
    +\frac18\varepsilon_{\hat\alpha\hat\beta}P^{\hat\alpha\hat\beta \,m(s-2)}
    \varepsilon_{\hat\gamma\hat\delta}P^{\hat\gamma\hat\delta}_{m(s-2)}
    +\frac14 P^{\hat\alpha\hat\beta\, m(s-2)}P_{\hat\beta\hat\alpha\, m(s-2)} \nonumber\\&
    +\frac19 D^{ij\,m(s-2)}D_{ij\,m(s-2)} 
  \bigg\}.
\end{align}
\end{subequations}
The auxiliary fields $V^{ij\,m(s-1)}$ and $D^{ij\,m(s-2)}$ have trivial equations of motion and completely decouple. The auxiliary field $P^{\hat\alpha\hat\beta \,m(s-2)}$, however, couples non-trivially with the physical components. This situation is common for supersymmetric higher spin field theories, see, e.g., \cite{KuzenkoSibiryakov94}. In the present case, elimination of this auxiliary field $P^{\hat\alpha\hat\beta \,m(s-2)}$ is a non-trivial task because this field is $\Gamma$-traceless, see Eq.~(\ref{P-Gamma-traceless}). As a consequence, the components of this field
\begin{equation}
  P^{\hat\alpha\hat\beta m(s-2)} = \varepsilon^{\hat\alpha\hat\beta}p_1^{m(s-2)} + \Gamma_{n}^{\hat\alpha\hat\beta}p_2^{n\,m(s-2)} + \frac12\Sigma_{rs}^{\hat\alpha\hat\beta}p_3^{[rs]m(s-2)}
\end{equation}
obey the following constraints for $s\geq3$:
\begin{subequations}
\label{p-constraints}
\begin{align}
  \eta_{nm_1}p_2^{n\,m_1\ldots m_{s-2}} = &\, 0\,,\\
  p_1^{m_1\ldots m_{s-2}} - \frac12\eta_{pq}p_3^{[p(m_1]m_2\ldots m_{s-2})q} =&\,0\,,\\
  p_2^{[p(k]m_2\ldots m_{s-2})} + \frac18 \varepsilon^{pk}{}_{rsn}p_3^{[rs](nm_2\ldots m_{s-2})} =&\,0\,.
\end{align}
\end{subequations}

For the spin-2 supermultiplet, the $\Gamma$-tracelessness (\ref{P-Gamma-traceless}) is not required, and the constraints (\ref{p-constraints}) are not imposed. In this case, the action (\ref{S-in-components}) simplifies to
\begin{align}
  S^{(s=2)}_\mathrm{bos} =& \int \mathrm{d}^5x\bigg\{
    -\frac12 B^{pm}(\partial_p \partial^n - \delta_p^n\Box)B_{nm} - 2\mathrm{i} p_2^p(\partial_p B^n{}_n - \partial_n B^n{}_p)\nonumber\\&
    +\frac32A^p(\partial_p \partial^n - \delta_p^n\Box)A_n -\frac{3\mathrm{i}}{2} p_3^{pq}(\partial_p A_q - \partial_q A_p)\nonumber\\&
    +2 p_1^2 + 2 p_2^m p_{2\,m} - \frac12 p_3^{mn}p_{3\,mn}
  \bigg\}.
\end{align}
By eliminating the remaining auxiliary fields,
\begin{align}
  p_1 =&\,0\,,\\
  p_{2\,m} =&\, \frac{\mathrm{i}}2 (\partial_m B^n{}_n - \partial_nB^n{}_m)\,,\\
  p_{3\,mn} = & -\frac{3\mathrm{i}}2(\partial_m A_n - \partial_n A_m)\,,
\end{align}
we obtain the quadratic action for the physical fields of spins 1 and 2:
\begin{align}
  S^{(s=2)}_\mathrm{bos} =& \int \mathrm{d}^5x \bigg\{ 
    \partial_p B^{pm} \partial^n B_{nm} - \partial_n B^{np}\partial_p B^m{}_m - \frac12\partial^p B^{mn} \partial_p B_{mn} - \frac12\partial^p B^m{}_m \partial_p B^n{}_n \nonumber\\&
    - \frac38 F_{mn}F^{mn} 
  \bigg\},
\label{s=2-component-action}
\end{align}
where $F_{mn} = \partial_m A_n - \partial_n A_m$ is the Maxwell field strength. Note that the action for $B_{mn}$ is canonically normalized while the Maxwell term requires a re-scaling
$A_m \to \sqrt{2/3} A_m$.


\subsection{Quadratic action for fermionic components}
\label{sec:s=3/2-component-action}
Recall that the fermionic contributions to the prepotentials 
\begin{equation}
h_A^{++ m (s-1)}\,, \quad h_A^{++ 5 m (s-2)} \,, \quad h^{++\hat\alpha+ m (s-2)}
\label{e0}
\end{equation}
are given in Eqs.~\eqref{fermionic-component-structure-a}, \eqref{fermionic-component-structure-b} and \eqref{fermionic-component-structure-c}, respectively. Let us denote
\begin{align}
\tilde{\psi}^{m_1\ldots m_{s-1} \hat\alpha i} = &\, \psi^{m_1\ldots m_{s-1} \hat\alpha i} +\frac{1}{2}
(\Gamma^{(m_1})^{\hat\alpha}{}_{\hat\beta} \chi^{m_2\ldots m_{s-1})\hat\beta i} \nonumber\\
 &-\frac{s-2}{2s-1}\eta^{(m_1m_2} (\Gamma_{n})^{\hat\alpha}{}_{\hat\beta} \chi^{m_3\ldots m_{s-1})n\hat\beta i} \,. 
\label{e1}
\end{align}
Then $h_A^{++ m (s-1)}$ is given by 
\begin{equation}
h_A^{++ m (s-1)}= -2 (\theta^+)^2 \theta_{\hat\alpha}^+ \tilde{\psi}^{m (s-1) \hat\alpha i} u^-_i\,. 
\label{e2}
\end{equation}
Just like in the bosonic case, one can use the zero-curvature equations \eqref{zero-curvature-general} to find the fermionic component structure of the superfields $h^{-- m (s-1)}$, $h^{-- 5 m (s-2)}$. Performing Grassmann integration using Eq.~\eqref{e00} and harmonic integration using 
\begin{equation}
\int \mathrm{d}u\, f^{i}u^+_i  g^{j}u^-_j  =  \frac{1}{2} f^{i}g^{j} \varepsilon_{ij}=\frac{1}{2} f^{i}g_{i} \,,
\label{e3}
\end{equation}
we obtain the action for the fermionic components
\begin{align}
S_{\mathrm{ferm}}  = &\, S_1- \frac{2s-1}{s-1} S_2 \,, \label{e4} \\
S_1 = &\, \frac{\mathrm{i}}{8}  \int {\mathrm{d}}^5 x \bigg\{ \tilde{\psi}^{m(s-1) \hat\alpha i}(\Gamma^n)_{\hat\alpha}{}^{\hat\beta} \partial_n \tilde{\psi}_{m(s-1) \hat\beta i}  -   \tilde{\psi}^{m_1 \dots m_{s-1} \hat\alpha i}(\Gamma_{m_1})_{\hat\alpha}{}^{\hat\beta}  \rho_{m_2 \dots m_{s-1} \hat\beta i} \bigg\}\,, 
 \nonumber \\
S_2 = &\, \frac{\mathrm{i}}{32}  \int {\mathrm{d}}^5 x \bigg\{ \chi^{m(s-2) \hat\alpha i}(\Gamma^n)_{\hat\alpha}{}^{\hat\beta} \partial_n \chi_{m(s-2) \hat\beta i} - 2  \chi^{m(s-2) \hat\alpha i}  \rho_{m (s-2) \hat\alpha i} \bigg\} \,. \nonumber
\end{align}
The component fields $ \rho^{m (s-2) \hat\alpha i}$ and the $\Gamma$-traceless part of $\chi^{m(s-2) \hat\alpha i}$, denoted by $\chi'^{m(s-2) \hat\alpha i}$ in Section \ref{sec:prepferm}, are auxiliary. The component fields $\psi^{m(s-1) \hat\alpha i}$ and the $\Gamma$-trace of $\chi^{m(s-2) \hat\alpha i}$, denoted by $\phi^{m(s-3) \hat\alpha i}$ in Section \ref{sec:prepferm}, combine into the Fang--Fronsdal field~\eqref{e000}. 

Let us consider the case $s=2$ in more detail. In this case $\tilde{\psi}^{m \hat\alpha i}$ is given by 
\begin{equation}
\tilde{\psi}^{m \hat\alpha i} =\psi^{m \hat\alpha i}+ \frac{1}{2} (\Gamma^m )^{\hat\alpha}{}_{\hat\beta} \chi^{\hat\beta i}\,,
\label{e5}
\end{equation}
and the action simplifies to yield
\begin{align}
S_{\mathrm{ferm}} = &\int {\mathrm{d}}^5 x 
\bigg\{ \frac{\mathrm{i}}{8} \psi^{m \hat\alpha i}(\Gamma^n)_{\hat\alpha}{}^{\hat\beta}\partial_n 
\psi_{m \hat\beta i} + \frac{\mathrm{i}}{8} \chi_{\hat\beta i} \partial_n \psi^{n \hat\beta i}
 +  \frac{\mathrm{i}}{4} (\Sigma^{mn})_{\hat\alpha}{}^{\hat\beta} \chi^{\hat\alpha i}\partial_n \psi_{m \hat\beta i}
 \nonumber \\
 &-\frac{\mathrm{i}}{8} \psi^{m \hat\alpha i} (\Gamma_m)_{\hat\alpha}{}^{\hat\beta} \rho_{\hat\beta i} 
 -\frac{\mathrm{i}}{8}  \chi^{\hat\alpha i}\rho_{\hat\alpha i}   \bigg\}\,.
 \label{e6}
\end{align}
From the equations of motion for  $\rho_{\hat\beta i} $ and $\chi_{\hat\beta i} $ we find that these fields are solved in terms $ \psi^{m \hat\alpha i}$ as follows
\begin{align}
\chi^{\hat\alpha i}= &\,(\Gamma_m)^{\hat\alpha}{}_{\hat\beta} \psi^{m \hat\beta i}\,,
 \\
\rho_{\hat\alpha i} =&\,\partial^n \psi_{n \hat\alpha i} +2 (\Sigma^{mn})_{\hat\alpha}{}^{\hat\beta} \partial_n \psi_{m \hat\beta i}\,. 
\label{e7}
\end{align}
Substituting these expressions back into the action~\eqref{e6} gives
\begin{equation} 
S_{\mathrm{ferm}} = - \frac{\mathrm{i}}{8} 
\int {\mathrm{d}}^5 x \,\psi^{m\hat\alpha i}\Big[
2 (\Gamma_m \Sigma^{pn})_{\hat\alpha}{}^{\hat\beta} \partial_n \psi_{p \hat\beta i} 
- (\Gamma^n)_{\hat\alpha}{}^{\hat\beta}\partial_n \psi_{m \hat\beta i}  +(\Gamma^n)_{\hat\alpha}{}^{\hat\beta}\partial_m \psi_{n \hat\beta i} \Big]\,. 
\label{e7.1}
\end{equation}
One can show that Eq.~\eqref{e7.1} can also be presented in the form 
\begin{equation}
S_{\mathrm{ferm}} = - \frac{\mathrm{i}}{8} \int {\mathrm{d}}^5 x\, \psi^{m \hat\alpha i} \eta_{ms} 
(\Gamma^{[s} \Gamma^n \Gamma^{p]})_{\hat\alpha}{}^{\hat\beta} \partial_n \psi_{ p \hat\beta i}\,, 
\label{e8}
\end{equation}
which is the standard action for a spin-$3/2$ field.


\section{Prepotentials for 5D conformal supergravity}
\label{AppendixC}

Prepotentials for 4D conformal supergravity in the harmonic superspace framework were studied in Ref.~\cite{GIOS, Galperin:1987ek,KT,Butter:2010sc, Butter:2015nza}. In this Appendix, we uplift this construction to the 5D harmonic superspace and introduce the prepotentials describing 5D Weyl supermultiplet, although the action for 5D conformal supergravity is not known.

In this Appendix, unlike the main body of this paper, we use abbreviations $D_M = (\partial_m ,D^-_{\hat\alpha}, D^{--})$ for superspace derivatives and $\lambda^M = (\lambda^m, \lambda^{\hat\alpha+},\lambda^{++})$ for superfields in the harmonic superspace. The superscript $M$ denotes now $M=(m,\hat\alpha+,++)$.

\subsection{Conformal transformations in harmonic superspace}

Conformal transformations in the 5D harmonic superspace were studied in Ref.~\cite{Kuzenko2006}. In this subsection, we collect basic relations for the superconformal Killing vector and the corresponding transformations of the superfields. 

By definition, superconformal transformations in 5D superspace with coordinates $z^A = (x^m,\theta_i^{\hat\alpha})$ are generated by a real vector field
\begin{equation}
  \xi = \bar\xi = \xi^m(z)\partial_m + \xi_i^{\hat\alpha}(z) D^i_{\hat\alpha}
\end{equation}
subject to the equation
\begin{equation}
  [\xi, D^i_{\hat\alpha}] = -(D^i_{\hat\alpha}\xi_j^{\hat\beta})D^j_{\hat\beta}\,.
  \label{master-eq}
\end{equation}
This equation implies a number of constraints on the components of the superconformal Killing vector field $(\xi^m,\xi_i^{\hat\alpha})$. Among them, are the following \cite{Kuzenko2006}
\begin{align}
  \partial^m \xi^n + \partial^n \xi^m = &\,\frac25 \eta^{mn}\partial_p \xi^p\,,\label{Killing-vector-eq}\\
  D^{(i}_{\hat\alpha} \xi^{j)}_{\hat\beta} = &\, \frac14\varepsilon_{\hat\alpha\hat\beta} D^{\hat\gamma(i}\xi^{j)}_{\hat\gamma}\,,\label{B34}\\
  \Gamma^m_{\hat\alpha\hat\beta} D^{i\hat\alpha}\xi_i^{\hat\beta} =&\,0\,.
  \label{B35}
\end{align}
Eq.~(\ref{Killing-vector-eq}) shows that $\xi^m$ is an ordinary conformal Killing vector with $\theta$-dependent coefficients.

The right-hand side of Eq.~(\ref{master-eq}) may be represented in a more informative form:
\begin{equation}
  [\xi,D^i_{\hat\alpha}] = \tilde\omega_{\hat\alpha}{}^{\hat\beta} D^i_{\hat\beta} - \tilde\sigma D^i_{\hat\alpha} - \tilde\Lambda^i_j D^j_{\hat\alpha}\,,
  \label{xi-D-commutator}
\end{equation}
where
\begin{equation}
  \tilde\omega^{\hat\alpha\hat\beta} = -\frac12 D^{k(\hat\alpha}\xi_k^{\hat\beta)}\,,\quad
  \tilde\sigma = \frac18 D^k_{\hat\gamma} \xi_k^{\hat\gamma}\,,\quad
  \tilde\Lambda^{ij} = \frac14 D^{(i}_{\hat\gamma}\xi^{j)\hat\gamma}\,
  \label{B37}
\end{equation}
are superfield parameters of Lorentz transformations, dilatations and $SU(2)_R$ transformations. These parameters obey the following identities
\begin{align}
  D^i_{\hat\alpha}\tilde\omega_{\hat\beta\hat\gamma} =&\, 2(\varepsilon_{\hat\alpha\hat\beta}D^i_{\hat\gamma}\tilde\sigma + \varepsilon_{\hat\alpha\hat\gamma}D^i_{\hat\beta}\tilde\sigma)\,,\\
  D^i_{\hat\alpha}\tilde\Lambda^{jk} =& \,3(\varepsilon^{ik}D^j_{\hat\alpha}\tilde\sigma + \varepsilon^{ij}D^k_{\hat\alpha}\tilde\sigma)\,,\label{B38}\\
  \partial_m \xi^m - D^i_{\hat\alpha}\xi^{\hat\alpha}_i =&\, 2\tilde\sigma\,.
\end{align}

By definition, an iso-tensor superfield $H^{i_1\ldots i_n} = H^{(i_1\ldots i_n)}$ is said to be superconformal primary of weight $p/2$ if it transforms by the rule
\begin{equation}
  \delta_\mathrm{conf} H^{i_1\ldots i_n} = -\xi H^{i_1\ldots i_n} - p\tilde\sigma H^{i_1\ldots i_n} - \tilde\Lambda_k^{(i_1}H^{i_2\ldots i_n)k}\,.
\end{equation}

To extend the conformal transformations to the harmonic superspace, we introduce the following analytic superfields \cite{Kuzenko2006}
\begin{equation}
  \Sigma = \tilde\Lambda^{ij}u^+_i u^-_j + 3 \tilde\sigma\,,\qquad
  \tilde\Lambda^{++} = D^{++}\Sigma = \tilde\Lambda^{ij}u^+_i u^+_j\,.
\end{equation}
It is possible to show that the operator $\xi-\tilde\Lambda^{++}D^{--}$ maps every analytic superfield into an analytic one and obeys the commutation relation
\begin{equation}
  [\xi - \tilde\Lambda^{++}D^{--},D^+_{\hat\alpha}] = \tilde\omega_{\hat\alpha}{}^{\hat\beta} D^+_{\hat\beta} - \tilde\sigma D^+_{\hat\alpha} - \tilde\Lambda^{+-}D^+_{\hat\alpha}\,,\label{B45}
\end{equation}
where $\tilde\Lambda^{+-} = \tilde\Lambda^{ij}u^+_i u^-_j$. Next, making use of the identities \eqref{B34}, \eqref{B35} and (\ref{B37}), we get the following identity
\begin{equation}
  [\xi^{\hat\beta-}D^+_{\hat\beta},D^+_{\hat\alpha} ] = \tilde\omega_{\hat\alpha}{}^{\hat\beta} D^+_{\hat\beta} - \tilde\sigma D^+_{\hat\alpha} - \tilde\Lambda^{+-}D^+_{\hat\alpha}\,,
\end{equation}
where $\xi^{\hat\alpha\pm} = u^\pm_i \xi^{\hat\alpha i}$. Comparing this with Eq.~(\ref{B45}), we conclude
\begin{equation}
  [\xi - \xi^{\hat\beta-}D^+_{\hat\beta} - \tilde\Lambda^{++}D^{--} , D^+_{\hat\alpha}] = 0\,.
\label{C.15}
\end{equation}
Recalling that $\xi = \xi^m\partial_m - \xi^{\hat\alpha+}D^-_{\hat\alpha} + \xi^{\hat\alpha-}D^+_{\hat\alpha}$, we can identify
\begin{equation}
  \xi - \xi^{\hat\beta-}D^+_{\hat\beta} - \tilde\Lambda^{++}D^{--} := \tilde\Lambda = \tilde\lambda^m \partial_m + \tilde\lambda^{\hat\alpha+}D^-_{\hat\alpha} + \tilde\lambda^{++} D^{--}\,,
\label{Lambda-definition}
\end{equation}
where
\begin{equation}
  \tilde\lambda^m := \xi^m \,,\quad
  \tilde\lambda^{+\hat\alpha} := -\xi^{+\hat\alpha}\,,\quad
  \tilde\lambda^{++} := -\tilde\Lambda^{++}\,.
  \label{lambda-xi-relation}
\end{equation}
In terms of these superfields, the superconformal weight factor reads
\begin{equation}
  \tilde\Omega = \partial_m\tilde\lambda^m - D^-_{\hat\alpha}\tilde\lambda^{\hat\alpha+} + D^{--}\tilde\lambda^{++} = 2\Sigma\,.
\label{Omega-conformal-definition}
\end{equation}

By definition, a superconformal transformation of a primary analytic superfield $\Phi^{(q)}$ with conformal weight $n$ and $U(1)$ charge $q$ is
\begin{equation}
  \delta_\mathrm{conf}\Phi^{(q)} = -\left(\tilde\Lambda+ \frac n2\tilde\Omega\right)\Phi^{(q)}\,.
\label{general-conformal-transformation}
\end{equation}

\subsection{Superconformal properties of the free \texorpdfstring{$q$}{q}-hypermultiplet model}

In the previous subsection, we defined rigid superconformal transformations in 5D harmonic superspace. In this section, we begin by considering transformations of the hypermultiplet superfield under the \emph{local} superconformal group. Invariance of the free hypermultiplet model under rigid superconformal symmetry will be discussed at the end of this subsection.

By analogy with the 4D case \cite{BIZ24, KT}, we introduce local superconformal transformations of the $q$-hypermultiplet in the form
\begin{equation}
  \delta q^{+a} = -\left(\Lambda  + \frac12 \Omega\right) q^{+a}\,,
  \label{delta-q-conformal}
\end{equation}
where 
\begin{align}
\Lambda := &\,\lambda^M D_M = \lambda^m \partial_m +\lambda^{\hat\alpha+} D^-_{\hat\alpha} + \lambda^{++}D^{--}\,,\\
\Omega := &\,(-1)^{\varepsilon_M} D_M\lambda^M = \partial_m \lambda^m - D^-_{\hat\alpha}\lambda^{\hat\alpha+} + D^{--}\lambda^{++}\,,
\end{align}
with some superfield parameters $\lambda^M$. The operator $\Lambda$ acting on the $q$-hypermultiplet should preserve analyticity, $D^+_{\hat\alpha}(\Lambda q^{+a})=0$. As a corollary, the superfields $\lambda^M$ obey the constraints
\begin{subequations}
\label{lambda-conf-analyticity}
\begin{align}
  D^+_{\hat\alpha} \lambda^m - 2\mathrm{i}(\Gamma^m)_{\hat\alpha\hat\beta}\lambda^{\hat\beta+} =&\, 0\,, \\
  D^+_{\hat\alpha} \lambda^{\hat\beta+} -\delta_{\hat\alpha}{}^{\hat\beta} \lambda^{++} = &\,0\,, \\
  D^+_{\hat\alpha} \lambda^{++} =&\, 0\,.
\end{align}
\end{subequations}
The general solution to these constraints is expressed via a single unconstrained gauge parameter $l^{--}$:
\begin{equation}
  \lambda^m = -\frac{\mathrm{i}}{4}(\Gamma^m)^{\hat\alpha\hat\beta}D^+_{\hat\alpha}D^+_{\hat\beta} l^{--}\,, \quad
  \lambda^{\hat\alpha+} = \frac18 D^{+\hat\alpha}(D^+)^2 l^{--}\,, \quad
  \lambda^{++} = (D^+)^4 l^{--}\,.
  \label{lambda-l-conformal}
\end{equation}
The operators $\Lambda$ and $\Omega$ in Eq.~(\ref{delta-q-conformal}) may be cast in the following form
\begin{align}
  \Lambda q^{+a} = &\,(D^+)^4 (l^{--} D^{--}q^{+a})\,,
  \label{Lambda-l}\\
  \Omega q^{+a} = &\,(D^+)^4[(D^{--}l^{--})q^{+a}]\,.
\end{align}

The free $q$-hypermultiplet action (\ref{Sq-free}) varies under the local superconformal transformations (\ref{delta-q-conformal}) as
\begin{equation}
      \delta S_q = \frac12 \int \mathrm{d}\zeta^{(-4)} q^{+a} [D^{++},\Lambda] q^+_a\,.
\end{equation}
In general, this variation is non-vanishing, unless we impose an additional constraint on the operator $\Lambda$:
\begin{equation}
  [D^{++},\Lambda] = \lambda^{\hat\alpha+}D^+_{\hat\alpha} + \lambda^{++}D^0\,.
\end{equation}
This equation implies the harmonic shortness constraints for the superfield $\lambda^M$:
\begin{equation}
  D^{++}\lambda^m = 0\,, \quad
  D^{++}\lambda^{+\hat\alpha} = 0\,, \quad
  D^{++}\lambda^{++} = 0\,.
\label{harmonic-shortness}
\end{equation}
As a result, the superfield $\lambda^m = \xi^m$ is harmonic-independent, $\lambda^{\hat\alpha+} = -\xi^{\hat\alpha+} = -u^+_i\xi^{\hat\alpha i}$ is linear in harmonic variables and $\lambda^{++} = -\tilde\Lambda^{ij}u^+_i u^+_j$ is just quadratic in $u^+_i$. Thus, equations (\ref{harmonic-shortness}) reduce the local superconformal transformations (\ref{delta-q-conformal}) to the rigid ones (\ref{general-conformal-transformation}). The action of the free $q$-hypermultiplet is invariant under the rigid superconformal transformations,
\begin{equation}
    \delta_\mathrm{conf} S_q =0\,.
\end{equation}

Below, we will show that the invariance of the $q$-hypermultiplet model under the local superconformal transformations (\ref{delta-q-conformal}) is achieved through coupling to the conformal supergravity background.

\subsection{Prepotentials for conformal supergravity}
\label{App:C3}

According to Refs.~\cite{Galperin:1987ek,KT}, prepotentials for conformal supergravity are identified with vielbein components of the covariant harmonic derivative 
\begin{equation}
    {\cal D}^{++} = D^{++} + H^{++}\,,
    \label{covariant-derivative}
\end{equation}
where
\begin{equation}
    H^{++} := h^{++M}D_M = h^{++ m}\partial_m + h^{++\hat\alpha +}D^-_{\hat\alpha} + h^{(+4)}D^{--}\,.
    \label{H++conformal}
\end{equation}
By definition, when acting on an analytic superfield $\Phi_A$, this operator should preserve analyticity, $D^+_{\hat\alpha}(H^{++}\Phi_A) = 0$. As a corollary, the prepotentials $h^{++M}$ obey the constraints 
\begin{subequations}
  \label{h-analyticity-conformal}
\begin{align}
  D^+_{\hat\alpha} h^{++ m} - 2\mathrm{i}(\Gamma^m)_{\hat\alpha\hat\beta}h^{++\hat\beta+} &= 0\,, \\
  D^+_{\hat\alpha} h^{++\hat\beta+} - \delta_{\hat\alpha}{}^{\hat\beta} h^{(+4)} &= 0\,,\\
  D^+_{\hat\alpha} h^{(+4)} &= 0\,.
\end{align}
\end{subequations}
The general solution to these constraints may be expressed via an unconstrained superfield $U$:
\begin{subequations}
\label{G-prepotential}
\begin{align}
  h^{++m} = &-\frac{\mathrm{i}}{4}(\Gamma^m)^{\hat\alpha\hat\beta}D^+_{\hat\alpha}D^+_{\hat\beta}U\,,\\
  h^{++\hat\alpha+} = &\,\frac18 D^{+\hat\alpha}(D^+)^2 U\,,\\
  h^{(+4)} = &\,(D^+)^4 U\,.
  \label{C30c}
\end{align}
\end{subequations}
This solution is given modulo the following gauge transformations of the unconstrained prepotential
\begin{equation}
  \delta_\rho U = (D^+)^2 \rho^{--}\,,
  \label{delta-rho-G}
\end{equation}
with some unconstrained superfield $\rho^{--}$. The prepotentials $h^{++M}$ remain invariant under (\ref{delta-rho-G}).

$\lambda$-gauge transformations of the supergravity prepotentials are derived from a supergravity gauge transformation of the covariant harmonic derivative, $\delta {\cal D}^{++} = [{\cal D}^{++},\Lambda] - \lambda^{++}D^0$:
\begin{equation}
  \delta_\lambda h^{++M} = {\cal D}^{++} \lambda^M -\Lambda h^{++M}\,,
  \label{h-lambda-conformal}
\end{equation}
where $\Lambda = \lambda^M D_M$, and the gauge parameters $\lambda^M$ obey the constraints (\ref{lambda-conf-analyticity}), but the harmonic shortness (\ref{harmonic-shortness}) is not imposed. The gauge transformations (\ref{h-lambda-conformal}) are generated by the following $\lambda$-gauge transformation of the unconstrained prepotential $U$:
\begin{equation}
  \delta_\lambda U = {\cal D}^{++} l^{--} - \Lambda U\,,
  \label{delta-lambda-G}
\end{equation}
where $l^{--}$ is an unconstrained gauge superfield introduced in Eqs.~(\ref{lambda-l-conformal}).
Relations (\ref{delta-rho-G}) and (\ref{delta-lambda-G}) determine the full conformal supergravity gauge group.

By analogy with the 4D case \cite{Galperin:1987ek,KT}, it is possible to impose the gauge 
\begin{equation}
    h^{(+4)} = 0\,,
\label{h-gauge}
\end{equation}
which restricts the residual gauge transformations in Eq.~(\ref{h-lambda-conformal}):
\begin{equation}
    {\cal D}^{++}\lambda^{++} = 0\,.
\end{equation}
In this gauge, equation (\ref{C30c}) implies
\begin{equation}
    U = D^+_{\hat\alpha}\Psi^{\hat\alpha-}\,,
\end{equation}
where $\Psi^{\hat\alpha-}$ is an unconstrained spinor superfield with $U(1)$ charge $-1$. This superfield coincides with one of the unconstrained prepotential introduced in Eqs.~(\ref{pre-prepotentials}) for Poincar\'e supergravity. In the case of 4D $\cN=2$ Poincar\'e supergravity in harmonic superspace this prepotential was introduced by Zupnik in Ref.~\cite{Zupnik:1998td}.

It is also possible to show that the gauge transformations (\ref{delta-lambda-G}) are sufficient for elimination of the infinite tower of auxiliary fields in $U$. This may be most easily seen in the linearized case, when the gauge transformation (\ref{delta-lambda-G}) is simply 
\begin{equation}
    \delta U = D^{++}l^{--}\,,
    \label{G-l-linear}
\end{equation}
and $D^{++}$ is given by Eq.~(\ref{D-harmonic-central}) in the central basis. In this basis, the superfields $U$ and $l^{--}$ are given by the following series over the harmonic variables:
\begin{align}
    U(x,\theta,u) = &\, {\mathfrak U}(x,\theta) + \sum_{k=1}^\infty U^{i_1\ldots i_k j_1\ldots j_k}(x,\theta) u^+_{(i_1}\ldots u^+_{i_k} u^-_{j_1}\ldots u^-_{j_k)}\,,\label{C.39}\\
    l^{--}(x,\theta,u) =& \sum_{k=1}^\infty l^{i_1\ldots i_{k-1}j_1\ldots j_{k+1}}(x,\theta) u^+_{(i_1}\ldots u^+_{i_{k-1}}u^-_{j_1}\ldots u^-_{j_{k+1})}\,.
\end{align}
Making use of Eq.~(\ref{G-l-linear}) it is possible to gauge away all the fields $U^{i_1\ldots i_k j_1\ldots j_k}$, $k=1,2,\ldots$, in the right-hand side of Eq.~(\ref{C.39}), and arrive at the gauge condition
\begin{equation}
D^{++}U = 0\,.
\label{G-gauge}
\end{equation}
In this gauge, the prepotential $U$ is harmonic-independent,
\begin{equation}
    U(x,\theta,u) = {\mathfrak U}(x,\theta)\,,
\end{equation}
with the residual gauge freedom
\begin{equation}
    \delta {\mathfrak U}(x,\theta) = \frac13 D^{i\hat\alpha}D^j_{\hat\alpha} \rho_{ij}(x,\theta)\,,
    \label{G-residual-freedom}
\end{equation}
where $\rho_{ij}(x\,\theta)$ is the lowest component in the harmonic decomposition of the gauge parameter $\rho^{--}$ in Eq.~(\ref{delta-rho-G}). The gauge transformation (\ref{G-residual-freedom}) is a linear combination of (\ref{delta-rho-G}) and (\ref{G-l-linear}) which preserves the gauge (\ref{G-gauge}).

The scalar superfield prepotential $\mathfrak U$ with gauge freedom of the form (\ref{G-residual-freedom}) was introduced in Ref.~\cite{Butter:2014xxa} for 5D conformal supergravity.

If one chooses to work with the gauge condition \eqref{h-gauge}, the residual gauge freedom makes it possible to bring Zupnik's prepotential $\Psi^{\hat\alpha-}$ to the form
\begin{equation}
\Psi^{\hat\alpha-}(x,\theta, u) = \psi^{\hat\alpha i}(x,\theta) u_i^-~,
\label{spinor gauge}
\end{equation}
where $\psi^{\hat\alpha i}$ is an unconstrained spinor superfield, which proves to be a 5D analogue of the Gates-Siegel prepotential for 4D $\cN=2$ supergravity \cite{Gates:1981qq}.


\subsection{Hypermultiplet coupling and supercurrent}
\label{AppendixC4}

Minimal coupling of the $q$-hypermultiplet to the conformal supergravity background has standard form: one replaces the flat harmonic derivative in Eq.~(\ref{Sq-free}) with the gauge-covariant one (\ref{covariant-derivative}). The corresponding cubic interaction vertex reads
\begin{equation}
    S_\mathrm{int} = -\frac12 \int \mathrm{d}\zeta^{(-4)} q^{+a} H^{++} q^+_a = -\frac12 \int \mathrm{d}\zeta^{(-4)} q^{+a} h^{++M}D_M q^+_a\,.
\label{Sint-conformal}
\end{equation}
Under the hypermultiplet transformations (\ref{delta-q-conformal}), the actions (\ref{Sq-free}) and (\ref{Sint-conformal}) vary as follows
\begin{align}
  \delta S_q = &\, \frac12 \int \mathrm{d}\zeta^{(-4)} q^{+a} [D^{++},\Lambda] q^+_a\,,\\
  \delta S_\mathrm{int} = & -\frac1{2} \int \mathrm{d}\zeta^{(-4)} q^{+a} \left( \delta H^{++} -\frac12 [H^{++},\Lambda]  \right)q^+_a\,.
\end{align}
These two variations cancel one another in the total action $S=S_q + S_\mathrm{int}$ provided the supergravity prepotentials transform by the rule (\ref{h-lambda-conformal}). Note that the harmonic shortness constraints (\ref{harmonic-shortness}) are not imposed.

Recall that the harmonic prepotentials $h^{++M}$ are expressed via the unconstrained prepotential $U$ as in Eqs.~(\ref{G-prepotential}). These equations allow us to prove the identity
\begin{equation}
  H^{++} q^+_a = (D^+)^4(U D^{--}q^+_a)\,,
\end{equation}
and represent the action (\ref{Sint-conformal}) in the form of the integral over full superspace in the case of linearized supergravity (further we restrict ourselves to the case when the prepotential $U$ describes the Weyl multiplet in flat harmonic superspace)
\begin{equation}
  S_\mathrm{int} = -\frac12 \int \mathrm{d}\zeta^{(-4)} q^{+a} (D^+)^4(U D^{--}q^+_a) = -\frac12 \int \mathrm{d}^{5|8}z\, U q^{+a} D^{--}q^+_a\,.
\label{S-int-G}
\end{equation}
Variation of this action with respect to $U$ yields the conserved supercurrent in the free hypermultiplet model:
\begin{equation}
  {\cal J} = \frac{\delta S_\mathrm{int}}{\delta U} = -\frac12 q^{+a} D^{--} q^+_a\,.
  \label{J-current-conformal}
\end{equation}
The same supercurrent was obtained in Sec.~\ref{Sec:supercurrent}, see Eq.~(\ref{J}). The hypermultiplet coupling to the conformal supergravity background with the scalar superfield prepotential $U$ given in this section explains why the master supercurrent in the hypermultiplet model is a scalar superfield $\cal J$.

In the linearized case, the $\lambda$-gauge transformation of the prepotential (\ref{delta-lambda-G}) is simply $\delta_\lambda U = D^{++} l^{--}$ with an unconstrained superfield parameter $l^{--}$. This gauge symmetry implies one of the conservation laws for the supercurrent,
\begin{equation}
    D^{++}{\cal J} = 0\,.
\end{equation}
The other conservation law follows from the $\rho$-gauge symmetry (\ref{harmonic-shortness}):
\begin{equation}
    (D^+)^2 {\cal J} =0\,.
\end{equation}
For the massive hypermultiplet model, this conservation law gets deformed as in Eq.~(\ref{J-T-conservation}).



\begin{thebibliography}{102}

\bibitem{Pashnev:1989}
A.~I.~Pashnev,
``Composite systems and field theory for a free Regge trajectory,''
\href{https://doi.org/10.1007/BF01017664}{Theor.\ Math.\ Phys.\ \textbf{78}, 272 (1989)} [Teor.\ Mat.\ Fiz.\ [\textbf{78}, 384 (1989)].

\bibitem{Klishevich:1997pd}
S.~M.~Klishevich and Y.~M.~Zinoviev,
``On electromagnetic interaction of massive spin-2 particle,''
Phys. Atom. Nucl. \textbf{61}, 1527 (1998)
[\href{https://arxiv.org/abs/hep-th/9708150}{arXiv:hep-th/9708150}].

\bibitem{Klishevich:1998yt}
S.~M.~Klishevich,
``Massive fields of arbitrary half integer spin in constant electromagnetic field,''
\href{https://doi.org/10.1142/S0217751X00000306}{Int. J. Mod. Phys. A \textbf{15}, 609 (2000)}
[\href{https://arxiv.org/abs/hep-th/9811030}{arXiv:hep-th/9811030}].

\bibitem{Zinoviev:2001}
Y.~M.~Zinoviev,
``On massive high spin particles in (A)dS,''
[\href{https://arxiv.org/abs/hep-th/0108192}{arXiv:hep-th/0108192}].

\bibitem{Buchbinder:2006nu}
I.~L.~Buchbinder, V.~A.~Krykhtin, L.~L.~Ryskina and H.~Takata,
``Gauge invariant Lagrangian construction for massive higher spin fermionic fields,''
\href{https://doi.org/10.1016/j.physletb.2006.08.060}{Phys. Lett. B \textbf{641}, 386 (2006)}
[\href{https://arxiv.org/abs/hep-th/0603212}{arXiv:hep-th/0603212}].

\bibitem{Metsaev}
R.~R.~Metsaev,
``Gauge-invariant formulation of massive totally symmetric fermionic fields in (A)dS space,''
\href{https://doi.org/10.1016/j.physletb.2006.11.002}{Phys. Lett. B \textbf{643}, 205 (2006)}
[\href{https://arxiv.org/abs/hep-th/0609029}{arXiv:hep-th/0609029}].

\bibitem{Buchbinder:2008ss}
I.~L.~Buchbinder and A.~V.~Galajinsky,
``Quartet unconstrained formulation for massive higher spin fields,''
\href{https://doi.org/10.1088/1126-6708/2008/11/081}{JHEP \textbf{11}, 081 (2008)}
[\href{https://arxiv.org/abs/0810.2852}{arXiv:0810.2852 [hep-th]}].

\bibitem{Lindwasser}
L.~W.~Lindwasser,
``Covariant actions and propagators for all spins, masses, and dimensions,''
\href{https://doi.org/10.1103/PhysRevD.109.085010}{Phys. Rev. D \textbf{109}, 085010 (2024)}
[\href{https://arxiv.org/abs/2307.11750}{arXiv:2307.11750 [hep-th]}].

\bibitem{Zinoviev:2007js}
Y.~M.~Zinoviev,
``Massive N=1 supermultiplets with arbitrary superspins,''
\href{https://doi.org/10.1016/j.nuclphysb.2007.06.008}{Nucl. Phys. B \textbf{785}, 98 (2007)}
[\href{https://arxiv.org/abs/0704.1535}{arXiv:0704.1535 [hep-th]}].

\bibitem{Buchbinder:2015mta}
I.~L.~Buchbinder, T.~V.~Snegirev and Y.~M.~Zinoviev,
``Lagrangian formulation of the massive higher spin supermultiplets in three dimensional space-time,''
\href{https://doi.org/10.1007/JHEP10(2015)148}{JHEP \textbf{10}, 148 (2015)}
[\href{https://arxiv.org/abs/1508.02829}{arXiv:1508.02829 [hep-th]}].

\bibitem{Buchbinder:2017izy}
I.~L.~Buchbinder, T.~V.~Snegirev and Y.~M.~Zinoviev,
``Lagrangian description of massive higher spin supermultiplets in AdS$_{3}$ space,''
\href{https://doi.org/10.1007/JHEP08(2017)021}{JHEP \textbf{08}, 021 (2017)}
[\href{https://arxiv.org/abs/1705.06163}{arXiv:1705.06163 [hep-th]}].

\bibitem{Buchbinder:2019dof}
I.~L.~Buchbinder, M.~V.~Khabarov, T.~V.~Snegirev and Y.~M.~Zinoviev,
``Lagrangian formulation of the massive higher spin N=1 supermultiplets in $AdS_4$ space,''
\href{https://doi.org/10.1016/j.nuclphysb.2019.03.011}{Nucl. Phys. B \textbf{942}, 1 (2019)}
[\href{https://arxiv.org/abs/1901.09637}{arXiv:1901.09637 [hep-th]}].

\bibitem{WB} J.~Wess and J.~Bagger,
{\it Supersymmetry and Supergravity},
Princeton Univ. Press, Princeton,1992.

\bibitem{GIOS}
A.~S.~Galperin, E.~A.~Ivanov, V.~I.~Ogievetsky and E.~S.~Sokatchev,
{\it Harmonic Superspace}, Cambridge University Press, Cambridge, 2001.

\bibitem{FVP}
D.~Z.~Freedman and A.~Van~Proeyen,
\emph{Supergravity}, Cambridge University Press (2012).

\bibitem{Lauria:2020rhc}
E.~Lauria and A.~Van Proeyen,
``${\cal N}=2$ Supergravity in $D=4,5,6$ Dimensions,''
\href{https://doi.org/10.1007/978-3-030-33757-5}{Lect. Notes Phys. \textbf{966} (2020)}
[\href{https://arxiv.org/abs/2004.11433}{arXiv:2004.11433 [hep-th]}].

\bibitem{Aragone:1987dtt}
C.~Aragone, S.~Deser and Z.~Yang,
``Massive higher spin from dimensional reduction of gauge fields,''
\href{https://doi.org/10.1016/S0003-4916(87)80005-2}{Annals Phys. \textbf{179}, 76 (1987).}

\bibitem{Fronsdal:1978}
C.~Fronsdal,
``Massless fields with integer spin,''
\href{https://doi.org/10.1103/PhysRevD.18.3624}{Phys. Rev. D \textbf{18}, 3624 (1978).}

\bibitem{Curtright:1979uz}
T.~Curtright,
``Massless field supermultiplets with arbitrary spin,''
\href{https://doi.org/10.1016/0370-2693(79)90583-5}{Phys. Lett. B \textbf{85}, 219 (1979).}

\bibitem{FF}
J.~Fang and C.~Fronsdal,
``Massless fields with half-integral spin,''
\href{https://doi.org/10.1103/PhysRevD.18.3630}{Phys.\ Rev.\  D {\bf 18}, 3630 (1978).}

\bibitem{Rindani:1988gb}
S.~D.~Rindani, D.~Sahdev and M.~Sivakumar,
``Dimensional reduction of symmetric higher spin actions. 1. Bosons,''
\href{https://doi.org/10.1142/S0217732389000332}{Mod. Phys. Lett. A \textbf{4}, 265 (1989).}

\bibitem{Rindani:1989ym}
S.~D.~Rindani, M.~Sivakumar and D.~Sahdev,
``Dimensional reduction of symmetric higher spin actions. 2: Fermions,''
\href{https://doi.org/10.1142/S0217732389000344}{Mod. Phys. Lett. A \textbf{4}, 275 (1989).}

\bibitem{Asano:2019smc}
M.~Asano,
``Minimal gauge invariant and gauge fixed actions for massive higher-spin fields,''
\href{https://doi.org/10.1007/JHEP04(2019)051}{JHEP \textbf{04}, 051 (2019)}
[\href{https://arxiv.org/abs/1902.05685}{arXiv:1902.05685 [hep-th]}].

\bibitem{Fegebank:2024yft}
A.~J.~Fegebank and S.~M.~Kuzenko,
``Equivalence of gauge-invariant models for massive integer-spin fields,''
\href{https://doi.org/10.1103/PhysRevD.110.105014}{Phys. Rev. D \textbf{110}, no.10, 105014 (2024)}
[\href{https://arxiv.org/abs/2406.02573}{arXiv:2406.02573 [hep-th]}].

\bibitem{KSP}
S.~M.~Kuzenko,  V.~V.~Postnikov and A.~G.~Sibiryakov,
``Massless gauge superfields of higher half-integer superspins,''
JETP Lett.\  {\bf 57},    534 (1993) 
[Pisma Zh.\ Eksp.\ Teor.\ Fiz.\  {\bf 57},  521 (1993)].
  
\bibitem{KS}
S.~M.~Kuzenko and A.~G.~Sibiryakov,
``Massless gauge superfields of higher integer superspins,''
JETP Lett.\ {\bf 57},   539 (1993)  
[Pisma Zh.\ Eksp.\ Teor.\ Fiz.\  {\bf 57}, 526 (1993)].

\bibitem{BK} I.~L.~Buchbinder and S.~M.~Kuzenko,
{\it Ideas and Methods of Supersymmetry and Supergravity or a Walk Through Superspace},
IOP, Bristol, 1998.

\bibitem{KuzenkoSibiryakov94}
S.~M.~Kuzenko and A.~G.~Sibiryakov,
``Free massless higher superspin superfields on the anti-de Sitter superspace,''
Phys. Atom. Nucl. \textbf{57}, 1257 (1994)
[\href{https://arxiv.org/abs/1112.4612}{arXiv:1112.4612 [hep-th]}].

\bibitem{Buchbinder:2018nkp}
E.~I.~Buchbinder, J.~Hutomo and S.~M.~Kuzenko,
``Higher spin supercurrents in anti-de Sitter space,''
\href{https://doi.org/10.1007/JHEP09(2018)027}{JHEP \textbf{09}, 027 (2018)}
[\href{https://arxiv.org/abs/1805.08055}{arXiv:1805.08055 [hep-th]}].

\bibitem{Gates:1996my}
S.~J.~Gates Jr., S.~M.~Kuzenko and A.~G.~Sibiryakov,
``N=2 supersymmetry of higher superspin massless theories,''
\href{https://doi.org/10.1016/S0370-2693(97)01037-X}{Phys. Lett. B \textbf{412}, 59 (1997)}
[\href{https://arxiv.org/abs/hep-th/9609141}{arXiv:hep-th/9609141}].

\bibitem{Gates:1996xs}
S.~J.~Gates Jr., S.~M.~Kuzenko and A.~G.~Sibiryakov,
``Towards a unified theory of massless superfields of all superspins,''
\href{https://doi.org/10.1016/S0370-2693(97)00034-8}{Phys. Lett. B \textbf{394}, 343 (1997)}
[\href{https://arxiv.org/abs/hep-th/9611193}{arXiv:hep-th/9611193}].

\bibitem{Butter:2011zt}
D.~Butter and S.~M.~Kuzenko,
``N=2 supersymmetric sigma-models in AdS,''
\href{https://doi.org/10.1016/j.physletb.2011.08.043}{Phys. Lett. B \textbf{703}, 620 (2011)}
[\href{https://arxiv.org/abs/1105.3111}{arXiv:1105.3111 [hep-th]}].

\bibitem{Butter:2011kf}
D.~Butter and S.~M.~Kuzenko,
``The structure of $\cN=2$ supersymmetric nonlinear sigma models in AdS$_4$,''
\href{https://doi.org/10.1007/JHEP11(2011)080}{JHEP \textbf{11}, 080 (2011)}
[\href{https://arxiv.org/abs/1108.5290}{arXiv:1108.5290 [hep-th]}].

\bibitem{Buchbinder:2021ite}
I.~Buchbinder, E.~Ivanov and N.~Zaigraev,
``Unconstrained off-shell superfield formulation of 4D, $\mathcal{N} $=2 supersymmetric higher spins,''
\href{https://doi.org/10.1007/JHEP12(2021)016}{JHEP \textbf{12}, 016 (2021)}
[\href{https://arxiv.org/abs/2109.07639}{arXiv:2109.07639 [hep-th]}].

\bibitem{Buchbinder:2019yhl}
E.~I.~Buchbinder, D.~Hutchings, J.~Hutomo and S.~M.~Kuzenko,
``Linearised actions for $\mathcal{N}$-extended (higher-spin) superconformal gravity,''
\href{https://doi.org/10.1007/JHEP08(2019)077}{JHEP \textbf{08}, 077 (2019)}
[\href{https://arxiv.org/abs/1905.12476}{arXiv:1905.12476 [hep-th]}].

\bibitem{PopeTownsend} 
C.~N.~Pope and P.~K.~Townsend,
``Conformal higher spin in (2+1) dimensions,''
\href{https://doi.org/10.1016/0370-2693(89)90813-7}{Phys.\ Lett.\ B {\bf 225}, 245 (1989)}.

\bibitem{Kuzenko:2016qdw}
S.~M.~Kuzenko,
``Higher spin super-Cotton tensors and generalisations of the linear-chiral duality in three dimensions,'' \href{https://doi.org/10.1016/j.physletb.2016.10.071}{Phys.\ Lett.\ B {\bf 763}, 308 (2016)}
[\href{https://arxiv.org/abs/1606.08624}{arXiv:1606.08624 [hep-th]}].

\bibitem{Kuzenko:2016qwo}
S.~M.~Kuzenko and M.~Tsulaia,
``Off-shell massive $\cN=1$ supermultiplets in three dimensions,''
\href{https://doi.org/10.1016/j.nuclphysb.2016.10.023}{Nucl. Phys. B \textbf{914}, 160 (2017)}
[\href{https://arxiv.org/abs/1609.06910}{arXiv:1609.06910 [hep-th]}].

\bibitem{KO} 
S.~M.~Kuzenko and D.~X.~Ogburn,
``Off-shell higher spin $\cN=2$ supermultiplets in three dimensions,''
\href{https://doi.org/10.1103/PhysRevD.94.106010}{Phys.\ Rev.\ D {\bf 94}, no. 10, 106010 (2016)}
[\href{https://arxiv.org/abs/1603.04668}{arXiv:1603.04668 [hep-th]}].
        
\bibitem{WS}
W.~Siegel,
``Unextended superfields in extended supersymmetry,"
\href{https://doi.org/10.1016/0550-3213(79)90498-X}{Nucl. Phys. B {\bf 156}, 135 (1979).}

\bibitem{JS}
J.~F.~Schonfeld, 
``A mass term for three-dimensional gauge fields,"
\href{https://doi.org/10.1016/0550-3213(81)90369-2}{Nucl. Phys. B {\bf 185}, 157 (1980).}
		
\bibitem{DJT1}
S.~Deser, R.~Jackiw and S.~Templeton,
``Three-dimensional massive gauge theories,''
\href{https://doi.org/10.1103/PhysRevLett.48.975}{Phys.\ Rev.\ Lett.\  {\bf 48}, 975 (1982).}

\bibitem{DJT2}
S.~Deser, R.~Jackiw and S.~Templeton,
``Topologically massive gauge theories,''
\href{https://doi.org/10.1016/0003-4916(82)90164-6}{Annals Phys.\ {\bf 140}, 372 (1982)}
[\href{https://doi.org/10.1016/0003-4916(88)90053-X}{Erratum-ibid.\ {\bf 185}, 406 (1988)}].

\bibitem{DeserKay}
S.~Deser and J.~H.~Kay
``Topologically massive supergravity,''
\href{https://doi.org/10.1016/0370-2693(83)90631-7}{Phys.\ Lett.\ B {\bf 120}, 97 (1983).}


  
\bibitem{Kuzenko:2018lru}
S.~M.~Kuzenko and M.~Ponds,
``Topologically massive higher spin gauge theories,''
\href{https://doi.org/10.1007/JHEP10(2018)160}{JHEP \textbf{10}, 160 (2018)}
[\href{https://arxiv.org/abs/1806.06643}{arXiv:1806.06643 [hep-th]}].
		
\bibitem{HK19}
J.~Hutomo and S.~M.~Kuzenko, ``Field theories with (2,0) AdS supersymmetry in ${\cal N}=1$ AdS superspace,''
\href{https://doi.org/10.1103/PhysRevD.100.045010}{Phys. Rev. D \textbf{100}, 045010 (2019)}
[\href{https://arxiv.org/abs/1905.05050}{arXiv:1905.05050 [hep-th]}].

\bibitem{Kuzenko:2021hyd}
S.~M.~Kuzenko and M.~Ponds,
``Higher-spin Cotton tensors and massive gauge-invariant actions in AdS$_3$,''
\href{https://doi.org/10.1007/JHEP05(2021)275}{JHEP \textbf{05}, 275 (2021)}
[\href{https://arxiv.org/abs/2103.11673}{arXiv:2103.11673 [hep-th]}].

\bibitem{HKO}
J.~Hutomo, S.~M.~Kuzenko, D.~Ogburn,
``${\cal N} = 2$ supersymmetric higher spin gauge theories and current multiplets in three dimensions,'' 
\href{https://doi.org/10.1103/PhysRevD.98.125004}{Phys.\ Rev. \ D {\bf 98}, 125004 (2018)} [\href{https://arxiv.org/abs/1807.09098}{arXiv:1807.09098 [hep-th]}].

\bibitem{HK18}
J.~Hutomo and S.~M.~Kuzenko,
``Higher spin supermultiplets in three dimensions: (2,0) AdS supersymmetry,''
\href{https://doi.org/10.1016/j.physletb.2018.10.060}{Phys. Lett. B \textbf{787}, 175 (2018)}
[\href{https://arxiv.org/abs/1809.00802}{arXiv:1809.00802 [hep-th]}].



\bibitem{KMT} 
S.~M.~Kuzenko, R.~Manvelyan and S.~Theisen,
``Off-shell superconformal higher spin multiplets in four dimensions,''
\href{https://doi.org/10.1007/JHEP07(2017)034}{JHEP {\bf 1707}, 034 (2017)}
[\href{https://arxiv.org/abs/1701.00682}{arXiv:1701.00682 [hep-th]}].

\bibitem{KP}
S.~M.~Kuzenko and M.~Ponds,
``Conformal geometry and (super)conformal higher-spin gauge theories,''
\href{https://doi.org/10.1007/JHEP05(2019)113}{JHEP \textbf{05}, 113 (2019)}
[\href{https://arxiv.org/abs/1902.08010}{arXiv:1902.08010 [hep-th]}].

\bibitem{Kuzenko:2020jie}
S.~M.~Kuzenko, M.~Ponds and E.~S.~N.~Raptakis,
``New locally (super)conformal gauge models in Bach-flat backgrounds,''
\href{https://doi.org/10.1007/JHEP08(2020)068}{JHEP \textbf{08}, 068 (2020)}
[\href{https://arxiv.org/abs/2005.08657}{arXiv:2005.08657 [hep-th]}].

\bibitem{Kuzenko:2021pqm}
S.~M.~Kuzenko and E.~S.~N.~Raptakis,
``Extended superconformal higher-spin gauge theories in four dimensions,''
\href{https://doi.org/10.1007/JHEP12(2021)210}{JHEP \textbf{12}, 210 (2021)}
[\href{https://arxiv.org/abs/2104.10416}{arXiv:2104.10416 [hep-th]}].

\bibitem{HKR}
D.~Hutchings, S.~M.~Kuzenko and E.~S.~N.~Raptakis,
``The $\cN=2$ superconformal gravitino multiplet,''
\href{https://doi.org/10.1016/j.physletb.2023.138132}{Phys. Lett. B \textbf{845}, 138132 (2023)}
[\href{https://arxiv.org/abs/2305.16029}{arXiv:2305.16029 [hep-th]}].

\bibitem{Kuzenko:2024vms}
S.~M.~Kuzenko and E.~S.~N.~Raptakis,
``Towards $\mathcal{N}$=2 superconformal higher-spin theory,''
\href{https://doi.org/10.1007/JHEP11(2024)013}{JHEP \textbf{11}, 013 (2024)}
[\href{https://arxiv.org/abs/2407.21573}{arXiv:2407.21573 [hep-th]}].

\bibitem{BIZ24}
I.~Buchbinder, E.~Ivanov and N.~Zaigraev,
``$\mathcal{N}$=2 superconformal higher-spin multiplets and their hypermultiplet couplings,''
\href{https://doi.org/10.1007/JHEP08(2024)120}{JHEP \textbf{08}, 120 (2024)}
[\href{https://arxiv.org/abs/2404.19016}{arXiv:2404.19016 [hep-th]}].

\bibitem{Ivanov:2024bsb}
E.~Ivanov and N.~Zaigraev,
``$\cN=2$ superconformal gravitino in harmonic superspace,''
\href{https://doi.org/10.1016/j.physletb.2025.139333}{Phys. Lett. B \textbf{862}, 139333 (2025)}
[\href{https://arxiv.org/abs/2412.14822}{arXiv:2412.14822 [hep-th]}].

\bibitem{Koutrolikos:2020tel}
K.~Koutrolikos,
``Superspace formulation of massive half-integer superspin,''
\href{https://doi.org/10.1007/JHEP03(2021)254}{JHEP \textbf{03}, 254 (2021)}
[\href{https://arxiv.org/abs/2012.12225}{arXiv:2012.12225 [hep-th]}].

\bibitem{Singh:1974}
L.~P.~S.~Singh and C.~R.~Hagen,
``Lagrangian formulation for arbitrary spin. 1. The boson case,''
\href{https://doi.org/10.1103/PhysRevD.9.898}{Phys. Rev. D \textbf{9}, 898 (1974).}

\bibitem{SinghHagen2}
L.~P.~S.~Singh and C.~R.~Hagen,
``Lagrangian formulation for arbitrary spin. 2. The fermion case,''
\href{https://doi.org/10.1103/PhysRevD.9.910}{Phys. Rev. D \textbf{9}, 910 (1974).}

\bibitem{Ohashi1}
T.~Kugo and K.~Ohashi,
``Supergravity tensor calculus in 5D from 6D,''
\href{https://doi.org/10.1143/PTP.104.835}{Prog.\ Theor.\ Phys.\  {\bf 104}, 835 (2000)}
[\href{https://arxiv.org/abs/hep-ph/0006231}{arXiv:hep-ph/0006231}].

\bibitem{Ohashi2}
T.~Kugo and K.~Ohashi,
``Off-shell d=5 supergravity coupled to matter-Yang-Mills system,''
\href{https://doi.org/10.1143/PTP.105.323}{Prog.\ Theor.\ Phys.\  {\bf 105}, 323 (2001)}
[\href{https://arxiv.org/abs/hep-ph/0010288}{arXiv:hep-ph/0010288}].

\bibitem{Ohashi3} 
T.~Fujita and K.~Ohashi,
``Superconformal tensor calculus in five dimensions,''
\href{https://doi.org/10.1143/PTP.106.221}{Prog.\ Theor.\ Phys.\  {\bf 106}, 221 (2001)}
[\href{https://arxiv.org/abs/hep-th/0104130}{arXiv:hep-th/0104130}].

\bibitem{Ohashi4} 
 T.~Kugo and K.~Ohashi,
``Gauge and non-gauge tensor multiplets in 5D conformal supergravity,''
\href{https://doi.org/10.1143/PTP.108.1143}{Prog.\ Theor.\ Phys.\  {\bf 108}, 1143 (2003)}
[\href{https://arxiv.org/abs/hep-th/0208082}{arXiv:hep-th/0208082}].
  
\bibitem{Bergshoeff1}
E.~Bergshoeff, S.~Cucu, M.~Derix, T.~de Wit, R.~Halbersma and A.~Van Proeyen,
``Weyl multiplets of N=2 conformal supergravity in five-dimensions,''
\href{https://doi.org/10.1088/1126-6708/2001/06/051}{JHEP {\bf 0106}, 051 (2001)}
[\href{https://arxiv.org/abs/hep-th/0104113}{arXiv:hep-th/0104113}].
  
\bibitem{Bergshoeff2}
E.~Bergshoeff, S.~Cucu, T.~de Wit, J.~Gheerardyn, R.~Halbersma, 
S.~Vandoren and A.~Van Proeyen,
``Superconformal N=2, D=5 matter with and without actions,''
\href{https://doi.org/10.1088/1126-6708/2002/10/045}{JHEP {\bf 0210}, 045 (2002)}
[\href{https://arxiv.org/abs/hep-th/0205230}{arXiv:hep-th/0205230}].

\bibitem{Bergshoeff3}  
E.~Bergshoeff, S.~Cucu, T.~de Wit, J.~Gheerardyn, 
S.~Vandoren and A.~Van Proeyen,
``N=2 supergravity in five dimensions revisited,''
\href{https://doi.org/10.1088/0264-9381/23/23/C01}{Class.\ Quant.\ Grav.\  {\bf 21}, 3015 (2004)}
[\href{https://arxiv.org/abs/hep-th/0403045}{arXiv:hep-th/0403045}]. 

\bibitem{KT-M_5D2}
S.~M.~Kuzenko and G.~Tartaglino-Mazzucchelli,
``Five-dimensional superfield supergravity,''
\href{https://doi.org/10.1016/j.physletb.2008.01.055}{Phys.\ Lett.\  B {\bf 661}, 42 (2008)}
[\href{https://arxiv.org/abs/0710.3440}{arXiv:0710.3440 [hep-th]}];
 
\bibitem{KT-M_5D3} 
S.~M.~Kuzenko and G.~Tartaglino-Mazzucchelli,
``5D supergravity and projective superspace,''
\href{https://doi.org/10.1088/1126-6708/2008/02/004}{JHEP {\bf 0802}, 004 (2008)}
[\href{https://arxiv.org/abs/0712.3102}{arXiv:0712.3102 [hep-th]}].

\bibitem{KT-M08}
S.~M.~Kuzenko and G.~Tartaglino-Mazzucchelli,
``Super-Weyl invariance in 5D supergravity,''
\href{https://doi.org/10.1088/1126-6708/2008/04/032}{JHEP {\bf 0804}, 032 (2008)}
[\href{https://arxiv.org/abs/0802.3953}{arXiv:0802.3953 [hep-th]}].

\bibitem{K07}
S.~M.~Kuzenko, ``On superconformal projective hypermultiplets,''
\href{https://doi.org/10.1088/1126-6708/2007/12/010}{JHEP {\bf 0712}, 010 (2007)} [\href{https://arxiv.org/abs/0710.1479}{arXiv:0710.1479 [hep-th]}].

\bibitem{Kuzenko2006}
S.~M.~Kuzenko,
``On compactified harmonic/projective superspace, 5D superconformal theories, and all that,''
\href{https://doi.org/10.1016/j.nuclphysb.2006.03.019}{Nucl. Phys. B \textbf{745}, 176 (2006)}
[\href{https://arxiv.org/abs/hep-th/0601177}{arXiv:hep-th/0601177}].

\bibitem{KLR}
A. Karlhede, U. Lindstr\"om and M. Ro\v cek,
``Self-interacting tensor multiplets in N=2 superspace,''
\href{https://doi.org/10.1016/0370-2693(84)90120-5}{Phys.\ Lett.\ B {\bf 147}, 297 (1984).}

\bibitem{LR1}
U.~Lindstr\"om and M.~Ro\v{c}ek,
``New hyperk\"ahler  metrics  and new supermultiplets,''
\href{https://doi.org/10.1007/BF01238851}{Commun.\ Math.\ Phys.\  {\bf 115}, 21 (1988).}
  
\bibitem{LR2}
U.~Lindstr\"om and M.~Ro\v{c}ek,
``N=2 super Yang-Mills theory in projective superspace,''
\href{https://doi.org/10.1007/BF02097052}{Commun.\ Math.\ Phys.\  {\bf 128}, 191 (1990).}

\bibitem{KL}
S.~M.~Kuzenko and W.~D.~Linch III,
``On five-dimensional superspaces,''
\href{https://doi.org/10.1088/1126-6708/2006/02/038}{JHEP {\bf 0602}, 038 (2006)}
[\href{https://arxiv.org/abs/hep-th/0507176}{arXiv:hep-th/0507176}].
  
\bibitem{Kuzenko:2007aj}
S.~M.~Kuzenko and G.~Tartaglino-Mazzucchelli,
``Five-dimensional $\cN=1$ AdS superspace: Geometry, off-shell multiplets and dynamics,''
\href{https://doi.org/10.1016/j.nuclphysb.2007.06.014}{Nucl. Phys. B \textbf{785}, 34 (2007)}
[\href{https://arxiv.org/abs/0704.1185}{arXiv:0704.1185 [hep-th]}].

\bibitem{Butter:2014xxa}
D.~Butter, S.~M.~Kuzenko, J.~Novak and G.~Tartaglino-Mazzucchelli,
``Conformal supergravity in five dimensions: New approach and applications,''
\href{https://doi.org/10.1007/JHEP02(2015)111}{JHEP \textbf{02}, 111 (2015)}
[\href{https://arxiv.org/abs/1410.8682}{arXiv:1410.8682 [hep-th]}].

\bibitem{Hutomo:2022hdi}
J.~Hutomo, S.~Khandelwal, G.~Tartaglino-Mazzucchelli and J.~Woods,
``Hyperdilaton Weyl multiplets of 5D and 6D minimal conformal supergravity,''
\href{https://doi.org/10.1103/PhysRevD.107.046009}{Phys. Rev. D \textbf{107}, no.4, 046009 (2023)}
[\href{https://arxiv.org/abs/2209.05748}{arXiv:2209.05748 [hep-th]}].

\bibitem{Gold:2023dfe}
G.~Gold, J.~Hutomo, S.~Khandelwal and G.~Tartaglino-Mazzucchelli,
``Curvature-squared invariants of minimal five-dimensional supergravity from superspace,''
\href{https://doi.org/10.1103/PhysRevD.107.106013}{Phys. Rev. D \textbf{107}, no.10, 106013 (2023)}
[\href{https://arxiv.org/abs/2302.14295}{arXiv:2302.14295 [hep-th]}].

\bibitem{Gold:2023ymc}
G.~Gold, J.~Hutomo, S.~Khandelwal, M.~Ozkan, Y.~Pang and G.~Tartaglino-Mazzucchelli,
``All gauged curvature-squared supergravities in five dimensions,''
\href{https://doi.org/10.1103/PhysRevLett.131.251603}{Phys. Rev. Lett. \textbf{131}, no.25, 251603 (2023)}
[\href{https://arxiv.org/abs/2309.07637}{arXiv:2309.07637 [hep-th]}].

\bibitem{Gold:2023ykx}
G.~Gold, J.~Hutomo, S.~Khandelwal and G.~Tartaglino-Mazzucchelli,
``Components of curvature-squared invariants of minimal supergravity in five dimensions,''
\href{https://doi.org/10.1007/JHEP07(2024)221}{JHEP \textbf{07}, 221 (2024)}
[\href{https://arxiv.org/abs/2311.00679}{arXiv:2311.00679 [hep-th]}].

\bibitem{Gold:2025ttt}
G.~Gold, P.~J.~Hu, J.~Hutomo, S.~Khandelwal, M.~Ozkan, Y.~Pang and G.~Tartaglino-Mazzucchelli,
``On five-dimensional curvature squared supergravity and holography,''
[\href{https://arxiv.org/abs/2504.10856}{arXiv:2504.10856 [hep-th]}].

\bibitem{HST}
P.~S.~Howe, K.~S.~Stelle and P.~K.~Townsend,
``Supercurrents,''
\href{https://doi.org/10.1016/0550-3213(81)90429-6}{Nucl. Phys. B \textbf{192}, 332 (1981).}

\bibitem{KT}
S.~M.~Kuzenko and S.~Theisen,
``Correlation functions of conserved currents in $\cN=2$ superconformal theory,''
\href{https://doi.org/10.1088/0264-9381/17/3/307}{Class.\ Quant.\ Grav.\ {\bf 17}, 665 (2000)} [\href{https://arxiv.org/abs/hep-th/9907107}{arXiv:hep-th/9907107}].

\bibitem{Butter:2010sc}
D.~Butter and S.~M.~Kuzenko,
``$\cN=2$ supergravity and supercurrents,''
\href{https://doi.org/10.1007/JHEP12(2010)080}{JHEP \textbf{12}, 080 (2010)}
[\href{https://arxiv.org/abs/1011.0339}{arXiv:1011.0339 [hep-th]}].

\bibitem{GIKOS}
A.~Galperin, E.~Ivanov, S.~Kalitsyn, V.~Ogievetsky 
and E.~Sokatchev,
``Unconstrained $\cN=2$ matter, Yang-Mills and supergravity theories in harmonic superspace,''
\href{https://doi.org/10.1088/0264-9381/1/5/004}
{Class.\ Quant.\ Grav.\ {\bf 1}, 469 (1984)}.

\bibitem{Galperin:1987em}
A.~S.~Galperin, N.~A.~Ky and E.~Sokatchev,
``$\cN=2$ supergravity in superspace: Solution to the constraints,''
\href{https://doi.org/10.1088/0264-9381/4/5/022}{Class. Quant. Grav. \textbf{4}, 1235 (1987).}

\bibitem{Galperin:1987ek}
A.~S.~Galperin, E.~A.~Ivanov, V.~I.~Ogievetsky and E.~Sokatchev,
``$\cN=2$ supergravity in superspace: Different versions and matter couplings,''
\href{https://doi.org/10.1088/0264-9381/4/5/023}{Class. Quant. Grav. \textbf{4}, 1255 (1987).}

\bibitem{Buchbinder:2022vra}
I.~Buchbinder, E.~Ivanov and N.~Zaigraev,
``$\mathcal{N}$=2 higher spins: superfield equations of motion, the hypermultiplet supercurrents, and the component structure,''
\href{https://doi.org/10.1007/JHEP03(2023)036}{JHEP \textbf{03}, 036 (2023)}
[\href{https://arxiv.org/abs/2212.14114}{arXiv:2212.14114 [hep-th]}].

\bibitem{Buchbinder:2022kzl}
I.~Buchbinder, E.~Ivanov and N.~Zaigraev,
``Off-shell cubic hypermultiplet couplings to $ \mathcal{N} $=2 higher spin gauge superfields,''
\href{https://doi.org/10.1007/JHEP05(2022)104}{JHEP \textbf{05}, 104 (2022)}
[\href{https://arxiv.org/abs/2202.08196}{arXiv:2202.08196 [hep-th]}].

\bibitem{Butter:2015nza}
D.~Butter,
``On conformal supergravity and harmonic superspace,''
\href{https://doi.org/10.1007/JHEP03(2016)107}{JHEP \textbf{03}, 107 (2016)}
[\href{https://arxiv.org/abs/1508.07718}{arXiv:1508.07718 [hep-th]}].

\bibitem{deWvanHvanP}
B.~de Wit, J.~W.~van Holten and A.~Van Proeyen,
``Structure of N=2 Supergravity,''
\href{https://doi.org/10.1016/0550-3213(83)90548-5}{Nucl. Phys. B \textbf{184}, 77 (1981)}
[erratum: \href{https://doi.org/10.1016/0550-3213(81)90211-X}{Nucl. Phys. B \textbf{222}, 516 (1983)}].

\bibitem{Z2}
B.~M.~Zupnik,
``The action of the supersymmetric $\cN=2$ gauge theory in harmonic superspace,''
\href{https://doi.org/10.1016/0370-2693(87)90433-3}{Phys.\ Lett.\ B {\bf 183}, 175 (1987).}

\bibitem{Kuzenko:2025bud}
S.~M.~Kuzenko and E.~S.~N.~Raptakis,
``$\mathcal{N}=3$ nonlinear multiplet and supergravity,''
\href{https://doi.org/10.1007/JHEP07(2025)262}{JHEP \textbf{07}, 262 (2025)}
[\href{https://arxiv.org/abs/2501.11339}{arXiv:2501.11339 [hep-th]}].


\bibitem{Castellani:1980cu}
L.~Castellani, P.~van Nieuwenhuizen and S.~J.~Gates, Jr.,
``Constraints for $N$=2 Superspace From Extended Supergravity in Ordinary Space,''
Phys. Rev. D \textbf{22}, 2364 (1980).

\bibitem{Gates}
S.~J.~Gates, Jr.,
``Supercovariant Derivatives, Super Weyl Groups, and $N$=2 Supergravity,''
Nucl. Phys. B \textbf{176}, 397 (1980).

\bibitem{Howe}
P.~S.~Howe,
``Supergravity in Superspace,''
Nucl. Phys. B \textbf{199}, 309 (1982).

\bibitem{Zupnik:1998td}
B.~M.~Zupnik,
``Background harmonic superfields in N=2 supergravity,''
\href{https://doi.org/10.1007/BF02557138}{Theor. Math. Phys. \textbf{116}, 964 (1998)}
[\href{https://arxiv.org/abs/hep-th/9803202}{arXiv:hep-th/9803202}].

\bibitem{Galperin:1985bj}
A.~Galperin, E.~A.~Ivanov, V.~Ogievetsky and E.~Sokatchev,
``Harmonic supergraphs. Green functions,''
\href{https://doi.org/10.1088/0264-9381/2/5/004}{Class. Quant. Grav. \textbf{2}, 601 (1985).}

\bibitem{Z}
B.~Zupnik,
``Harmonic superpotentials and symmetries in gauge theories with eight supercharges,''
\href{https://doi.org/10.1016/S0550-3213(99)00267-9}{Nucl.\ Phys.\ B {\bf 554},  365 (1999)}
[\href{https://doi.org/10.1016/S0550-3213(02)00840-4}{Erratum-ibid.\ B {\bf 644},  405  (2002)}]
[\href{https://arxiv.org/abs/hep-th/9902038}{arXiv:hep-th/9902038}].



\bibitem{Ivanov:2024gjo}
E.~Ivanov and N.~Zaigraev,
``Off-shell invariants of linearized 4D, $\cN=2$ supergravity in the harmonic approach,''
\href{https://doi.org/10.1103/PhysRevD.110.066020}{Phys. Rev. D \textbf{110}, no.6, 066020 (2024)}
[\href{https://arxiv.org/abs/2407.08524}{arXiv:2407.08524 [hep-th]}].

\bibitem{Sohnius-supercurrent}
M.~F.~Sohnius,
``The multiplet of currents for $\cN=2$ extended supersymmetry,''
\href{https://doi.org/10.1016/0370-2693(79)90703-2}{Phys. Lett. B \textbf{81}, 8 (1979).}

\bibitem{Zucker:1999ej}
M.~Zucker,
``Minimal off-shell supergravity in five-dimensions,''
\href{https://doi.org/10.1016/S0550-3213(99)00750-6}{Nucl. Phys. B \textbf{570}, 267 (2000)}
[\href{https://arxiv.org/abs/hep-th/9907082}{arXiv:hep-th/9907082}].

\bibitem{Kuzenko:2023vgf}
S.~M.~Kuzenko and E.~S.~N.~Raptakis,
``On higher-spin $ \mathcal{N} = 2$ supercurrent multiplets,''
\href{https://doi.org/10.1007/JHEP05(2023)056}{JHEP \textbf{05}, 056 (2023)}
[\href{https://arxiv.org/abs/2301.09386}{arXiv:2301.09386 [hep-th]}].

\bibitem{Ivanov:2025jdp}
E.~Ivanov and N.~Zaigraev,
``$\mathcal{N}=2$ AdS hypermultiplets in harmonic superspace,''
[\href{https://arxiv.org/abs/2509.01406}{arXiv:2509.01406 [hep-th]}].

\bibitem{Mezincescu}
L.~Mezincescu,
``On the superfield formulation of O(2) supersymmetry,'' Dubna preprint JINR-P2-12572 (June, 1979).

\bibitem{Butter:2011ym}
D.~Butter and S.~M.~Kuzenko,
``$\cN=2$ AdS supergravity and supercurrents,''
\href{https://doi.org/10.1007/JHEP07(2011)081}{JHEP \textbf{07}, 081 (2011)}
[\href{https://arxiv.org/abs/1104.2153}{arXiv:1104.2153 [hep-th]}].

\bibitem{Gates:1981qq}
S.~J.~Gates, Jr. and W.~Siegel,
``Linearized N=2 superfield supergravity,''
\href{https://doi.org/10.1016/0550-3213(82)90047-5}{Nucl. Phys. B \textbf{195}, 39 (1982).}

\end{thebibliography}
\end{document}